\documentclass{thesis}
\usepackage{psfig,latexsym,mathrsfs,axodraw}
\newcommand{\Be}{{\mathcal B}}
\newcommand{\para}{\parallel}
\newcommand{\pr}{\perp}
\makeatletter
\@addtoreset{equation}{chapter}

\makeatother
\def\Tr{\mathop{\rm Tr}\nolimits}
\def\omit#1{_{\!\rlap{$\scriptscriptstyle \backslash$}
{\scriptscriptstyle #1}}}
\def\vec#1{\mathchoice 
	{\mbox{\boldmath $#1$}}
	{\mbox{\boldmath $#1$}}
	{\mbox{\boldmath $\scriptstyle #1$}}
	{\mbox{\boldmath $\scriptscriptstyle #1$}}
}
\def\eqn#1{Eq.\ (\ref{#1})}

\def\fig#1{Fig.~\ref{#1}}
\renewcommand{\bar}{\overline}
\newlength{\Notewidth}
\def\Note#1{\framebox{$\clubsuit$}%
\setlength{\Notewidth}{\marginparwidth}%
\setbox0=\hbox{#1}%
\ifdim \wd0 < \marginparwidth%
\setlength{\Notewidth}{\wd0}%
\fi%
\marginpar{\framebox{\parbox{\Notewidth}{\raggedright #1}}}%
\typeout{}%
\typeout{**************************}%
\typeout{!!! Marginal note here !!!}%
\typeout{**************************}%
\typeout{}%
}
\begin{document}
\ifx\href\undefined\else\hypersetup{linktocpage=true}\fi
\submissionmonth{October}
\submissionyear{2004} \author{\bf {KAUSHIK BHATTACHARYA} \\ THEORY
GROUP \\ SAHA INSTITUTE OF NUCLEAR PHYSICS \\ KOLKATA} \title{ELEMENTARY
PARTICLE INTERACTIONS \\IN A  BACKGROUND \\MAGNETIC FIELD} \maketitle

%
%
%
\begin{acknowledgements}
{\it I am indebted to Professor\, Palash Baran Pal for guidance,
encouragement, collaboration and numerous stimulating discussions. I
am also grateful to Dr.\,Avijit Kumar Ganguly and Dr.\, Sushan
Konar for collaboration and many valuable discussions.

Interactions with many other members, both past and present, of the
Theory Group of SINP, including Professor\, Binayak Dutta-Roy,
Professor\, Kumar Shankar Gupta, Professor\, Gautam Bhattacharya,
Professor\,Debadesh Bandyopadhya and Professor\, Samir Mallik, have helped
me to grasp the subtleties of the various phenomenological models and
some theoretical ideas employed in this thesis. In this regard I must
acknowledge my academic debts to Professor\, Amitava Roy Chowdhury of
Calcutta University. The weekly informal talks and seminars arranged
in his office room gave me glimpses of serious science and enough fun
to enjoy serious work.

I express my gratitude to all members of the Theory Group for creating
a congenial atmosphere of collective work and lively discussions.}
\end{acknowledgements}

%

\tableofcontents
\listoffigures

%
%

\chapter{Introduction and overview}\label{chap1}
\section{A brief overview on magnetic fields in the cosmos, their
observation, their origin and particle physics}
Magnetic fields are most frequently found in various length scales and
various positions of our universe. The average magnitude of the
Earth's magnetic field is about $10^{-1}$G. This magnetic field
encircles the Earth and produces a magnetosphere surrounding it. The
field is not uniform. Charged elementary particles from the cosmos
sometimes gets trapped in this magnetic field and hover over our Earth
in a belt shaped region which is popularly called the Van Allen
belt. Also the solar storms produce electromagnetic disturbances on
the magnetic field of the Earth and the particles trapped in the Van
Allen belt, resulting in `aurora borealis', one of the strangest
things seen by human beings.  In the solar black spots magnetic fields
can go up to $10^3$G. Very high magnetic fields are found in the cores
of supernovas, and this magnetic field remains frozen on the
proto-neutron star remnant whose surface magnetic field has been
measured to be $10^{13}$G. Not only the planets and the stars have
magnetic fields associated with them there are magnetic fields in the
galactic and intergalactic mediums. In Milky Way the average galactic
magnetic field is of the order of $3-4$G with a length scale of a few
Kilopersecs. Average intergalactic magnetic fields can be of order
$10^{-11}$G~\cite{obs1} with a length scale of about $1$
Megaparsec. Except gravitational interactions which becomes important
at astronomical distances (some Megaparsecs) at average energies (much
lesser than the Planck scale $10^{19}$Gev) the only other field which
is omnipresent in the universe in various shapes, sizes and magnitudes
is the magnetic field.

The main observational tracers of galactic and extra-galactic magnetic
fields are~\cite{obs1,obs2} the Zeeman splitting of spectral lines, the
intensity and polarization of synchrotron emission from free
relativistic electrons and Faraday rotations of polarized
electromagnetic radiation passing through an ionized medium. The
Zeeman splitting is too small to be useful outside our own galaxy. The
synchrotron emission method and Faraday rotation measurements require
an independent estimate of the local electron density. If the magnetic
field to be measured is far away one relies on Faraday rotation. The
agreement generally found between the strength of the field determined
by Faraday rotation technique and that inferred from the analysis of
the synchrotron emission in relatively close objects gives reasonable
confidence on the reliability of the first method also for far away
systems.

The origin of the magnetic fields observed in the galaxies and in the
cluster of galaxies is unknown. This is an outstanding problem in
modern astrophysics and historically it was the first motivation to
look for a primordial origin of the magnetic
fields~\cite{Grasso:2000wj}. The general trend is to use
magnetohydrodynamic methods to amplify very weak magnetic fields into
the $\mu$G fields generally observed in galaxies. Today the efficiency
of such a mechanism for production of magnetic fields is in question
from new observations of magnetic fields in high redshifted
galaxies. Furthermore, even if the magnetohydrodynamical calculations
are taken into account, the origin of the seed fields which initiated
the process has still to be identified. It is understood that
somewhere elementary particle physics will enter the scene to explain
the formation of the elusive seed fields. With the cosmological
observational evidences, as cosmic micro wave background radiation
(CMBR) and nucleosynthesis data, big bang theory is a cosmological
reality now and so is standard big bang cosmology.  In near future
perhaps we will have a standard mechanism for production of the seed
fields coming out from big bang cosmology. 

In the previous paragraph we discussed how models of particle physics
can be used to produce the seed fields. But the magnetic fields thus
produced can themselves affect various elementary particle
interactions. Magnetic fields can enhance the scattering
cross-sections or decay rates from their vacuum values.  The action of
magnetic fields on elementary particle interactions can produce non
trivial results and can explain some of the interesting observations
in astrophysics.  The main portion of this thesis deals with
elementary particle interactions, of which the neutrino is of prime
concern, in presence of a uniform background magnetic field. To put
the theme in perspective the following section gives an overview of
the nature of work done in this field by previous workers and the main
points around which this thesis evolves.
\section{Some well known observations, results and motivation for this
thesis}
\label{intneu}
Neutrinos have no electric charge. So they do not have any direct
coupling to photons in any renormalizable quantum field theory.  The
standard Dirac contribution to the magnetic moment, which comes from
the vector coupling of a fermion to the photon, is therefore absent
for the neutrino. In the standard model of electroweak interactions,
the neutrinos cannot have any anomalous magnetic moment either. The
reason is: anomalous magnetic moment comes from chirality-flipping
interactions $\bar\psi\sigma_{\mu\nu}\psi F^{\mu\nu}$, and neutrinos
cannot have such interactions because there are no right-chiral
neutrinos in the standard model. Other particles as the electron and
muon have normal and anomalous magnetic moments and so can interact
with external magnetic fields. 

\begin{figure}[btp]
\begin{center}
%
%
\begin{picture}(180,120)(-90,-35)
\Text(0,-30)[ct]{\large\bf (a)}
\ArrowLine(80,0)(40,0)
\Text(60,-10)[c]{$\nu$}
\Photon(40,0)(-40,0)37
\Text(0,-10)[c]{$W$}
\ArrowLine(-40,0)(-80,0)
\Text(-60,-10)[c]{$\nu$}
\ArrowArc(0,0)(40,0,90)
\Text(35,40)[l]{$\ell$}
\ArrowArc(0,0)(40,90,180)
\Text(-35,40)[r]{$\ell$}
\Photon(0,40)(0,80){2}{4}
\end{picture}
%
%
\begin{picture}(180,120)(-90,-35)
\Text(0,-30)[ct]{\large\bf (b)}
\ArrowLine(80,0)(40,0)
\Text(60,-10)[c]{$\nu$}
\ArrowLine(40,0)(-40,0)
\Text(0,-10)[c]{$\ell$}
\ArrowLine(-40,0)(-80,0)
\Text(-60,-10)[c]{$\nu$}
\PhotonArc(0,0)(40,0,90){3}{6}
\Text(43,40)[r]{$W$}
\PhotonArc(0,0)(40,90,180){3}{6}
\Text(-43,40)[l]{$W$}
\Photon(0,40)(0,80){2}{4}
\end{picture}
\caption[One-loop diagrams that give rise to neutrino magnetic moment.]{\sf 
One-loop diagrams that give rise to neutrino magnetic moment in
standard model aided with right-handed neutrinos.  The lines marked
$\nu$ are generic neutrino lines, whereas those marked $\ell$ are
generic charged leptons.  The external vector boson line is the
photon.  In renormalizable gauges, there are extra diagrams where any
of the $W$ lines can be replaced by the corresponding unphysical Higgs
scalar. 
\label{f:magmom}}
\end{center}
\end{figure}
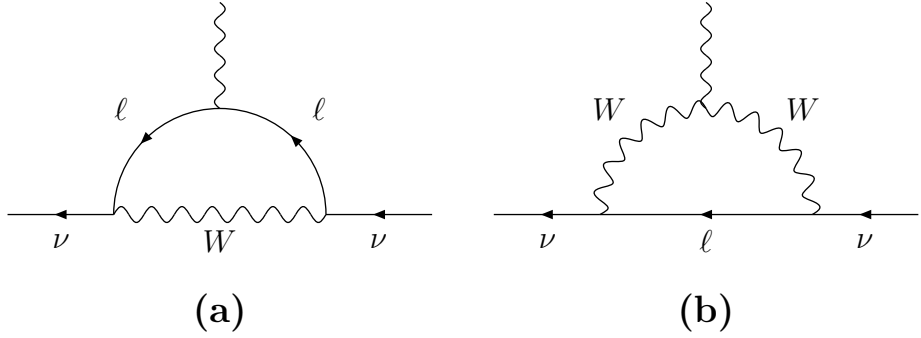

Inclusion of neutrino mass naturally takes us beyond the standard
model, where the issue of neutrino interactions with a magnetic field
must be reassessed. If the massive neutrino turns out to be a Dirac
fermion, its right-chiral projection must be included in the fermion
content of the theory, and in that case an anomalous magnetic moment
of a neutrino automatically emerges when quantum corrections are taken
into account.  In the simplest extension of the standard model
including right-chiral neutrinos, the magnetic moment arises from the
diagrams in \fig{f:magmom} and is given by~\cite{Fujikawa:yx}
\begin{eqnarray}
\mu_\nu = {3eG_F m_\nu \over 8\surd 2 \pi^2} = 3 \times 10^{-19} \mu_B
\times {m_\nu \overwithdelims() 1\,{\rm eV}} \,,
\label{munu}
\end{eqnarray}
where $m_\nu$ is the mass of the neutrino and $\mu_B$ is the Bohr
magneton.
If, on the other hand, neutrinos have Majorana masses,\footnote{A 
detailed discussion on Dirac and Majorana masses of neutrinos is present
in \cite{Mohapatra:rq}.}  i.e., they are their own
antiparticles, they cannot have any magnetic moment at all, because
$\bf{CPT}$ symmetry implies that the magnetic moments of a particle
and its antiparticle should be equal and opposite.  However, even in
this case there can be transition magnetic moments, which are
co-efficients of effective operators of the form
$\bar\psi_1\sigma_{\mu\nu}\psi_2F^{\mu\nu}$, where $\psi_1$ and
$\psi_2$ denote two different fermion fields.  These will also
indicate some sort of interaction with the magnetic field, associated
with a change of the fermion flavour. The question of the neutrino
magnetic moment assumed importance when it was suggested that
it can be a potential solution for the solar neutrino puzzle
\cite{Cisneros:1970nq,Okun:na,Okun:hi}.  A viable solution required a
neutrino magnetic moment around $10^{-10}\mu_B$, orders of magnitude
larger than that given by Eq.~(\ref{munu}), knowing that the neutrino
masses cannot be very large. At present the situation has
changed and neutrino magnetic moment is not a lucrative object to look
for to explain the solar neutrino data. The Sudbury neutrino
observatory results~\cite{Ahmad:2002jz} indicate no deficiency of left
handed neutrinos in neutral-current interactions.

If we leave out non-standard neutrinos and take its magnetic moment to
be zero, then also magnetic fields can alter their properties. This is
the prime concern of the present thesis. One of the simplest things
that gets affected is the neutrino self-energy in a magnetic
field. The magnetic field enters the scene through interaction with
the charged leptons in the loops of the self-energy diagram.  Erdas
and Feldman calculated the dispersion relation of neutrinos in a
uniform background magnetic field. To lowest order in the magnetic
field ${\mathcal B}$ they obtained the dispersion relation
\cite{Erdas:gy}
\begin{eqnarray}
\Omega^2 = \vec q^2 + {eg \overwithdelims() 2\pi M_W^2}^2
\left( \frac13 \ln {M_W \over m} + \frac18 \right) 
{\mathcal B}^2 q_\perp^2 \,.
\label{intro-nself}
\end{eqnarray}
Here $\Omega$ and $\vec q$ are the neutrino energy and 3-momentum. 
$q_\perp$ is the magnitude of that component of the
neutrino 3-momentum which is perpendicular to the magnetic field
3-vector. $m$ is the mass of the charged fermion which has the same
flavour as that of the neutrino, $M_W$ the W boson mass and $g$ the
weak coupling constant.
%
%
The origin of such a term like $q_\perp^2$ will be discussed in 
chapter \ref{chap3}.  The dispersion relation reduces to the
usual dispersion relation of a zero mass particle when we put
${\mathcal B}=0$. For strong magnetic fields, the dispersion relation
has been calculated more recently by Elizalde, Ferrer and de la
Incera~\cite{Elizalde:2000vz}.

There are other processes as neutrino neucleon scattering which are
also affected by magnetic fields. The charged current interaction
Lagrangian involving neutrinos and nucleons is given by
\begin{eqnarray}
\mathscr L_{\rm int} = \sqrt 2 G_\beta \left[ \overline
\psi_{(e)} \gamma^\mu L \psi_{(\nu_e)} \right] \; 
\left[ \overline
\psi_{(p)} \gamma_\mu (G_V + G_A \gamma_5) \psi_{(n)} \right] \,,
\label{chargecurrent}
\end{eqnarray}
where $L = \frac12(1 - \gamma_5)$ and 
$G_\beta = G_F \cos\theta_C$, $\theta_C$ being the Cabibbo angle, and
$G_A/G_V=-1.26$.  This can be used to find the cross-section for
various neutrino-nucleon scattering processes. 

The above interaction Lagrangian is most vital for processes involving
neutrinos or antineutrinos. In a star, when such reactions occur, the
final neutrino or the antineutrino escapes and the star loses energy.
Such processes are collectively known as URCA processes, named after a
casino in Rio de Janeiro where customers lose money little by little
\cite{Clayton}.  These processes are,
\begin{eqnarray}
n &\to& p + e^- +\bar\nu \,,\\*
p + e^- &\to& n + \nu \,,\\*
n + e^+ &\to& p + \bar\nu \,.
\end{eqnarray}
The calculations of the cross-sections and decay rates in presence of
a magnetic field uses a benchmark value of the background field: 
\begin{eqnarray}
{\mathcal B}_e &=& \frac{m^2}{e}\,,\nonumber\\
               &=& 4.41 \times 10^{13} {\rm G}\,,
\label{Bc}
\end{eqnarray}
where $m$ is the electron mass, and $e$ is the charge of the
proton. ${\mathcal B}_e$ is sometimes called the `critical field'
although nothing that critical happens at this field strength. The
only condition which defines the critical field is that when the
magnitude of the magnetic field reaches the value ${\mathcal B}_e$ the
electron cyclotron frequency equals its rest mass in the natural
units. If the magnitude of the magnetic field is above
the critical field the electronic wave functions get considerably
modified. It is assumed henceforth that while discussing scattering or
decay processes the magnitude of the external fields will be around
${\mathcal B}_e$. If the magnitude of the field is comparable to
$m_p^2/e$ then the proton wave functions can also get affected. The
calculations in this thesis will not involve such high fields.

The rate of the neutron beta decay process in a magnetic field was
calculated by various authors. An early paper by Fassio-Canuto
\cite{Can69} derived the rate in a background of degenerate
electrons. The work considers exact wave functions of the electrons in
a background of a uniform magnetic field.  Contemporary papers by
Matese and O'Connell~\cite{MOc69,MOc70} derived the rate where the
background did not contain any matter, but included the effects of the
polarization of neutrons due to the magnetic field. Protons and
neutrons were assumed to be non-relativistic in the calculations.
Further, the magnetic field was assumed to be much smaller than
$m_p^2/e$ so that its effect on the proton wave function could be
neglected.  Various calculations of the other processes exist in the
literature. Some calculations take the background matter density into
account~\cite{Dorofeev:az, DRT85, Baiko:1998jq, Bandyopadhyay:1998qs,
Leinson:2001ei}, some include magnetic effects on the proton
wavefunction as well~\cite{Leinson:2001ei}. In all of these
calculations the cross-section or the decay rate is sensitive to the
direction between the neutrino 3-momentum and the external uniform
magnetic field.

Neutrinos can also interact with photons, but this interaction 
is an effective one mediated by
virtual charged particles.  In a magnetic field the propagators of
these charged particles are affected by the external field. Previously
it has been shown how the neutrino self-energy is modified in an
external magnetic field. The properties of photons also get modified
in an external uniform magnetic field due to their interaction with
virtual charged particles. It was shown by Adler~\cite{Adler} that when
the background magnetic field magnitude is around ${\mathcal B}_e$
many interesting things can happen. The photon dispersion relation in
presence of a magnetic field changes from that in vacuum $|\vec{k}| =
\omega$. There are actually two different indices of refraction
corresponding to the two photon propagation modes. The modes are
linearly polarized, with the magnetic field of the mode either
parallel or perpendicular to the plane containing the external
magnetic field and the direction of propagation. Most of the
calculations done by Adler uses the effective Euler-Heisenberg
Lagrangian~\cite{Heisenberg:1935qt}. Later Schwinger, while trying to
derive the Euler-Heisenberg Lagrangian, derived the form of the
propagator of the charged particles in an uniform background magnetic
field~\cite{Schwinger:nm}. After Schwinger's work most of the
calculations, as the self-energy of the photon in a magnetic field,
are done using the Schwinger propagator.
  
Two second rank tensors play an important part in building up the
effective neutrino photon vertex function. They are the photon
self-energy $\Pi_{\mu \nu}$ and another tensor $\Pi^5_{\mu\nu}$.  To
one loop the tensors $\Pi^5_{\mu\nu}$ and $\Pi_{\mu \nu}$ are similar
except one point. In $\Pi_{\mu \nu}$ both the vertices, corresponding
to the two tensor indices, are of vector type whereas in $\Pi^5_{\mu
\nu}$ one of the vertices is of the axial-vector type~\cite{pal1}.
$\Pi^5_{\mu\nu}$ is called the axialvector-vector amplitude in this
thesis. In a magnetic field background $\Pi_{\mu \nu}$ was first
calculated by Tsai~\cite{Tsai:1}. In a different work DeRaad, Milton
and Hari Dass~\cite{DeRaad:1976kd} calculated $\Pi^5_{\mu\nu}$. Both
these calculations were to one loop but to all orders in the external
magnetic field.  Using the expression of the above two tensors in a
magnetic field the rate of the neutrino Cherenkov process,
\begin{eqnarray}
\nu(q) \to \nu(q') + \gamma(k)\,,
\end{eqnarray}
where $q$, $q'$ and $k$ are the four momenta of the particles as
shown, was calculated by Raffelt and Ioannisian~\cite{Raffelt:1997br,
Ioannisian:1996pn}. The neutrino Cherenkov process is kinematically
forbidden in vacuum. But in a magnetic field as the photon dispersion
relation changes so this process becomes viable.  The photon 
self-energy $\Pi_{\mu \nu}$~\cite{D'Olivo:2002sp, Ganguly:1999ts} and
$\Pi^5_{\mu \nu}$~\cite{Bhattacharya:2003hq, Bhattacharya:2001nm,
Nieves:2003kw} both have been calculated in a magnetized medium to one
loop.

In this thesis the main emphasis has been given on neutrino scattering
processes in presence of a magnetic field and the electromagnetic
vertex of neutrinos in a magnetized medium. $\Pi^5_{\mu \nu}$,
which is related to the electromagnetic vertex of neutrinos, has been
calculated in the background of a magnetized medium. Using the
expression of $\Pi^5_{\mu \nu}$ attempt has been made to find out a
specific limit of this quantity, which gives the `effective electric
charge' of the neutrino.  A general form factor analysis of
$\Pi^5_{\mu \nu}$ for various backgrounds, as in vacuum, thermal
medium, magnetic field and a magnetized medium is also given in
chapter \ref{chap4}. Neutrino interactions in presence of a classical
background magnetic field is a subject which has been studied for a
long time and most of the topics mentioned above are briefly discussed
in the review by Bhattacharya and Pal~\cite{Bhattacharya:2002aj}.
\section{Notations and conventions}\label{conv}
Before going into the second chapter, this section summarizes the
mathematical notations and conventions we use throughout this thesis.
In future the magnitude of the uniform background magnetic field will
always be denoted by the symbol ${\mathcal B}$ and the magnetic field
vector pointing in the $z$ direction will be given by 
\begin{eqnarray}
\vec{{\mathcal B}} = {\hat z} {\mathcal B}\,,
\end{eqnarray}
where ${\hat z}$ is the unit vector along the $z$-axis.

The first thing to note is that a uniform classical magnetic field
chooses a certain preferred direction in space.  Consequently the
Lorentz invariance of the system is restricted, arbitrary boosts will
not preserve a pure magnetic field. A boost along the magnetic field
direction or a rotation about the external field direction, are the
only Lorentz transformations allowed now. As a result the 4-vector
structure breaks down into a perpendicular part and a parallel
part. If $a^{\mu}$ is a 4-vector, then
\begin{eqnarray}
a^{\mu}_\para &=& (a^0, 0,  0, a^3)\,, \label{defpara}\\
a^{\mu}_\pr &=& (0, a^1,  a^2, 0)\,,
\label{defpr}
\end{eqnarray}
such that%
\begin{eqnarray}
a^{\mu} = a^{\mu}_\para + a^{\mu}_\pr\,.
\label{fvectordec}
\end{eqnarray}
Also in our convention
\begin{eqnarray}
g_{\mu \nu} = g^{\para}_{\mu \nu} + g^{\pr}_{\mu \nu}\,,
\label{Gmunupa}
\end{eqnarray}
where
\begin{eqnarray}
g^{\para}_{\mu \nu} & = & (1,0,0,-1)\,,\\
g^{\pr}_{\mu \nu} & = & (0,-1,-1,0)\,.
\end{eqnarray}
Some times instead of $a^{\mu}_\para$ and $a^{\mu}_\pr$ use has been
made of $a^{\mu_\para}$ and $a^{\mu_\pr}$, they are the same things
written in two different ways. 

Frequently we will come across terms as $a^2_{\para}$ and
$a^2_{\pr}$. These terms stands for
\begin{eqnarray}
a^2_\para &=& g^{\para}_{\mu \nu} a^\mu a^\nu\,,\\
          &=& (a^1)^2  -  (a^3)^2\,,
\label{defparasq}\\
a^2_\pr   &=& - g^{\pr}_{\mu \nu} a^\mu a^\nu\,,\\
          &=&  (a^1)^2 +  (a^2)^2\,,
\label{defprsq}
\end{eqnarray}
such that 
\begin{eqnarray}
a^2 = a_\para^2 - a_\pr^2\,.
\end{eqnarray}

Sometimes the equations will involve terms as $(a \cdot {\widetilde
b})_\para$, where $a$ and $b$ are two 4-vectors. This dot product
stands for 
\begin{eqnarray}
(a \cdot {\widetilde b})_\para = (a^0 b^3 - a^3 b^0)
= -(a_0 b_3 - a_3 b_0)\,, \label{defpartd} 
\end{eqnarray}
which implies
\begin{eqnarray}
(a \cdot {\widetilde b})_\para = - (b \cdot {\widetilde a})_\para\,, \nonumber
\end{eqnarray}
and as a result
\begin{eqnarray}
(a \cdot {\widetilde a})_\para = 0\,. \nonumber
\end{eqnarray}
%
\section{Scheme of things to come}
This last section is about the scheme of the things to come. Chapter
\ref{chap2} focuses on the quantum field theoretical description of
scattering processes which involves charged particles, in presence of
a background uniform magnetic field. The field theoretical techniques
are then applied to find out the cross-section of the inverse beta
decay process. Using the expression of the inverse beta decay
cross-section an attempt has been made to explain the puzzle regarding
the high velocities of pulsars.

Chapter \ref{chap3} discusses about the charged particle propagators
in an uniform background magnetic field using Schwinger's
method~\cite{Schwinger:nm}. The Schwinger propagator is introduced and
then by using the methods of quantum statistical field theory it is
expressed in a thermal background. A general analysis on neutrino
self-energy in a magnetic field and in a magnetized medium
follows. The chapter ends with a discussion on the effects of a
background magnetic field on neutrino oscillation phenomenology.

Chapter \ref{chap4} deals with the topic of neutrino photon
interactions. It is shown that the photon vacuum polarization tensor
$\Pi_{\mu \nu}$ and the axialvector-vector amplitude $\Pi^5_{\mu \nu}$
are essential ingredients of the neutrino electromagnetic vertex. Some
comments on $\Pi_{\mu \nu}$ and a discussion on $\Pi^5_{\mu \nu}$ is
presented in this chapter. The chapter ends with an expression of
the effective electric charge of neutrinos in a magnetized medium. 

The last chapter concludes by summarizing the points which are
discussed in the various chapters of this thesis. The thesis ends with
a set of appendices containing some detailed calculations which were
put at the end so as not to block the readers mind with lengthy
calculations.

\chapter{Inverse beta decay
in a constant background magnetic field}\label{chap2}
\section{Introduction} 
The interactions of elementary particles show novel features when they
occur in non-trivial backgrounds. Study of particle propagation in
matter has proved pivotal in the understanding of the solar neutrino
problem. A typical example of this is the neutrino-electron scattering
in presence of a medium which culminated in the now famous
MSW~\cite{Wolfenstein:1977ue, Mikheev:wj} effect. 
The cross-section for the elastic
neutrino-electron scattering in presence of a magnetic field has been
studied~\cite{KP2}. Bezchastnov and Haensel~\cite{Bezchastnov:1996gy}
has also calculated the neutrino-electron scattering cross-section in
presence of a magnetized medium.

There are processes related to the neutrino-electron scattering,
obtained by crossing, which contain a neutrino-antineutrino pair in
the final state.  For example, one can have the pair annihilation of
electron and positron into neutrino-antineutrino pair, or the neutrino
synchrotron radiation: 
%
%
%
\begin{eqnarray}
e^- \to e^- + \nu + \bar\nu \,.
\label{eenunu}
\end{eqnarray}
This is similar to a the normal synchrotron radiation reaction, the
difference being that a neutrino-antineutrino pair is produced instead
of a photon.  
It should be noted that this process cannot occur in the
vacuum.  However, in the presence of a background magnetic field and
background matter, the dispersion relation of the electron changes so
that it becomes kinematically feasible.  These processes provide
important mechanism for stellar energy loss, and the rates of these
processes have been calculated~\cite{Kaminker:su, Kaminker:ic,
Vidaurre:1995iv}.  The reverse of these processes are important for
neutrino absorption, and have also been studied~\cite{Hardy:2000gg}.
Moreover studies of scattering cross-sections and decay rates in
background magnetic fields are important since stellar objects like
neutron stars are expected to possess very high magnetic fields, of
the order of $10^{12}$\,G or higher.  Analysis of these processes
might be crucial for obtaining a proper understanding of the
properties of these stars. As the preferred direction of the
background magnetic fields break the isotropy of space the scattering
cross-sections of various processes calculated in such a background
also becomes anisotropic.  This anisotropy of the cross-sections of
processes involving neutrinos as initial particles or final particles
can give rise to an overall asymmetric emission of neutrinos from a
newly born neutron star and as a result the star may get an overall
momentum thrust.

As discussed in chapter \ref{chap1}, there are various possible
ways to introduce the external uniform magnetic field in the
calculation of processes including elementary particles. In this
chapter we explore the exact solutions of the Dirac equation in the
presence of such an external field, and then apply the results to
obtain the physically interesting numbers as scattering
cross-sections~\cite{Bhattacharya:2002qf, KP2}. The
initial sections of this chapter develops a consistent quantum field
theoretical method for calculations of cross-sections in a background
magnetic field. The discussions up to section \ref{fo} is general, the
results discussed can be appropriately used in any scattering
cross-section calculation in a background magnetic field.  The
particularities of the inverse beta decay process cross-section
calculation enters from section \ref{ib}. Based upon the form of the
scattering cross-section, section \ref{asymnem} elucidates the point
about asymmetric emission of neutrinos from a proto-neutron star.  The
chapter concludes by summarizing the ideas exposed.
\section{Solutions of the Dirac equation in a uniform magnetic
field}\label{so}
For a particle of mass $m$ and charge $eQ$, the Dirac equation in
presence of a magnetic field is given by
\begin{eqnarray}
i {\partial\psi \over \partial t} = \big[ \vec \alpha \cdot
(-i\vec\nabla - eQ\vec A) + \beta m \big] \psi \,,
\label{DiracEq}
\end{eqnarray}
where $\vec\alpha$ and $\beta$ are the Dirac matrices, and $\vec A$ is
the vector potential.  In our convention, $e$ is the positive unit of
charge, taken as usual to be equal to the proton charge.

For stationary states, we can write
\begin{eqnarray}
\psi = e^{-iEt} \left( \begin{array}{c} \phi \\ \chi \end{array}
\right) \,,
\end{eqnarray}
where $\phi$ and $\chi$ are 2-component objects.  We use the Pauli-Dirac
representation of the Dirac matrices, in which
\begin{eqnarray}
\vec \alpha = \left( \begin{array}{cc} 
		0 & \vec\sigma \\
		\vec\sigma & 0
	      \end{array} \right) \,, \qquad
\beta = \left( \begin{array}{cc} 
		1 & 0 \\
		0 & 1 
	      \end{array} \right)
\end{eqnarray}
where each block represents a $2\times2$ matrix, and $\vec\sigma$ are
the Pauli matrices.  With this notation, we can write Eq.\
(\ref{DiracEq}) as
\begin{eqnarray}
(E-m)\phi &=& \vec \sigma \cdot (-i\vec\nabla - eQ\vec A) \chi \,, 
\label{eq1}\\*
(E+m) \chi &=& \vec \sigma \cdot (-i\vec\nabla - eQ\vec A) \phi \,.
\label{eq2}
\end{eqnarray}
Eliminating $\chi$, we obtain 
\begin{eqnarray}
(E^2 - m^2)\phi &=& \Big[ \vec \sigma \cdot (-i\vec\nabla - eQ\vec A) 
\Big]^2 \phi \,.
\label{phieq1} 
\end{eqnarray}
We will work with a constant magnetic field $\vec {\mathcal B}$.
Without loss of generality, it can be taken along the
$z$-direction. The vector potential can be chosen in many equivalent
ways. We take
\begin{eqnarray}
A_0 = A_y = A_z = 0 \,, \qquad A_x = -y{\mathcal B}\,.
\label{A}
\end{eqnarray}
With this choice, Eq.\ (\ref{phieq1}) reduces to the form
\begin{eqnarray}
(E^2 - m^2)\phi 
&=& \Big[ -\vec\nabla^2 + (eQ{\mathcal B})^2 y^2 - eQ{\mathcal B}(2iy
{\partial\over 
\partial x} + \sigma_3) \Big] \phi \,. 
\label{phieq} 
\end{eqnarray}
Here $\sigma_3$ is the diagonal Pauli matrix. Noticing that the
co-ordinates $x$ and $z$ do not appear in the equation except through
the derivatives, we can write the solutions as
\begin{eqnarray}
\phi = e^{i \vec {\scriptstyle p} \cdot \vec{\scriptstyle X} \omit y}
f(y) \,, 
\label{phiform}
\end{eqnarray}
where $f(y)$ is a 2-component matrix which depends only on the
$y$-coordinate, and possibly some momentum components, as we will see
shortly. We have also introduced the notation $\vec X$ for the spatial
co-ordinates (in order to distinguish it from $x$, which is one of the
components of $\vec X$), and $\vec X\omit y$ for the vector $\vec X$
with its $y$-component set equal to zero. In other words, $\vec p\cdot
\vec X{\omit y} \equiv p_xx+p_zz$, where $p_x$ and $p_z$ denote the
eigenvalues of momentum in the $x$ and $z$ directions.\footnote{It is
to be understood that whenever we write the spatial component of any
vector with a lettered subscript, it would imply the corresponding
contravariant component of the relevant 4-vector.}

There will be two independent solutions for $f(y)$, which can be
taken, without any loss of generality, to be the eigenstates of
$\sigma_3$ with eigenvalues $s=\pm 1$. This means that we choose the
two independent solutions in the form
\begin{eqnarray}
f_+ (y) = \left( \begin{array}{c} F_+(y) \\ 0 \end{array} \right) \,,
\qquad 
f_- (y) = \left( \begin{array}{c} 0 \\ F_-(y) \end{array} \right) \,.
\end{eqnarray}
Since $\sigma_3 f_s = sf_s$, the differential equations satisfied by
$F_s$ is
\begin{eqnarray}
{d^2F_s \over dy^2} - (eQ{\mathcal B}y + p_x)^2 F_s + (E^2 - m^2 -
p_z^2 + eQ{\mathcal B}s) 
F_s = 0 \,,
\label{Fseqn}
\end{eqnarray}
which is obtained from Eq.\ (\ref{phieq}).  The solution is obtained
by using the dimensionless variable
\begin{eqnarray}
\xi = \sqrt{e |Q| {\mathcal B}} \left( y + {p_x \over eQ{\mathcal B}}
\right) \,, 
\label{xi}
\end{eqnarray}
which transforms Eq.\ (\ref{Fseqn}) to the form
\begin{eqnarray}
\left[ {d^2 \over d\xi^2} -\xi^2 + a_s \right] F_s = 0 \,,
\end{eqnarray}
where
\begin{eqnarray}
a_s = {E^2 - m^2 - p_z^2 + eQ{\mathcal B}s \over e|Q|{\mathcal B}} \,.
\end{eqnarray}
This is a special form of Hermite's equation, and the solutions exist
provided $a_s=2\nu+1$ for $\nu=0,1,2,\cdots$. This provides the energy
eigenvalues 
\begin{eqnarray}
E^2 = m^2 + p_z^2 + (2\nu+1)e|Q|{\mathcal B} - eQ{\mathcal B}s \,,
\label{E}
\end{eqnarray}
and the solutions for $F_s$ are
\begin{eqnarray}
N_{\nu} e^{-\xi^2/2} H_{\nu}(\xi) \equiv I_{\nu}(\xi) \,,
\label{In}
\end{eqnarray}
where $H_\nu$ are Hermite polynomials of order $\nu$, and $N_\nu$ are
normalizations which we take to be
\begin{eqnarray}
N_\nu = 
\left( {\sqrt{e|Q|{\mathcal B}} \over \nu! \, 2^\nu \sqrt{\pi}} \,
\right)^{1/2} \,.  
\label{Nn}
\end{eqnarray}
We stress that the choice of normalization can be arbitrarily made, as
will be clarified later.  With our choice, the functions $I_\nu$
satisfy the completeness relation
\begin{eqnarray}
\sum_\nu I_\nu(\xi) I_\nu(\xi_\star) = \sqrt{e|Q|{\mathcal B}} \;
\delta(\xi-\xi_\star) = \delta (y-y_\star) \,, 
\label{completeness}
\end{eqnarray}
where $\xi_\star$ is obtained by replacing $y$ by $y_\star$ in Eq.\
(\ref{xi}).

So far, $Q$ was arbitrary.  We now specialize to the case of
electrons, for which $Q=-1$.  The solutions are then conveniently
classified by the energy eigenvalues
\begin{eqnarray}
E_n^2 = m^2 + p_z^2 + 2ne{\mathcal B} \,,
\label{En}
\end{eqnarray}
which is the relativistic form of Landau energy levels. The solutions
are two fold degenerate in general: for $s=1$, $\nu=n-1$ and for
$s=-1$, $\nu=n$.  In the case of $n=0$, only the second solution is
available since $\nu$ cannot be negative.  The solutions can have
positive or negative energies. We will denote the positive square root
of the right side by $E_n$. Representing the solution corresponding to
this $n$-th Landau level by a superscript $n$, we can then write for
the positive energy solutions,
\begin{eqnarray}
f_+^{(n)} (y) = \left( \begin{array}{c} 
I_{n-1}(\xi) \\ 0 \end{array} \right) \,,
\qquad 
f_-^{(n)} (y) = \left( \begin{array}{c} 
0 \\ I_n (\xi) \end{array} \right) \,.
\label{fsolns}
\end{eqnarray}
For $n=0$, the solution $f_+$ does not exist. We will consistently
incorporate this fact by defining
\begin{eqnarray}
I_{-1} (y) = 0 \,,
\label{I_-1}
\end{eqnarray}
in addition to the definition of $I_n$ in Eq.\ (\ref{In}) for
non-negative integers $n$.

The solutions in Eq.\ (\ref{fsolns}) determine the upper components of
the spinors through Eq.\ (\ref{phiform}). The lower
components, denoted by $\chi$ earlier, can be solved using
Eq.\ (\ref{eq2}), and finally the positive energy solutions of the
Dirac equation can be written as
\begin{eqnarray}
e^{-ip\cdot X {\omit y}} U_s (y,n,\vec p \omit y) \,,
\end{eqnarray}
where $X^\mu$ denotes the space-time coordinate. And $U_s$ are given
by
\begin{eqnarray}
U_+ (y,n,\vec p \omit y) = \left( \begin{array}{c} 
I_{n-1}(\xi) \\[2ex] 0 \\[2ex] 
{\strut\textstyle p_z \over \strut\textstyle E_n+m} I_{n-1}(\xi) \\[2ex]
-\, {\strut\textstyle \sqrt{2ne{\mathcal B}} \over \strut\textstyle 
E_n+m} I_n (\xi) 
\end{array} \right) \,, \qquad 
U_- (y,n,\vec p \omit y) = \left( \begin{array}{c} 
0 \\[2ex] I_n (\xi) \\[2ex]
-\, {\strut\textstyle \sqrt{2ne{\mathcal B}} \over \strut\textstyle E_n+m}
I_{n-1}(\xi) \\[2ex] 
-\,{\strut\textstyle p_z \over \strut\textstyle E_n+m} I_n(\xi) 
\end{array} \right) \,. 
\label{Usoln}
\end{eqnarray}

A similar procedure can be adopted for negative energy spinors which
have energy eigenvalues $E=-E_n$. In this case, it is easier to start
with the two lower components first and then find the upper components
from Eq.\ (\ref{eq1}). The solutions are
\begin{eqnarray}
e^{ip\cdot X{\omit y}} V_s (y,n, \vec p\omit y) \,,
\end{eqnarray}
where 
\begin{eqnarray}
V_+ (y,n,\vec p\omit y) = \left( \begin{array}{c} 
{\strut\textstyle p_z \over \strut\textstyle E_n+m}
I_{n-1}(\widetilde\xi) \\[2ex] 
{\strut\textstyle \sqrt{2ne{\mathcal B}} \over \strut\textstyle E_n+m} 
I_n (\widetilde\xi)  \\[2ex] 
I_{n-1}(\widetilde\xi) \\[2ex] 0
\end{array} \right) \,, \qquad 
V_- (y,n,\vec p\omit y) = \left( \begin{array}{c} 
{\strut\textstyle \sqrt{2ne{\mathcal B}} \over \strut\textstyle E_n+m}
I_{n-1}(\widetilde\xi) \\[2ex] 
-\,{\strut\textstyle p_z \over \strut\textstyle E_n+m}
I_n(\widetilde\xi)  \\[2ex] 
0 \\[2ex] I_n (\widetilde\xi)
\end{array} \right) \,.
\label{Vsoln}
\end{eqnarray}
where $\widetilde\xi$ is obtained from $\xi$ by changing the sign of
the $p_x$-term.

For future use, we note down a few identities involving the spinors
which can be obtained by direct substitutions of the solutions
obtained above. The details of the calculation has been worked out in
appendix \ref{assum}. The spin sum for the $U$-spinors is
\begin{eqnarray}
P_U (y,y_\star ,n,\vec p\omit y) &\equiv&
\sum_s U_s (y,n,\vec p\omit y) \overline U_s (y_\star ,n,\vec p\omit
y) \nonumber\\*  
& = &
{1\over 2(E_n+m)} \times 
\begin{array}[t]{l}
\bigg[ \left\{ m(1+\sigma_z) +
\rlap/p_\parallel - 
\widetilde{\rlap/p}_\parallel \gamma_5 \right\} I_{n-1}(\xi)
I_{n-1}(\xi_\star) \\ 
+ \left\{ m(1-\sigma_z) + \rlap/p_\parallel +
\widetilde{\rlap/p}_\parallel \gamma_5 \right\} I_n(\xi)
I_n (\xi_\star) \\ 
- \sqrt{2ne{\mathcal B}} (\gamma_1 - i\gamma_2) I_n(\xi) I_{n-1}(\xi_\star) \\
- \sqrt{2ne{\mathcal B}} (\gamma_1 + i\gamma_2) I_{n-1}(\xi) I_n(\xi_\star) 
\bigg] \,,
\end{array}
\nonumber\\* 
\label{PU}
\end{eqnarray}
where the above notations for the 4-vectors and the $\para$ and $\pr$
notations has been defined in chapter \ref{chap1}, and
$\sigma_z=i\gamma^1\gamma^2$. The symbol
$\widetilde{\rlap/p}_\parallel = p^0 \gamma_3 + p^3 \gamma_0$.
%
%
Similarly, the spin sum for the $V$-spinors can also be calculated,
and we obtain
\begin{eqnarray}
P_V (y,y_\star ,n,\vec p\omit y) &\equiv&
\sum_s V_s (y,n,\vec p\omit y) \overline V_s (y,n,\vec p\omit y) 
\nonumber\\* 
& = &
{1\over 2(E_n+m)} \times 
\begin{array}[t]{l}
\Bigg[ \left\{ -m(1+\sigma_z) +
\rlap/p_\parallel - 
\widetilde{\rlap/p}_\parallel \gamma_5 \right\} I_{n-1}(\widetilde\xi)
I_{n-1} (\widetilde\xi _\star) \\ 
+ \left\{ -m(1-\sigma_z) + \rlap/p_\parallel +
\widetilde{\rlap/p}_\parallel \gamma_5 \right\} I_n(\widetilde\xi)
I_n(\widetilde\xi _\star) \\ 
+ \sqrt{2ne{\mathcal B}} (\gamma_1 - i\gamma_2) I_n(\widetilde\xi)
I_{n-1}(\widetilde\xi _\star) \\ 
+ \sqrt{2ne{\mathcal B}} (\gamma_1 + i\gamma_2) I_{n-1}(\widetilde\xi)
I_n(\widetilde\xi 
_\star) \Bigg] \,. 
\end{array} \nonumber\\*
\label{PV}
\end{eqnarray}
%

\section{The fermion field operator}\label{fo}
Since we have found the solutions to the Dirac equation, we can now
use them to construct the fermion field operator in the second
quantized version. For this, we write
\begin{eqnarray}
\psi(X) = \sum_{s=\pm} \sum_{n=0}^\infty \int {dp_x \, dp_z \over D}
\left[ f_s (n,\vec p\omit y) e^{-ip\cdot X {\omit y}} U_s (y,n,\vec
p\omit y) + 
\widehat f_s^\dagger (n,\vec p\omit y) e^{ip\cdot X {\omit y}} V_s
(y,n,\vec p\omit y) \right] \,.
\label{2ndquant}
\end{eqnarray}
Here, $f_s(n,\vec p\omit y)$ is the annihilation operator for the
fermion, and $\widehat f_s^\dagger(n,\vec p\omit y)$ is the creation
operator for the antifermion in the $n$-th Landau level with given
values of $p_x$ and $p_z$. The creation and annihilation operators
satisfy the anticommutation relations
\begin{eqnarray}
\left[ f_s (n,\vec p\omit y), f_{s'}^\dagger (n',\vec p'\omit y)
\right]_+ = 
\delta_{ss'} \delta_{nn'} \delta(p_x-p'_x) \delta (p_z - p'_z) \,,
\label{freln}
\end{eqnarray}
and a similar one with the operators $\widehat f$ and $\widehat
f^\dagger$, all other anticommutators being zero. The quantity $D$
appearing in Eq.\ (\ref{2ndquant}) depends on the normalization of the
spinor solutions, and this is why the normalization of the spinors
could have been chosen arbitrarily, as remarked after Eq.\ (\ref{Nn}).
Once we have chosen the spinor normalization, the factor $D$ appearing
in Eq.\ (\ref{2ndquant}) is however fixed, and it can be determined
from the equal time anticommutation relation
\begin{eqnarray}
\left[ \psi(X), \psi^\dagger(X_\star) \right]_+ = \delta^3 (\vec X - \vec
X_\star) \,.
\label{anticomm}
\end{eqnarray}
Plugging in the expression given in
Eq.\ (\ref{2ndquant}) to the left side of this equation and using the
anticommutation relations of Eq.\ (\ref{freln}), we obtain
\begin{eqnarray}
\left[ \psi(X), \psi^\dagger(X_\star) \right]_+ = \sum_{s} \sum_{n} \int
{dp_x \, dp_z \over D^2} && 
\Big( e^{-ip_x(x-x_\star)} e^{-ip_z(z-z_\star)} 
U_s (y,n,\vec p\omit y) U_s^\dagger (y_\star ,n,\vec p\omit y)
\nonumber\\* 
&& +  e^{ip_x(x-x_\star)} e^{ip_z(z-z_\star)} 
V_s (y,n,\vec p\omit y) V_s^\dagger (y_\star ,n,\vec p\omit y) \Big)
\,.\nonumber\\
\end{eqnarray}
Changing the signs of the dummy integration variables $p_x$ and $p_z$
in the second term, we can rewrite it as
\begin{eqnarray}
\left[ \psi(X), \psi^\dagger(X_\star) \right]_+ = \sum_{s} \sum_{n} \int
{dp_x \, dp_z \over D^2} && e^{-ip_x(x-x_\star)}
e^{-ip_z(z-z_\star)} \Big( 
U_s (y,n,\vec p\omit y) U_s^\dagger (y_\star ,n,\vec p\omit y)
\nonumber\\* 
&& +  V_s (y,n,-\vec p\omit y) V_s^\dagger (y_\star ,n,-\vec p\omit y)
\Big) \,. 
\label{anticomm1}
\end{eqnarray}
Using now the solutions for the $U$ and the $V$ spinors from
Eqs. (\ref{Usoln}) and (\ref{Vsoln}), it is straight forward to verify
that 
\begin{eqnarray}
&& \sum_s \Big( U_s (y,n,\vec p\omit y) U_s^\dagger (y_\star ,n,\vec
p\omit y) 
+  V_s (y,n,-\vec p\omit y) V_s^\dagger (y_\star ,n,-\vec p\omit y) \Big)
\nonumber\\* 
&=& \left( 1 + {p_z^2 + 2ne{\mathcal B} \over (E_n+m)^2} \right) \times {\rm
diag} \; \Big [ I_{n-1}(\xi) I_{n-1}(\xi_\star), 
I_n(\xi) I_n(\xi_\star),
I_{n-1}(\xi) I_{n-1}(\xi_\star),  I_n(\xi) I_n(\xi_\star) 
\Big] \,,\nonumber\\
\label{ssum}
\end{eqnarray}
where `diag' indicates a diagonal matrix with the specified entries,
and $\xi$ and $\xi_\star$ involve the same value of $p_x$. At this stage,
we can perform the sum over $n$ in Eq.\ (\ref{anticomm1}) using the
completeness relation of Eq.\ (\ref{completeness}), which gives the
$\delta$-function of the $y$-coordinate that should appear in the
anticommutator.  Finally, performing the integrations over $p_x$ and
$p_z$, we can recover the $\delta$-functions for the other two
coordinates as well, provided
\begin{eqnarray}
{2E_n \over E_n+m} \; {1\over D^2} = {1\over (2\pi)^2} \,,
\end{eqnarray}
using the expression for the energy eigenvalues from Eq.\ (\ref{En}) to
rewrite the prefactor appearing on the right side of
Eq.\ (\ref{ssum}). Putting the solution for $D$, we can rewrite
Eq.\ (\ref{2ndquant}) as
\begin{eqnarray}
\psi(X) &=& \sum_{s=\pm} \sum_{n=0}^\infty \int {dp_x \, dp_z \over
2\pi} \sqrt {E_n+m \over 2E_n} \nonumber\\* && \times
\left[ f_s (n,\vec p\omit y) e^{-ip\cdot X {\omit y}} U_s (y,n,\vec
p\omit y) +  
\widehat f_s^\dagger (n,\vec p\omit y) e^{ip\cdot X {\omit y}} V_s
(y,n,\vec p\omit y) \right] \,.
\label{psi}
\end{eqnarray}

The one-fermion states are defined as
\begin{eqnarray}
\left| n,\vec p\omit y \right> = C f^\dagger (n,\vec p\omit y) \left|
0 \right> \,. 
\end{eqnarray}
The normalization constant $C$ is determined by the condition that the
one-particle states should be orthonormal. For this, we need to define
the theory in a finite but large region whose dimensions are  $L_x$,
$L_y$ and $L_z$ along the three spatial axes. This gives
\begin{eqnarray}
C = {2\pi \over \sqrt{L_x L_z}} \,.
\end{eqnarray}
Then
\begin{eqnarray}
\psi_U(X) \left| n,\vec p\omit y \right> = \sqrt {E_n+m \over 2E_n L_xL_z}
e^{-ip\cdot X {\omit y}} U_s (y,n,\vec p\omit y) \left| 0 \right> \,,
\label{psiket}
\end{eqnarray}
where $\psi_U$ denotes the term in Eq.\ (\ref{psi}) that contains the
$U$-spinors. Similarly, 
\begin{eqnarray}
\left< n,\vec p\omit y \left| \overline \psi_U(X) \right. \right.
= \sqrt {E_n+m \over 2E_n L_xL_z}
e^{ip\cdot X \omit y} \overline U_s (y,n,\vec p\omit y) \left< 0
\right| \,.
\label{brapsi}
\end{eqnarray}
%
\section{Inverse beta-decay}\label{ib} 
\subsection{Preliminary comments}
Now we have all the tools required for calculating the cross-section of
the inverse beta-decay process in a magnetic field.  We consider the
possibility that the neutrons may be totally or partially polarized in
the magnetic field, and find the cross-section as a function of this
polarization.  The neutrinos are assumed to be strictly standard model
neutrinos, without any mass and consequent properties.  The presence
of the magnetic field breaks the isotropy of the background, and a
careful calculation in this background reveals a dependence of the
cross-section on the incident neutrino direction with respect to the
magnetic field.

Considerable work has been done on the magnetic field dependence of
the URCA processes which have neutrinos in their final states
\cite{DRT85, MOc69, MOc70, Gvozdev:1999md, Arras:1998mv,
Baiko:1998jq}. An angular dependence obtained in the differential
cross-section of these reactions imply that in a star with high
magnetic field, neutrinos are created asymmetrically with respect to
the magnetic field direction.  The process that we consider, on the
other hand, have neutrinos in the initial state. So this process
influences the neutrino opacity in a star.

Some calculations of this process exist in the literature.  Roulet
\cite{Roulet:1997sw}, as well as Lai and Qian~\cite{Lai:1998sz}
performed the calculation by assuming that the magnetic field effects
enter only through the phase space integrals, whereas the matrix
element remains unaffected.  Gvozdev and Ognev~\cite{Gvozdev:1999md}
considered the final electron to be exclusively in the lowest Landau
level.  Arras and Lai~\cite{Arras:1998mv} calculated only the angular
asymmetry, and only to the first order in the background magnetic
field.  Some earlier calculations~\cite{KP2, Bhattacharya:1999bm} did
not take neutron polarization into account.

In this section the calculation of the cross-section for the inverse
beta-decay process $\nu_e+n\to p+e^-$ in a background magnetic field
is presented in full detail. The calculations  involve
evaluating the matrix element using spinor solutions of the electron
in a magnetic field, taking all possible final Landau levels into
account, including the possibility of neutron polarization, and
performing the calculations to all orders in the background field in
the 4-fermi interaction theory. The magnetic field might provide a net
polarization of the neutrons, which is taken into account. However,
the magnitude of the field is assumed to be much smaller than
$m_p^2/e$, so its effects on the proton 
spinors are ignored.  The electron spinors, on the other hand, are the
ones appropriate for the Landau levels. Thus, we can write the process
as
\begin{eqnarray}
\nu_e(q) + n(P) \to p(P') + e(\vec p'\omit y, n') \,.
\label{invbeta}
\end{eqnarray}
%
\subsection{The $S$-matrix element}
The charged current interaction Lagrangian for this process is given
by Eq.~(\ref{chargecurrent}). Using it in first order perturbation,
the $S$-matrix element between the final and the initial states of the
process in Eq.\ (\ref{invbeta}) is given by
\begin{eqnarray}
S_{fi} = \sqrt 2 \, G_\beta \int d^4X &&
\left< e(\vec p'\omit y, n') \left| \overline
\psi_{(e)} \gamma^\mu L \psi_{(\nu_e)} 
\right| \nu_e (q) \right> \nonumber\\*
&& \times  
\left< p(P') \left| \overline
\psi_{(p)} \gamma_\mu (G_V - G_A \gamma_5) \psi_{(n)}  \right| n(P)
\right> \,. 
\label{Sfi1}
\end{eqnarray}
For the hadronic part, we should use the usual solutions of the Dirac
field which are normalized within a box of volume $V$, and this gives
\begin{eqnarray}
\left< p(P') \left| \overline
\psi_{(p)} \gamma_\mu (G_V - G_A \gamma_5) \psi_{(n)}  \right| n(P)
\right> 
&=& {e^{i(P'-P)\cdot X} \over \sqrt{2{\cal E} V}
\sqrt{2{\cal E}' V}} \; 
\left[ \overline u_{(p)}(\vec P') \gamma_\mu (G_V - G_A \gamma_5)
u_{(n)}(\vec P) \right], \nonumber\\
\end{eqnarray}
using the notations ${\cal E}=P_0$ and ${\cal E}'=P'_0$. For the
leptonic part, we need to take into account the magnetic spinors for
the electron. Using Eq.\ (\ref{brapsi}), we obtain
\begin{eqnarray}
\left< e(\vec p'\omit y, n') \left| \overline
\psi_{(e)} \gamma^\mu L \psi_{(\nu_e)} 
\right| \nu_e (q) \right> 
&=& {e^{-iq\cdot X + ip'\cdot X\omit y} \over \sqrt{2\Omega V}}
\sqrt{E_{n'}+m \over 
2E_{n'}L_xL_z} 
\left[ \overline U_{(e)}(y,n',\vec p'\omit y) \gamma^\mu L
u_{(\nu_e)}(\vec q) \right]\,.\nonumber\\
\end{eqnarray}
Here $\Omega=q_0$. The symbol $m$ was used previously to denote the
mass of an arbitrary charged particle, from now onwards it will denote
the mass of an electron in this chapter.  Putting these back into Eq.\
(\ref{Sfi1}) and performing the integrations over all co-ordinates
except $y$, we obtain
\begin{eqnarray}
S_{fi} &=& (2\pi)^3 \delta^3 \omit y (P+q-P'-p')
\left[E_{n'}+m \over 2\Omega V \; 2{\cal E}V \; 2{\cal E}'V
2E_{n'}L_xL_z \right]^{1/2} {\mathscr M}_{fi} \,.
\label{Sfi2}
\end{eqnarray}
Here, $\delta^3\omit y$ implies, in accordance with the notation
introduced earlier, the $\delta$-function for all space-time
co-ordinates except $y$. Contrary to the field-free case, we do not
get 4-momentum conservation because the $y$-component of momentum is
not a good quantum number in this problem. The quantity ${\mathscr
M}_{fi}$ is the Feynman amplitude, given by
\begin{eqnarray}
{\mathscr M}_{fi} = \surd 2 G_\beta
\Big[ \overline u_{(p)}(\vec P') \gamma_\mu (G_V - G_A\gamma_5)
u_{(n)}(\vec P) \Big]  
\int dy \; e^{iu_y y}
\Big[ \overline U_{(e)} (y,n',\vec p'\omit y) \gamma^\mu L
u_{(\nu_e)}(\vec q) \Big] \,,\nonumber\\
\end{eqnarray}
using the shorthand
\begin{eqnarray}
u_y = P_y+q_y-P'_y \,.
\end{eqnarray}

The transition rate in a large time $T$ is given by
$|S_{fi}|^2/T$. {}From Eq.\ (\ref{Sfi2}), using the usual rules like
\begin{eqnarray}
\Big| \delta ({\cal E}+\Omega-{\cal E}' - E_{n'}) \Big|^2 
&=& {T\over 2\pi} \;
\delta ({\cal E}+\Omega-{\cal E}' - E_{n'})  \,,\nonumber\\*
\Big| \delta (P_x+q_x-P'_x-p'_x) \Big|^2 &=& {L_x\over 2\pi} \;
\delta (P_x+q_x-P'_x-p'_x) \,,\nonumber\\*
\Big| \delta (P_z+q_z-P'_z-p'_z) \Big|^2 &=& {L_z\over 2\pi} \;
\delta (P_z+q_z-P'_z-p'_z) \,,
\end{eqnarray}
we obtain
\begin{eqnarray}
|S_{fi}|^2/T &=& {1\over 16} (2\pi)^3 \delta^3 \omit y (P + q -P'- p')
{E_{n'}+m \over V^3 \Omega{\cal EE}' E_{n'}}
\Big| {\mathscr M}_{fi} \Big|^2 \,.
\end{eqnarray}
%

\subsection{The scattering cross-section}
Using unit flux $1/V$ for the incident particle as usual, we can write
the differential cross-section as
\begin{eqnarray}
d\sigma = V\, {|S_{fi}|^2\over T}d\rho \,,
\end{eqnarray}
where $d\rho$, the differential phase space for final particles, is
given in our case by
\begin{eqnarray}
d\rho = {L_x\over 2\pi} \, dp'_x \; {L_z\over 2\pi} \, dp'_z \;
{V\over (2\pi)^3} \, d^3P'
\,. 
\label{drho}
\end{eqnarray}
Therefore
\begin{eqnarray}
d\sigma &=& V\, {|S_{fi}|^2\over T} \; {L_xL_z\over (2\pi)^2} 
\, dp'_x \, dp'_z \;
{V\over (2\pi)^3} \, d^3P' \nonumber\\*
&=& {1\over 64\pi^2} \delta^3 \omit y (P+q-P'-p') \;
{E_{n'}+m \over \Omega {\cal E}{\cal E}' E_{n'}} \;
\Big| {\mathscr M}_{fi} \Big|^2 {L_xL_z\over V} 
\, dp'_x \, dp'_z \; d^3P' \,.
\label{dsigma}
\end{eqnarray}
The square of the matrix element is
\begin{eqnarray}
\Big| {\mathscr M}_{fi} \Big|^2 = 2G_\beta^2 \ell^{\mu\nu} H_{\mu\nu} \,,
\end{eqnarray}
where $H_{\mu\nu}$ is the hadronic part and $\ell^{\mu\nu}$ the
leptonic part, whose calculation we outline now.

For the hadronic part, we can use the usual Dirac spinors because of
our assumption that the magnetic field is much smaller than $m_p^2/e$.
We will work in the rest frame of the neutron.  Due to the presence of
the background magnetic field, the neutrons may be totally or
partially polarized.  We define the quantity 
\begin{eqnarray}
S \equiv {N_n^{(+)} - N_n^{(-)} \over N_n^{(+)} + N_n^{(-)}} \,,
\end{eqnarray}
where $N_n^{(\pm)}$ denote the number of neutrons parallel and
antiparallel to the magnetic field.  Then
\begin{eqnarray}
H_{\mu\nu} = \frac12 (1+S) H_{\mu\nu}^{(+)} + 
\frac12 (1-S) H_{\mu\nu}^{(-)} \,,
\end{eqnarray}
where $H_{\mu\nu}^{(\pm)}$ denotes the contribution calculated with
spin-up and spin-down neutrons respectively.  Either of these
contributions can be calculated by using the spin projection operator,
which is $\frac12(1\pm\gamma_5\gamma_3)$ for up and down spins.  A
straight forward calculation then yields
\begin{eqnarray}
H_{\mu\nu} &=& 2(G_V^2 + G_A^2) (P_\mu P'_\nu + P_\nu P'_\mu -
g_{\mu\nu} P \cdot P') \nonumber\\*
&& + 2 (G_V^2-G_A^2) m_n m_p g_{\mu\nu} + 4i G_V
G_A \varepsilon_{\mu\nu\lambda\rho} P^\lambda P'^\rho \nonumber\\*
&-& S \Big[ 
4 G_VG_A m_n (P'_\mu g_{3\nu} + P'_\nu g_{3\mu} - P'_3
g_{\mu\nu})
+ 2 i \varepsilon_{\mu\nu3\alpha} R^\alpha \Big]  
\,,
\end{eqnarray}
where we have introduced the shorthand
\begin{eqnarray}
R^\alpha = (G_V^2 + G_A^2) m_n P'^\alpha 
- (G_V^2-G_A^2) m_p P^\alpha \,.
\end{eqnarray}
We have omitted some terms in the expression for $H_{\mu\nu}$ that
involve spatial components of the neutron momentum, with the
anticipation that we will perform the calculation in the neutron rest
frame. 

In the leptonic part $\ell^{\mu\nu}$, we should use the magnetic
spinors given in section \ref{so}.  This gives
\begin{eqnarray}
\ell^{\mu\nu} = \int dy \int dy_\star \; e^{iu_y(y_\star-y)} \; {\rm Tr}
\Big[ P_U (y, y_\star, n', \vec p' \omit y) \gamma^\mu \rlap/q
\gamma^\nu L \Big] \,,
\end{eqnarray}
where $P_U$ denotes the spinor sum for the electrons, given in Eq.\
(\ref{PU}).  We now have to perform the integrations over $y$ and
$y_\star$.  Each of these variables should be integrated in the range
$-\frac12L_y$ to $+\frac12L_y$.  However, since we will take the
infinite volume limit at the end as usual, we let $L_y\to\infty$ and
use the result~\cite{GradRyzh}
\begin{eqnarray}
\int_{-\infty}^{+\infty} dy \; e^{ixy} I_n(y) 
= i^n \; \sqrt{2\pi} \; I_n (x) 
\,.
\end{eqnarray}
This gives
\begin{eqnarray}
\ell^{\mu\nu} = {2\pi\over e{\mathcal B}} \; {1 \over (E_{n'}+m)} (\Lambda^\mu
q^\nu + \Lambda^\nu q^\mu - q \cdot \Lambda g^{\mu\nu} - i
\varepsilon^{\mu\nu\alpha\beta} \Lambda_\alpha q_\beta) \,,
\end{eqnarray}
where
\begin{eqnarray}
\Lambda^\alpha &=& \left[I_{n'-1} 
\left({u_y \over \sqrt{e{\mathcal B}}} \right) \right]^2 
(p'^\alpha_\parallel - \widetilde p'^\alpha_\parallel) 
+ \left[ I_{n'} \left({u_y \over \sqrt{e{\mathcal B}}} \right) \right]^2 
(p'^\alpha_\parallel + \widetilde p'^\alpha_\parallel) \nonumber\\* && 
- 2 \sqrt{2n'e{\mathcal B}} g_2^\alpha I_{n'}
\left({u_y \over \sqrt{e{\mathcal B}}} \right) I_{n'-1}
\left({u_y \over \sqrt{e{\mathcal B}}} \right) \,.
\label{Lambda}
\end{eqnarray}
Thus,
\begin{eqnarray}
\Big| {\mathscr M}_{fi} \Big|^2 &=& 8G_\beta^2 \times {2\pi\over
e{\mathcal B}} \; 
{1 \over (E_{n'}+m)}  \Bigg[ (G_V^2+G_A^2) (P\cdot
\Lambda \;
P'\cdot q + P'\cdot \Lambda \; P\cdot q)\nonumber\\* 
&&  
- (G_V^2-G_A^2) m_nm_p q \cdot \Lambda - 2 G_VG_A (P\cdot \Lambda \;
P'\cdot q - P'\cdot \Lambda \; P\cdot q) \nonumber\\* 
&& + S \Big( 2 G_VG_A m_n
(P' \cdot \Lambda q_z + P' \cdot q \Lambda_z) - \Lambda_z q \cdot R +
q_z \Lambda \cdot R \Big) 
\Bigg]  \,. 
\end{eqnarray}

We now choose the axes such that the 3-momentum of the incoming
neutrino is in the $x$-$z$ plane.  We will also assume that $|\vec P'|
\ll m_p$ for the range of energies of interest to us. In that case, it
is easy to see that the terms involving $\sqrt{2n'e{\mathcal B}}$ drop out, and
we obtain
\begin{eqnarray}
\Big| {\mathscr M}_{fi} \Big|^2 = 8G_\beta^2 \times {2\pi\over
e{\mathcal B}} \; 
{m_nm_p \over E_{n'}+m} &\times& \Big[ (G_V^2+3G_A^2) \Omega \Lambda_0
+ (G_V^2-G_A^2) q_z \Lambda_z \nonumber\\*
&& + 2G_A S \Big( (G_V-G_A) \Omega \Lambda_z + (G_V+G_A) q_z\Lambda_0 
\Big) \Big] \,. 
\end{eqnarray}

We now put this expression into Eq.\ (\ref{dsigma}) and calculate the
total cross-section by performing the integrations over different
final state momenta appearing in that formula.  First we integrate
over $P'_x$ and $P'_z$.  These appear only in the momentum conserving
$\delta$-function.  Integration over them therefore just gets rid of
the corresponding $\delta$-functions.  For the integration over
$p_x'$, we refer to Eq.\ (\ref{xi}). Since the center of the
oscillator has to lie between $-\frac12 L_y$ and $\frac12 L_y$, we
conclude that $-\frac12 L_ye{\mathcal B}\leq p'_x\leq\frac12
L_ye{\mathcal B}$. Thus the 
integration over $p_x'$ gives a factor $L_ye{\mathcal B}$.

Putting back into Eq.\
(\ref{dsigma}) and using $V=L_xL_yL_z$, we obtain
\begin{eqnarray}
d\sigma 
= {G_\beta^2\over 4\pi} {\delta (D+\Omega-E_{n'})
 \over \Omega E_{n'}} &\times& 
\Big[ (G_V^2+3G_A^2) \Omega \Lambda_0
+ (G_V^2-G_A^2) q_z \Lambda_z \nonumber\\* 
&& + 2G_AS \Big( (G_V-G_A) \Omega \Lambda_z + (G_V+G_A) q_z\Lambda_0 
\Big) \Big] \; dP'_y dp'_z \,,\nonumber\\
\label{dsigma2}
\end{eqnarray}
where $D$ is the neutron-proton mass difference, $m_n-m_p$.

We next perform the integration over $P'_y$. In the integrand, it
occurs only as the argument of the functions $I_n$ and $I_{n-1}$.
The functions $I_n$ are orthogonal in the sense that
\begin{eqnarray}
\int_{-\infty}^{+\infty} da \; I_n(a) I_{n'}(a) 
= \sqrt{e{\mathcal B}} \; \delta_{nn'} \,.
\end{eqnarray}
This property can be used to perform the integration over $P'_y$.  We
have already remarked that the term proportional to
$\sqrt{2n'e{\mathcal B}}$ in 
Eq.\ (\ref{Lambda}) does not contribute.  {}From other two terms, we
obtain
\begin{eqnarray}
\int dP'_y \, \Lambda^\alpha &=& e{\mathcal B} \Big[ (p'^\alpha_\parallel -
\widetilde p\,'^\alpha_\parallel) (1 - \delta_{n',0}) 
+ (p'^\alpha_\parallel + \widetilde p\,'^\alpha_\parallel) \Big]
\nonumber\\* 
&=& e{\mathcal B} \Big[ g_{n'} p'^\alpha_\parallel + \delta_{n',0}
\widetilde p\,'^\alpha_\parallel \Big] \,,
\label{intdP'y}
\end{eqnarray}
where
\begin{eqnarray}
g_{n'} = 2 - \delta_{n',0}
\label{gn}
\end{eqnarray}
gives the degeneracy of the Landau level.  Notice the appearance of
the Kronecker delta, $\delta_{n',0}$, in the expression of Eq.\
(\ref{intdP'y}).  The reason for this is that, while two terms of Eq.\
(\ref{Lambda}) contribute in the integral for $n'\neq0$, only one of
them contributes for $n'=0$ since $I_{-1}=0$.

The final integration is over $p'_z$.  Writing the argument of the
remaining $\delta$-function in terms of $p'_z$, we find that the zeros
occur when
\begin{eqnarray}
p'_z = p'_\pm \equiv \pm \sqrt{(D+\Omega)^2 - m^2 - 2n'e{\mathcal B}} \,.
\end{eqnarray}
Therefore, 
\begin{eqnarray}
\delta (D+\Omega-E_{n'}) = {D+\Omega \over \sqrt{(D+\Omega)^2 - m^2 -
2n'e{\mathcal B}}} \Big( \delta (p'_z - p'_+) + \delta (p'_z - p'_-) \Big) \,.
\end{eqnarray}
In the integration, the terms proportional to $p'_z$ in the integrand
receive equal and opposite contributions from the two $\delta$
functions and cancel.  For the other terms, independent of $p'_z$,
both the contributions are equal.  So we obtain
\begin{eqnarray}
\sigma_{n'} 
= {e{\mathcal B}G_\beta^2\over 2\pi} && 
\bigg[ g_{n'} \Big\{ (G_V^2+3G_A^2)  + 2G_AS(G_V+G_A) \cos\theta \Big\}
\nonumber\\* 
&& +  \delta_{n',0} \Big\{ (G_V^2-G_A^2) \cos\theta 
+ 2G_AS (G_V-G_A) \Big\} \bigg] 
 {D+\Omega \over \sqrt{(D+\Omega)^2 - m^2 - 2n'e{\mathcal B}}} \,,\nonumber\\
\label{sigman'}
\end{eqnarray}
where we have defined the direction of the incoming neutrino by the
angle $\theta$, with
\begin{eqnarray}
q_z = \Omega \cos\theta \,.
\end{eqnarray}

In Eq.\ (\ref{sigman'}), we have denoted the cross-section by
$\sigma_{n'}$ because the electron ends up in a specific Landau level
$n'$.  The total cross-section is then given as a sum over all
possible values of $n'$, i.e.,
\begin{eqnarray}
\sigma = \sum_{n'=0}^{n'_{\rm max}} \sigma_{n'} 
&=& {e{\mathcal B}G_\beta^2\over
2\pi} \sum_{n'=0}^{n'_{\rm
max}} 
\bigg[ g_{n'} \Big\{ (G_V^2+3G_A^2)  + 2G_AS(G_V+G_A) \cos\theta \Big\}
\nonumber\\* 
&& +  \delta_{n',0} \Big\{ (G_V^2-G_A^2) \cos\theta 
+ 2G_AS (G_V-G_A) \Big\} \bigg] 
 {D+\Omega \over \sqrt{(D+\Omega)^2 - m^2 - 2n'e{\mathcal B}}}  \,.\nonumber\\
\label{sigma}
\end{eqnarray}
The possible allowed Landau level has a maximum, $n'_{\rm max}$, which
is given by the fact that the quantity under the square root sign in
the denominator of Eq.\ (\ref{sigma}) must be non-negative, i.e.,
\begin{eqnarray}
n'_{\rm max} = {\rm int} \left\{ {1\over 2e{\mathcal B}} \Big[(D+\Omega)^2-m^2 
\Big] \right\} \,.
\label{n'max}
\end{eqnarray}

Eq.\ (\ref{sigma}) gives our result for the cross-section of the
inverse beta decay process.  Some properties of this formula are worth
noting. 

For unpolarized neutrons, $S=0$, the cross-section for $n'\neq0$ does
not depend on the direction of the incoming neutrino.  The same is not
true if the electron ends up in the lowest Landau level.  The cross-
section will be asymmetric in this case.

All terms in the cross-section which depend on $S$ have a common
factor $G_A$.  The reason is that, if $G_A$ were equal to zero, the
interaction in the hadronic sector would have been spin-independent.

If the final electron is in the lowest Landau level and the initial
neutrino momentum is antiparallel to the magnetic field, Eq.\
(\ref{sigma}) shows that
\begin{eqnarray}
\sigma_0 = {e{\mathcal B}G_\beta^2\over 2\pi} 
\bigg[  4G_A^2 (1-S) \bigg] 
 {D+\Omega \over \sqrt{(D+\Omega)^2 - m^2}} \,. 
\end{eqnarray}
Note that the vector coupling $G_V$ does not contribute to the cross-
section in this limit.  This can be understood easily.  The neutrino
spin is along the $+z$ direction whereas the electron spin in the
lowest Landau level must be in the $-z$ direction.  Thus there is a
spin-flip in the leptonic sector.  Conservation of angular momentum
then implies that there must be a spin-flip in the hadronic sector as
well.  In the non-relativistic limit for hadrons that we have
employed, this can occur only through the axial coupling.

If further we consider totally polarized neutrons, i.e., $S=1$, we see
that $\sigma_0$ vanishes.  Again, this is a direct consequence of
angular momentum conservation.  Since both initial particles have spin
up, angular momentum conservation requires both final particles in
spin up states as well.  But the spin-up state is not available for
the electron in the lowest Landau level.

\begin{figure}[tb]
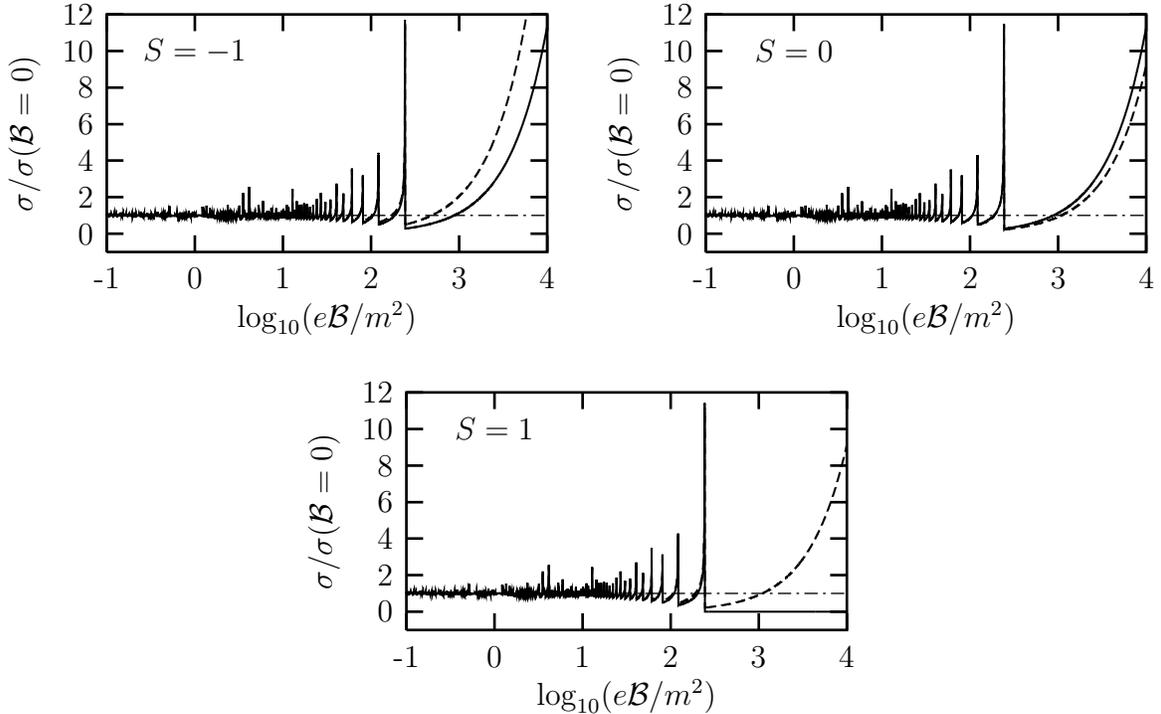

\begin{center}\input{sigma_Sneg.psl} \hfil 
\input{sigma_S0.psl} \\  \bigskip
\input{sigma_Spos.psl}
\end{center}
\caption[Enhancement of inverse beta decay cross-section in a magnetic field.
]  {\small\sf Enhancement of
cross-section in a magnetic field for an initial neutrino energy of
10\,MeV.  Different panels show the results for different net
polarizations of the neutrons.  The solid and the dashed lines
correspond to the initial neutrino momentum antiparallel and parallel
to the magnetic field.  Each curve has been normalized by the
cross-section in field-free case for the same values of $S$ and
$\cos\theta$.  The horizontal dashed lines represent unity in the
vertical scale.}
\label{f:sigma}
\end{figure}
It is instructive to check that the result obtained in Eq.\
(\ref{sigma}) reduces to the known result for the field-free case. The
contribution specific to the zeroth Landau level vanishes in the limit
${\mathcal B}\to0$ owing to the overall factor of $e{\mathcal B}$.
The other terms also have the factor $e{\mathcal B}$, but in this case
we also need to sum over infinitely many states. This gives
\begin{eqnarray}
\sigma &=& {e{\mathcal B}G_\beta^2\over \pi} 
\Big[ (G_V^2+3G_A^2) + 2G_A S (G_V+G_A) \cos\theta \Big] \nonumber\\*
&& \times \left( \sum_{n'=0}^{n'_{\rm max}} {D+\Omega \over
\sqrt{(D+\Omega)^2 - m^2 - 2n'e{\mathcal B}}} - {D+\Omega \over
2\sqrt{(D+\Omega)^2 - m^2}} \right)\,.
\end{eqnarray}
For ${\mathcal B}\to 0$, the last
term vanishes, and we can identify $n'_{\rm max}$ as the integer for
which the denominator of the summand vanishes. Thus we obtain
\begin{eqnarray}
\sigma &\longrightarrow& {e{\mathcal B}G_\beta^2\over \pi} 
\Big[ (G_V^2+3G_A^2) + 2G_A S (G_V+G_A) \cos\theta \Big]
\int_0^{n'_{\rm max}} dn'\; {D+\Omega \over
\sqrt{(D+\Omega)^2 - m^2 - 2n'e{\mathcal B}}} \nonumber\\*
&=& {G_\beta^2\over \pi}  
\Big[ (G_V^2+3G_A^2) + 2G_A S (G_V+G_A) \cos\theta \Big]
(D+\Omega) \sqrt{(D+\Omega)^2 - m^2} \,,
\label{b0inv}
\end{eqnarray}
which is the correct result in the field-free case. 

In \fig{f:sigma}, we have plotted the ratio of the cross-section
to its corresponding value at ${\mathcal B}=0$ as a function of the magnetic
field.  The plots have been done for unpolarized ($S=0$) as well as
totally polarized neutrons along ($S=1$) and opposite ($S=-1$) to the
magnetic field, with the initial neutrino momentum parallel and
antiparallel to the magnetic field.  For $S=-1$, we find that
neutrinos parallel to the magnetic field have a smaller cross-section
than those antiparallel to the field, and the difference is pronounced
for large fields.  For $S=0$, the situation is just reversed.  For
$S=1$, if the magnetic field is high enough so that $n'_{\rm max}=0$,
we see that the cross-section vanishes for neutrino momentum
antiparallel to the field.  The reason for this has already been
discussed.

For the inverse beta decay process, which is symbolically written as in
Eq.~(\ref{invbeta}), the cross-section has been calculated 
earlier by several authors~\cite{Roulet:1997sw, Lai:1998sz}.
They assumed that the matrix element remains unaffected by the
magnetic field, only the modified phase space integral makes the
difference in the cross-section. The magnetic field effects enters
only through the following modification of the phase space integral
and the spin factor of the electron:
\begin{eqnarray}
2\int {d^3p \over (2\pi)^3}  \longrightarrow {e{\mathcal B} \over (2\pi)^2}
\sum_{n'} g_{n'} \int dp_z \,,
\end{eqnarray}
where $g_{n'}$ is the degeneracy of the $n'$-th Landau level as
explained in Eq.~(\ref{gn}).  The sum over $n'$ is restricted to the
region
\begin{eqnarray}
n'< {(D+\Omega)^2 - m_e^2 \over 2e{\mathcal B}} \,,
\end{eqnarray}
where $\Omega$ is the neutrino energy. The results they obtained is
the same as the term proportional to $G_V^2+3G_A^2$ that we obtained.

Subsequent calculations incorporated the modification of wave
functions.  Arras and Lai~\cite{Arras:1998mv}, while still treating
the nucleons as non-relativistic, used the non-relativistic Landau
levels as well as the finiteness of the recoil energy for the proton.
They found the cross-section and went on to derive expressions for the
neutrino opacity.  From the final expressions, one can only recognize
the terms linear in ${\mathcal B}$.  The opacity was calculated also
by Chandra, Goyal and Goswami~\cite{Chandra:2001at}.  Like the
previous authors, they also considered the contribution to the opacity
from other reactions like neutrino-nucleon elastic scattering.  
\subsection{Consequences of neutrino energy spread}\label{es}
The enhancement factor in \fig{f:sigma} shows some spikes.  They
appear at values of the magnetic field for which the denominator of
Eq.\ (\ref{sigma}) vanishes for some $n'$.  For field values larger
than this, that particular Landau level does not contribute to the
cross-section.  To the right of the final spike that appears in the
figure, only the zeroth Landau level contributes.  In other words, the
final electron can go only to the lowest Landau level for such high
values of the magnetic field.  The exact value of ${\mathcal B}$ for
which this occurs depends of course on the energy of the initial
neutrino.

We need to make an important point about these spikes.  Each spike in
fact go all the way up to infinity.  The finite height of a spike in
the figure is an artifact of the finite step size taken in plotting
it.

In reality, of course, a cross-section cannot be infinite. In the
present case, the cross-section remains finite due to the fact that
the initial particles cannot be exactly monochromatic due to the
uncertainty relation. For an example if we concentrate on the initial
neutrino alone then there must be a spread in its energy, which can be
represented by a probability distribution $\Phi(\Omega)$, defined by
\begin{eqnarray}
\int d\Omega \; \Phi(\Omega) = 1 \,.
\end{eqnarray}
In that case, the cross-section in a real experiment should be written
in the form
\begin{eqnarray}
\sigma = \int d\Omega \; \Phi(\Omega) \sigma(\Omega) \,,
\label{sigmaint}
\end{eqnarray}
where $\sigma(\Omega)$ is the expression derived in Eq.\
(\ref{sigma}) for a single value of energy.

\begin{figure}[tbp]
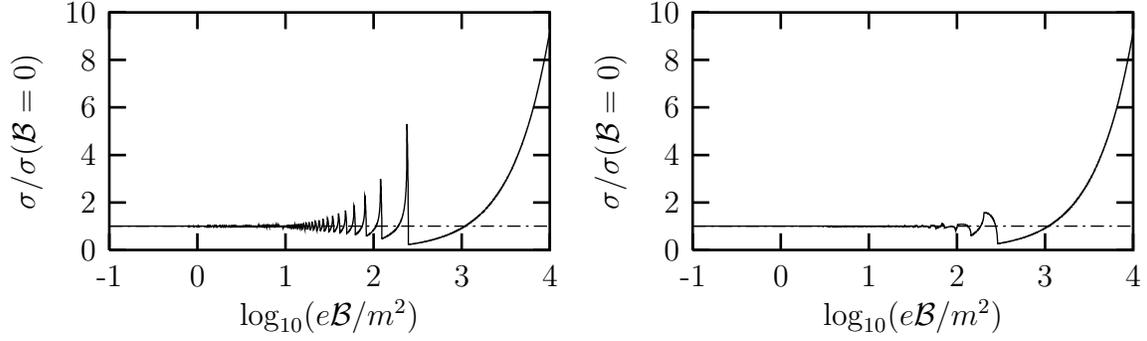

\centerline{\input{smooth1.psl} 
\input{smooth2.psl} }
\caption[The inverse beta decay cross-section in a magnetic field 
for a flat neutrino energy distribution.]{\small\sf Cross-section 
for unpolarized neutrons
as a function of the magnetic field for a flat energy distribution,
normalized to the cross-section in the field-free case. The initial
neutrino momentum is along the magnetic field, and energy is 10\,MeV.
The energy spread $\Omega_2-\Omega_1$ is $0.2$\,MeV for the left panel
and 2\,MeV for the right panel.}\label{f:smooth}
\end{figure}%
As an illustration, we consider the case of unpolarized neutrons
($S=0$), and take a flat probability distribution of initial neutrino
energy, viz.,
\begin{eqnarray}
\Phi(\Omega) = \cases{{\displaystyle 1\over \displaystyle
\Omega_2 - \Omega_1} & if $\Omega_1 \leq \Omega \leq \Omega_2$, \cr \cr
0 & otherwise.}
\label{Edistrn}
\end{eqnarray}
Then the integration of Eq.\ (\ref{sigmaint}) gives
\begin{eqnarray}
\sigma &=& {e{\mathcal B}G_\beta^2\over 2\pi (\Omega_2 - \Omega_1)} \Big[
F(\Omega_2) - F(\Omega_1) \Big] \,,
\end{eqnarray}
where
\begin{eqnarray}
F(\Omega) = \sum_{n'=0}^{n'_{\rm max}}
\Bigg[ g_{n'} (G_V^2+3G_A^2) 
+ \delta_{n',0} (G_V^2-G_A^2) \cos\theta 
\Bigg] 
\times{\sqrt{(D+\Omega)^2 - m^2 - 2n'e{\mathcal B}}} \,,\nonumber\\
\end{eqnarray}
with $n'_{\rm max}$ determined by Eq.\ (\ref{n'max}).
\fig{f:smooth} shows the variation of this quantity with the
magnetic field for $\cos\theta=1$.  In this figure, we normalize the
cross-section by ${\mathcal B}=0$ cross-section with the energy
distribution of Eq.\ (\ref{Edistrn}), which is
\begin{eqnarray}
\sigma ({\mathcal B}=0) = {G_\beta^2 (G_V^2+3G_A^2) \over 3\pi
(\Omega_2-\Omega_1)}  
\bigg( \Big[(D+\Omega_2)^2 - m^2 \Big]^{3/2}
- \Big[(D+\Omega_1)^2 - m^2 \Big]^{3/2} \bigg) \,.
\end{eqnarray}
Keeping the central value of neutrino energy as 10 MeV as before, we
have drawn these plots for two different values of the spread, as
mentioned in the caption.  For the smaller value of the spread in
particular, the graph looks very similar to that drawn in
\fig{f:sigma}, but the difference is that now the height of the
spikes denote the actual enhancement, and is not an artifact of the
plotting procedure.  For the higher value of the energy spread, we see
that the spikes have smoothened out.
\section{Asymmetric emission of neutrinos from a proto-neutron star}
\label{asymnem}
From what has been discussed so far we know that the inverse beta
decay cross-section is direction sensitive through its dependence on the
angle $\theta$, the angle between the neutrino 3-momentum and the
magnetic field direction. This anisotropic effect in the cross-section can have
interesting astrophysical applications in explaining the high
velocities of pulsars, of the order of $450\pm90\;{\rm Km\,s}^{-1}$.
Typical pulsars have masses between $1.0M_\odot$ and $1.5M_\odot$,
i.e., about $2\times10^{33}$g.  The momentum associated with the
proper motion of a pulsar would therefore be of order $10^{41}$
g\,cm/s. On the other hand the energy carried off by neutrinos in a
supernova explosion is about $3\times10^{53}$erg, which corresponds to
a sum of magnitudes of neutrino momenta of $10^{43}$g\,cm/s. Thus an
asymmetry of order of $1\%$ in the distribution of the outgoing
neutrinos would explain the kick of the pulsars. It has been argued
that an asymmetry of this order in the distribution of outgoing
neutrinos can be generated by the anisotropic cross-sections of the
various neutrino related processes in presence of a constant magnetic
field~\cite{Dorofeev:az, Bisnovatyi-Kogan:1997an, Goyal:1998nq,
Bhattacharya:1999bm, Bhattacharya:2002qf, KP2}.  
A simple calculation can make the point clear.

The typical size of a neutron star is about $R\approx 10\,$km. Thus,
when a neutrino, generated at the core reaches a density where its
mean free path is about $R$, it escapes from the star. Therefore the
condition for the neutrino to escape can be written as
\begin{eqnarray}
n_n \sigma = {1\over R} \,,
\end{eqnarray}
where $n_n$ is the neutron number density. We already observed that
$\sigma$ is direction dependent. Therefore, the value of $n_n$ on the
``neutrino sphere'' depends on the direction as well, and the surface
is no longer a sphere. Different values of $n_n$ will correspond to
different temperatures. Thus, neutrinos will be emitted with different
momenta in different directions. This can result in a kick to the
star.

To estimate the magnitude of the kick, let us abbreviate Eq.\
(\ref{sigma}) as
\begin{eqnarray}
\sigma = e{\mathcal B}G_\beta^2 (a + b \cos\theta) \,.
\end{eqnarray}
If we now consider the directions $\theta=0$ and
$\theta=\pi$, the difference in neutron density on the corresponding
points on the neutrino surface is given by
\begin{eqnarray}
\Delta n_n = {1\over e{\mathcal B}G_\beta^2 R}\left(\frac{1}{a-b} -
\frac{1}{a+b}\right) = {2b\over e{\mathcal B}G_\beta^2 a^2 R}\,, 
\end{eqnarray}
neglecting corrections of order $b/a$.

The neutron gas in a typical proto-neutron star can be considered to
be non-relativistic and degenerate. The number density of neutrons is
thus given by~\cite{statmech}
\begin{eqnarray}
n_n = {p_F^3 \over 3\pi^2} \left[ 1 + {\pi^2 m_n^2 T^2 \over 2 p_F^4}
+ \cdots \right] \,, 
\label{n_n}
\end{eqnarray}
where $p_F$ is the Fermi momentum, and we have neglected higher order
terms in the temperature. This gives
\begin{eqnarray}
{dn_n \over dT} = {m_n^2 \over 3} \left( {T\over p_F} \right) \,.
\end{eqnarray}
So the temperature difference between the points on the neutrino
surface in the $\theta=0$ and $\theta=\pi$ directions is
\begin{eqnarray}
\Delta T = {3\over m_n^2} {p_F \over T} {2b \over e{\mathcal B}G_\beta^2a^2R}
\end{eqnarray}

The momentum asymmetry can now be written as
\begin{eqnarray}
{\Delta |\vec{q}| \over |\vec{q}|} = {1\over 6} \cdot {4 \Delta T \over T} \,,
\end{eqnarray}
where we have assumed a black body radiation luminosity ($\propto
T^4$) for the effective neutrino surface. The factor $1/6$ comes in
because the asymmetry pertains only to $\nu_e$, whereas 6 types of
neutrinos and antineutrinos contribute to the energy emitted. This
gives
\begin{eqnarray}
{\Delta |\vec{q}| \over |\vec{q}|} = {4\over m_n^2} {p_F \over T^2} {b
\over e{\mathcal B}G_\beta^2a^2R}  \,,
\label{dk/k}
\end{eqnarray}

To find $p_F$, we use the leading term in Eq.\ (\ref{n_n})
and estimate $n_n$ from the equation
\begin{eqnarray}
n_n = {\rho (1-Y_e) \over m_p} \,,
\end{eqnarray}
where $Y_e$ is the electron fraction and $\rho$ is the mass
density. Taking $Y_e=1/10$, we obtain
\begin{eqnarray}
p_F = 24 \rho_{11}^{1/3}\; {\rm MeV} \,.
\end{eqnarray}
where $\rho_{11}$ is the mass density in units of $10^{11} {\rm g\,
cm}^{-3}$. Putting this back in Eq.\ (\ref{dk/k}), we obtain
\begin{eqnarray}
\left|{\Delta |\vec{q}| \over |\vec{q}|}\right| = 27 \rho_{11}^{1/3}\;
{\mathcal B}_{14}^{-1} 
T_{\rm MeV}^{-2}  {|b| \over a^2}  \,,
\end{eqnarray}
where ${\mathcal B}_{14}={\mathcal B}/(10^{14}\,{\rm Gauss})$ and $T_{\rm MeV}=T/(1\,{\rm
MeV})$. For $\Omega\gg m_e$ which is the relevant case and $S=0$ as a
typical value of neutron polarization we can find out the values of
the constants $a$ and $b$. The values of both of them will in general
depend upon $n'_{\rm max}$ but if we take only the contribution from 
the $n'=0$ level then
\begin{eqnarray}
a = {G_V^2 + 3G_A^2 \over 2\pi} = 9.2 \times 10^{-1} \,.
\end{eqnarray}
and
\begin{eqnarray}
b = {G_V^2 - G_A^2 \over 2\pi} =  - 9.3 \times 10^{-2} \,,
\end{eqnarray}
using $G_V=1$ and $G_A=-1.26$.  
%
%
This gives
\begin{eqnarray}
\left|{\Delta |\vec{q}| \over |\vec{q}|}\right| = 3 \rho_{11}^{1/3}\;
{\mathcal B}_{14}^{-1}  
T_{\rm MeV}^{-2}  \,.
\end{eqnarray}
Obviously, with reasonable choices of $\rho$, ${\mathcal B}$ and $T$, it is
possible to obtain a fractional momentum imbalance of the order of 1\%
which is necessary for explaining the pulsar kicks.
\section{Conclusion}\label{co}
This chapter started with a discussion on neutrino-electron scattering
and other processes related to it by crossing. The basic theory of
spinor solutions in an external magnetic field has been exposed with
some detail in next two sections. A formal field theoretical framework
has been presented based on which cross-sections of any process, which
contains charged particles in the initial or the final state, can be
calculated.  The main calculations concerning the inverse beta decay
cross-section start from the fourth section.

The calculations show that for the inverse beta-decay process, even
for unpolarized neutrons, the cross-section depends on the direction
of the neutrino momentum.  This asymmetry is not surprising since the
background magnetic field makes it an anisotropic problem.  A similar
asymmetry has been noted for URCA processes~\cite{Baiko:1998jq}, where
the neutrino is in the final state. From Eq.~(\ref{sigma}) it is seen
that the anisotropy 
in the cross-section comes only from the zeroth Landau level
contribution of the electron~\cite{Gvozdev:1999md}.  However, for the
higher levels, there is a cancellation between the two possible states
in a Landau level which washes out all angular dependence in these
levels, provided the neutrons are unpolarized.  The asymmetry in
cross-section will therefore come only from the the zeroth Landau
level state of the electrons and its amount will depend on the
relative contribution of this state to the total cross-section. If the
magnetic field is so high that only the zeroth Landau level state can
be obtained for the electron, the asymmetry will be large, about
18\%. For smaller and smaller magnetic fields, the asymmetry decreases
with new Landau levels contributing.  For polarized neutrons, however,
there is an asymmetry even in the field-free case as is evident from
Eq.~(\ref{b0inv}). In presence of a magnetic field, the asymmetry will
in general depend on the magnetic field, as it appears from various
plots in \fig{f:sigma}.

From Eq.~(\ref{b0inv}) it is seen that the inverse beta decay
cross-section has a smooth ${\mathcal B}\to 0$ limit. From the curves
in \fig{f:sigma} it is observed that the cross-section has spikes, which
go all the way to infinity, for specific values of the magnetic field
magnitude. Subsection \ref{es} deals with the issue of these infinite
spikes. It is shown that if instead of a monochromatic neutrino the
initial neutrino has an energy spread, as is expected in a real
situation, then the spikes gets smoothed up. This fact is evident from
\fig{f:smooth}.

The anisotropic nature of the inverse beta decay cross-section can
have far reaching consequences for neutrino emission from a
proto-neutron star as has been shown in the section concerning
asymmetric emission of neutrinos. It has been discussed in the
literature that the presence of asymmetric magnetic fields can cause
asymmetric neutrino emission from a proto-neutron star~\cite{Bisno}. 
However, the calculations in this chapter show that even
with a uniform magnetic field, neutrino emission would be asymmetric
because of the $\cos\theta$-dependent terms in the cross-section.

\chapter{Observable effects due to 
intermediate virtual charged particles in a magnetic field}\label{chap3}
\section{Introduction}
In quantum field theoretic calculations, the spinors given in the
previous chapter in Eq.~(\ref{Usoln}) and Eq.~(\ref{Vsoln}) should be
used if the charged particle appears in the initial or the final state
of a physical process. If, on the other hand, the charged particle
appears in the internal lines of the Feynman diagram of the process
then we should use its propagator.

This chapter is on propagators for charged particles, mainly 
fermions, in presence of a background uniform magnetic field. The
calculations will be extrapolated to the case where there is a
background medium also. For the medium modification of the propagators
a brief discussion on statistical field theory (commonly
called finite temperature field theory) is discussed in the appendix
\ref{rtf}. The Schwinger propagator is introduced in
this chapter and is followed by a discussion on the phase factor
accompanying it. A detailed discussion on the derivation of the
Schwinger propagator is supplied in \cite{Chyi:1999fc}. The phase
factor of the Schwinger propagator is dealt in some detail. In most
of the works this phase factor is treated with irrelevance but the
reason is rarely explained.

In the chapter \ref{chap1} we noticed that the standard model
neutrinos do not have any magnetic moment and so they cannot interact
with external magnetic fields. In going beyond standard model i.e. by
including right handed neutrinos we can generate a magnetic moment of
the neutrinos. Magnetic fields can also affect the neutrino properties
in a different way i.e., through the charged particle propagators as
discussed in this chapter. This indirect effect of the magnetic fields
from the propagators can affect neutrino self-energy calculations.
Without doing actual loop calculations it is possible to predict the
form of the self-energy expressions in a magnetic field and in a
magnetized medium using a purely general form-factor
analysis. Although such an analysis is purely formal but still it can
predict interesting effects, like the dependence of the neutrino
self-energy on the angle between the neutrino propagation direction
and the external magnetic field direction, in a magnetized medium. The
fact that the neutrino self-energy is sensitive to the angle between
the neutrino propagation direction and the magnetic field direction
can have far reaching consequences. In the field of neutrino
oscillations the angular dependence of the self-energy can modify the
resonant level crossing condition of the neutrinos. The resonant
condition itself becomes sensitive to the angle between the
propagating neutrinos and the external magnetic field which can have
an interesting astrophysical application like the asymmetric emission
of neutrinos from a newly formed neutron star. 

In the two following sections the fermion propagator in a magnetic
field in vacuum and in a thermal medium will be discussed. In the
remaining sections the topic of neutrino self-energy in a magnetic
field and the effect of magnetic fields on neutrino oscillations will
be discussed briefly. The chapter ends with a conclusion which
summarizes the various things discussed in this chapter.
\section{Propagators in vacuum}
\label{propv}
\subsection{Furry propagator}
There are two ways to write the propagator.  The first is to start
with the fermion field operator $\psi(X)$ written in terms of the
spinor solutions and the creation and annihilation operators, and
construct the time ordered product, as is usually done for finding the
propagator of a free fermion field in the vacuum.  The algebra is
straight forward and yields the result
\begin{eqnarray}
i S^V_B (X,X') = i \sum_N \int \frac{dp_0\, dp_x\, dp_z}{(2\pi)^3} \;
{E_n+m \over p_0^2 - E_n^2 + 
i\epsilon} e^{-ip \cdot (X\omit y - X'\omit y)} \nonumber\\*
\times \sum_s U_s (y,N,\vec p\omit y) \overline U_s (y', N,\vec p\omit
y) \,,
\label{Furryprop}
\end{eqnarray}
where $E_n$ is the positive root obtained from \eqn{E}. The notation
requires some explanation. Instead of $x$ use has been made of $X$ as
the coordinates in the left hand side of the above equation. This
style of writing is carried over from section \ref{so}. It is
motivated by the fact that in calculations where we employ these
propagators we often have to write down  $\vec p\cdot
\vec X{\omit y} \equiv p_xx+p_zz$ in the explicit form where the
quantity $x$ appears as a coordinate and not as the 4-vector
itself. The notation $S^V_B (X,X')$ specifies that the propagator is
in vacuum and in presence of a uniform background magnetic field whose
magnitude is ${\mathcal B}$. The spin sum appearing in
\eqn{Furryprop} is given in Eq.~(\ref{PU}). The resulting
propagator is called the propagator in the Furry picture.
\subsection{Schwinger propagator}\label{ss:sp}
Alternatively, one uses the propagators introduced by Schwinger
\cite{Schwinger:nm}. Schwinger's calculation of the fermion propagator
relies on a functional procedure and it is written in the form
\begin{eqnarray}
iS^V_B(x,x') = \kappa(x,x') \int \frac{d^4 p}{(2 \pi)^4} e^{-ip \cdot
(x-x')} iS^V_B (p) \,,
\label{schwingprop}
\end{eqnarray}
here $x$ stands for the coordinate 4-vector as usual.  $S^V_B(p)$
is expressed as an integral over a variable $s$, usually (though
confusingly) called the `proper time':
\begin{eqnarray}
i S^V_B (p) =\int_0^\infty ds\; e^{\Phi(p,s)} G(p,s) \,.
\label{SB}
\end{eqnarray}
The quantities $\Phi(p,s)$ and $G(p,s)$ can be written in the
following way :
\begin{eqnarray}
\Phi(p,s) &\equiv& is \left( p_\parallel^2 - {\tan
(e{\mathcal B}s) \over e{\mathcal B}s} \, p_\perp^2 - m^2 \right) - 
\epsilon|s| \,, 
\label{Phi}
\\
G(p,s) &\equiv&  {e^{ie{\mathcal B}s\sigma\!_z} \over \cos(e{\mathcal
B}s)} \;
\left( 
\rlap/p_\parallel + {e^{-ie{\mathcal B}s\sigma_z} \over \cos(e{\mathcal B}s)}
\rlap/ p_\perp + m \right) \nonumber\\*
&=& ( 1 + i\sigma_z \tan(e{\mathcal B}s) ) 
(\rlap/p_\parallel + m ) + \sec^2 (e{\mathcal B}s) \rlap/ p_\perp \,.
\label{G}
\end{eqnarray}
In the above expressions $\sigma_z=i\gamma^1\gamma^2$.
In a typical loop diagram, one therefore will have to perform not only
integrations over the loop momenta, but also over the proper time
variables.  $i S_B (p)$ is manifestly translation invariant. Most of
the calculations involving the Schwinger propagator use this
translation invariant part only, while the other term
%
$\kappa(x,x')$
%
appearing in
the propagator in Eq.~(\ref{schwingprop}) remains irrelevant. 

Not going into any detailed description of $\kappa(x,x')$ we can simply
understand its necessity in the propagator. A propagator connects two
points in space-time. In presence of a background gauge field the
gauge transformation property of the fields of the charged fermions
are different at two different space-time points. Unless there is some
factor in the propagator which can connect these two fields at
different space-time points with different gauge transformation
properties, the calculations involving charged fermion propagators
will not be manifestly gauge invariant.  In the limit the background
magnetic field goes to zero Eq.~(\ref{SB}) shows that $i S^V_B (p)$
reduces to the vacuum fermionic propagator. This indicates that $i
S^V_B (p)$ in Eq.~(\ref{schwingprop}) do not carry any information
about gauge invariance of the background gauge field, it is only the
translation invariant modified version of the vacuum fermionic
propagator. Therefore the gauge transformation property of the
propagator must be related to $\kappa(x,x')$, and its presence is
necessary for the overall gauge invariance of physical processes. The
gauge transformed fields comes with phase factors where the
phase depends upon the space-time point where the gauge transformation
is made. The fermionic fields at two different space-time points will
therefore have two different phase factors. To make a connection
between them $\kappa(x,x')$ must also be some form of a phase.
Conventionally it is named the phase-factor.

The phase factor is generally written as 
\begin{eqnarray}
\kappa(x,x') = \exp \left\{ ie I(x,x')\right\}
\label{PF}
\end{eqnarray}
where
\begin{eqnarray}
I(x,x') = \int_{x'}^x d\xi^\mu \left[A_\mu(\xi) +
\frac12 F_{\mu \nu} (\xi - x')^\nu\right]
\label{IPF}
\end{eqnarray}
where $A_\mu$ is the background gauge field, and $ F_{\mu\nu}$ is the
field strength tensor. From Eq.~(\ref{IPF}) we notice that the phase
factor breaks the translation invariance of the propagator.

For a constant background field we can always write the gauge field as
\begin{eqnarray}
A_\mu(\xi) = - \frac12 F_{\mu \nu} \xi^\nu  + \partial_\mu \lambda(\xi)\,,
\label{ConsF}
\end{eqnarray}
where $\lambda(\xi)$ is an arbitrary well behaved function and depends
upon our choice of gauge.
Using the above relation in conjunction with Eq.~(\ref{IPF}) we can
simplify the integration appearing in the phase factor as
\begin{eqnarray}
I(x,x') = \int_{x'}^x d\xi^\mu \left[- \frac12
F_{\mu \nu} x'^\nu + \partial_\mu \lambda(\xi) \right]\,.
\end{eqnarray}
Using the constancy of the field strength tensor the above expression
can be written as
\begin{eqnarray}
I(x,x')= \frac12 x'^\mu F_{\mu \nu} x^\nu + \lambda(x) - \lambda(x')\,. 
\label{compactIPF}
\end{eqnarray}
From Eq.~(\ref{compactIPF}) we can immediately see if we set $x = x'$,
in other words if we integrate over any closed contour in space-time
$I(x,x')$ vanishes. Thus $I(x,x')$ connecting two points in space-time
is independent of the path joining them, and as a result the phase
factor of the Schwinger propagator joining the points $x'$ and $x$ in
Eq.~(\ref{PF}) is also path independent.

Utilizing the path independence of the phase factor of the propagator 
the general convention is to choose a straight line path connecting
the two points $x'$ and $x$. Points on this path are represented by
\begin{eqnarray}
\xi^\mu = (1 - \zeta) x'^\mu + \zeta x^\mu\,,
\label{stpath}
\end{eqnarray}
where the parameter $\zeta$ ranges from $0$ to $1$. Using
Eq.~(\ref{IPF}) and the straight line path given in Eq.~(\ref{stpath})
we get
\begin{eqnarray}
I(x,x') &=& \int_{x'}^x d\xi^\mu A_\mu(\xi) + \frac{\zeta}{2}
\int_0^1 d\zeta (x^\mu - x'^\mu)F_{\mu \nu}(x^\nu - x'^\nu)\,,\\\nonumber
       &=&  \int_{x'}^x d\xi^\mu A_\mu(\xi)\,.
\label{AI}
\end{eqnarray}
%
%
%
Now using Eq.~(\ref{ConsF}) for the gauge field we can retrieve
Eq.~(\ref{compactIPF}). 

From Eq.~(\ref{compactIPF}) it is clear that the phase factor is
dependent on the form of the function $\lambda(\xi)$, that is to say
the fermion propagator is dependent on the gauge in which the constant
background magnetic field is specified. 
Suppose we are working in such a gauge that $\lambda(\xi) = 0$, and
then we make a gauge transformation of the background field as
\begin{eqnarray}
A_\mu \to A_\mu + \partial_\mu \lambda(\xi)
\end{eqnarray}
then from Eq.~(\ref{schwingprop}) it follows that the fermion
propagator will transform as 
\begin{eqnarray}
iS^V_B(x,x') \to \exp(ie\lambda(x))iS^V_B(x,x')\exp(-ie\lambda(x'))\,.
\label{GTP}
\end{eqnarray}
under the gauge transformation. As because we are working in presence
of a background gauge field, the fields of the charged particles and
their propagators both become background gauge dependent. This is the
reason why the phase factor arises in the expression of the
propagator.  A detailed study of the phase factor appears in appendix
\ref{s:LCP}.
\section{Schwinger propagator in presence of a medium}
\label{smed}
In presence of a medium the propagators of the elementary particles
get changed. There are various formulations of statistical field
theory, like the imaginary time method, the real time formalism, to
name the most prominent ones. In the present work we will be using the
real time formalism developed in the canonical approach~\cite{Nieves:1990ne}.
Some of the important points of the above formalism is discussed in
appendix \ref{rtf}. 

In this formalism it is not difficult to obtain the effects of a
magnetized medium on the Schwinger propagator.  In the real-time
formalism, the 1-1 component of the propagator $iS(p)$ involving the
time-ordered product\footnote{It should be mentioned here that other
orderings also appear in the evaluation of general Green's functions.
The other orderings give rise to other propagators as discussed in
appendix \ref{rtf}.  We will not need those other propagators
here.}  can be written in terms of the free propagator $iS_0(p)$:
\begin{eqnarray}
iS(p) = iS_0(p) - \eta_F(p\cdot u) \Big[ iS_0(p) - i\overline S_0(p) 
\Big] \,,
\label{vacrel}
\end{eqnarray}
where 
\begin{eqnarray}
\overline S_0(p) = \gamma_0 S_0^\dagger(p) \gamma_0 \,,
\end{eqnarray}
and $\eta_F(p\cdot u)$ contains the distribution function for particles and
antiparticles:
\begin{eqnarray}
\eta_F(p\cdot u) = \Theta(p\cdot u) f_F(p,\mu,\beta) 
+ \Theta(-p\cdot u) f_F(-p,-\mu,\beta) \,.
\label{eta}
\end{eqnarray}
Here, $\Theta$ is the step function which takes the value $+1$ for
positive values of its argument and vanishes for negative values of
the argument, $u^\mu$ is the 4-vector denoting the center-of-mass
velocity of the background plasma. Conventionally $u$ is normalized in
such a way that in the rest frame of the medium
\begin{eqnarray}
u_\mu = (1,\vec{0})\,.
\label{urf}
\end{eqnarray}
$f_F$ denotes the Fermi-Dirac
distribution function:
\begin{eqnarray}
f_F(p,\mu,\beta) = {1\over e^{\beta(p\cdot u - \mu)} + 1} \,.
\label{distrib}
\end{eqnarray}
In a similar manner, the 1-1 component of the propagator in a
magnetized plasma is given by~\cite{Elmfors:1996gy, ditt}
\begin{eqnarray}
iS_B(p) = iS^V_B(p) - \eta_F(p\cdot u) \Big[ iS_B(p) - i\overline S_B(p)
\Big] \,.
\end{eqnarray}
In the Schwinger proper-time representation, this can also be written
as an integral over the proper-time variable $s$:
\begin{eqnarray}
iS_B(p) = \int_0^\infty ds\; e^{\Phi(p,s)} G(p,s) 
- \eta_F(p\cdot u) \int_{-\infty}^\infty ds\; 
e^{\Phi(p,s)} G(p,s) \,,
\label{fullprop}
\end{eqnarray}
where $\Phi(p,s)$ and $G(p,s)$ are given by the expressions in
\eqn{Phi} and \eqn{G}. Symbolically the above equation can be written
as
\begin{eqnarray}
iS_B(p) = iS^V_B(p) + S^\eta_B(p)
\label{sbp}
\end{eqnarray}
where $iS^V_B(p)$ is the vacuum propagator in a uniform magnetic field
and is given by Eq.~(\ref{SB}). The other part $S^\eta_B(p)$ of the
propagator carries the medium effect, and is given by
\begin{eqnarray}
S^\eta_B(p) = - \eta_F(p\cdot u) \int_{-\infty}^\infty ds\; 
e^{\Phi(p,s)} G(p,s) \,.
\label{Seta}
\end{eqnarray}
It is straight forward to see that when ${\mathcal B}=0$, the
propagator in Eq.\ (\ref{SB}) reduces to 
\begin{eqnarray}
iS_0^V (p) &=& \int_0^\infty ds  \; \exp
\left[ is \left( p^2 - m^2 + i\epsilon \right) \right] \left( 
\rlap/p + m \right) \nonumber\\*
&=& i\, {\rlap/p + m \over p^2-m^2+i\epsilon} \,,
\end{eqnarray}
which is the normal vacuum Feynman propagator.  In the same limit, the
background dependent part reduces to
\begin{eqnarray}
S_0^\eta (p) = - 2\pi \, \delta(p^2-m^2) \eta_F(p\cdot u) (\rlap/p+m) \,,
\end{eqnarray}
which is what we expect in a medium in absence of a magnetic field.
\section{Neutrino self-energy}\label{vise}
\begin{figure}[btp]
\begin{center}
\begin{picture}(180,75)(-90,-20)
\ArrowLine(80,0)(40,0) 
\Text(60,-10)[b]{$\nu$} 
\ArrowLine(40,0)(-40,0)
\Text(0,-10)[b]{$\ell$} 
\ArrowLine(-40,0)(-80,0)
\Text(-60,-10)[b]{$\nu$} 
\PhotonArc(0,0)(40,0,180){2.5}{10.5}
\ArrowArcn(0,27)(20,120,60)
\Text(0,53)[b]{$W^-$}
\Text(0,-20)[]{\bf (a)}
\end{picture}
\qquad
\begin{picture}(180,75)(-90,-20)
\ArrowLine(50,0)(0,0) 
\Text(25,-10)[c]{$\nu$} 
\ArrowLine(0,0)(-50,0)
\Text(-25,-10)[c]{$\nu$} 
\Photon(0,0)(0,30)24
\Text(5,15)[l]{$Z$}
\ArrowArc(0,45)(15,-90,270)
\Text(0,-20)[]{\bf (b)}
\end{picture}
\caption[One-loop diagrams for neutrino self-energy in a
magnetized medium.]{\sf One-loop diagrams for neutrino self-energy in a
magnetized medium.  Diagram b is absent if the background contains
only a magnetic field but no matter.  For legends and related
diagrams, see the caption of \fig{f:magmom}.
\label{f:selfen}}
\end{center}
\end{figure}
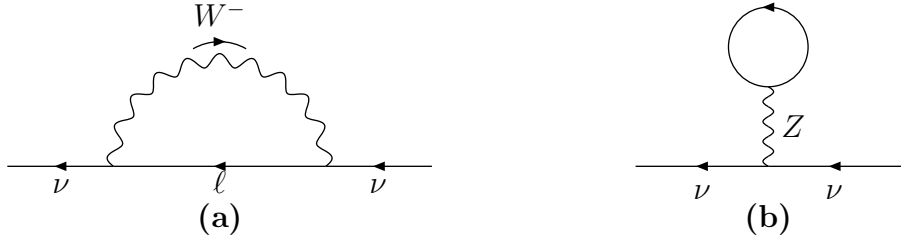
In chapter \ref{chap1} the self-energy of the neutrino was discussed in
a background magnetic field and an expression of the self-energy was
supplied in Eq.~(\ref{intro-nself}). Here we take up the issue and
briefly discuss about the form of the self-energy
expression~\cite{Bhattacharya:2002aj}.
Skipping detailed calculations using the Schwinger propagators 
a form-factor analysis of the neutrino self-energy is given in this
section and its implications are highlighted in the case of neutrino
oscillations.

It is already mentioned in chapter \ref{chap1} that the simplest
physical quantity where background magnetic field effects appear
through virtual lines of charged particles is the self-energy of the
neutrino. The 1-loop diagram for the self-energy is given in
\fig{f:selfen}. In the present discussion we take the case of neutrinos in a
medium with a magnetic field. From this general approach we will be
able to find the form of the self-energy of the neutrino in absence of
a medium also.
 
It is easy to see how the self-energy might be modified within a
magnetized plasma.  In the vacuum, the self-energy of a fermion has
the general structure
\begin{eqnarray}
\Sigma(q) = a \gamma^\mu q_\mu + b \,,
\end{eqnarray}
which is the most general form dictated by Lorentz covariance.  Here,
$a$ and $b$ are Lorentz invariant, and can therefore depend only on
$q^2$.  In the presence of a homogeneous medium, the self-energy will
involve the 4-vector $u^\mu$.  Further, if the medium contains a
background magnetic field, the background field $F_{\mu\nu}$ also
enters the general expression for the self-energy.  These new objects,
$u^\mu$ and $F_{\mu\nu}$, enter in two different ways.  First, any
form factor now can depend on more Lorentz invariants which are
present in the problem.  Second, the number of form factors also
increases, since it is possible to write some more Lorentz covariant
terms using $u^\mu$ and $F_{\mu\nu}$. There will in fact be a lot of
form factors in the most general case.  However, if we have chiral
neutrinos as in the standard electroweak theory, the
expression is not very complicated~\cite{Bhattacharya:2002aj}:
\begin{eqnarray}
\Sigma_B(q) = 
\Big( a_1 q_\mu + b_1 u_\mu 
+ a_2 q^\nu F_{\mu\nu} + b_2 u^\nu F_{\mu\nu} 
+ a_3 q^\nu \widetilde F_{\mu\nu} + b_3 u^\nu \widetilde F_{\mu\nu} 
\Big) \gamma^\mu L \,, 
\label{Sigma}
\end{eqnarray}
where $L$ is the left-chiral projection operator, and
\begin{eqnarray}
\widetilde F_{\mu\nu} = \frac12 \epsilon_{\mu\nu\lambda\rho}
F^{\lambda\rho} \,. 
\end{eqnarray}

We first consider the self-energy when the background consists of a
pure magnetic field, without any matter.  Then all $b$-type
form-factors disappear from the self-energy.  The dispersion relation
of neutrinos can then be obtained by the zeros of $\rlap/q-\Sigma_B$,
which gives
\begin{eqnarray}
\Big[ (1-a_1) q^\mu - a_2 q^\nu F_{\mu\nu} 
- a_3 q^\nu \widetilde F_{\mu\nu} \Big]^2 = 0 \,.
\end{eqnarray}
It is interesting to note that the terms linear in the background
field all vanish due to the antisymmetry of the field tensor.
Moreover, the $a_2a_3$ term is also zero for a purely magnetic
field. This gives
\begin{eqnarray}
(1-a_1)^2 q^\mu q_\mu 
+ a_2^2 q^\nu q_\lambda F_{\mu\nu} F^{\mu\lambda} 
+ a_3^2 q^\nu q_\lambda \widetilde F_{\mu\nu} \widetilde F^{\mu\lambda} 
+ 2a_2 a_3 q^\nu q_\lambda F_{\mu\nu} \widetilde F^{\mu\lambda} =0 \,.
\end{eqnarray}

The remaining terms can be most easily understood if we take the
$z$-axis along the direction of the magnetic field.  Then the
only non-zero components of the tensor $F_{\mu\nu}$ and $\widetilde
F_{\mu\nu}$ are given by
\begin{eqnarray}
F_{12} = - F_{21} = {\mathcal B} \,, \qquad
\widetilde F_{03} = - \widetilde F_{30} = {\mathcal B} \,,
\label{edual}
\end{eqnarray}
where we have adopted the convention
\begin{eqnarray}
\epsilon_{0123} = +1 \,.
\label{leviciv}
\end{eqnarray}
Thus
\begin{eqnarray}
q^\nu q_\lambda F_{\mu\nu} F^{\mu\lambda} &=& - (q_x^2 + q_y^2)
{\mathcal B}^2
= - q_\perp^2 {\mathcal B}^2 \,,
\nonumber\\*
q^\nu q_\lambda \widetilde F_{\mu\nu} \widetilde F^{\mu\lambda} &=& 
(\Omega^2 - q_z^2) {\mathcal B}^2 = q_\parallel^2 {\mathcal B}^2 \,,
\end{eqnarray}
where the notations for parallel and perpendicular products were
introduced in the chapter \ref{chap1}. The form factor $a_1$ can be
set equal to zero by a choice of the renormalization prescription.  So
the dispersion relation is now a solution of the equation
\begin{eqnarray}
q^2 - a_2^2 q_\perp^2 {\mathcal B}^2 + a_3^2 q_\parallel^2 {\mathcal
B}^2 =0 \,, 
\end{eqnarray}
which can also be written as
\begin{eqnarray}
\Omega^2 = q_z^2 + {1 + a_2^2 {\mathcal B}^2 \over 1 + a_3^2 {\mathcal
B}^2} \, q_\perp^2 \,. 
\end{eqnarray}
Of course, this should not be taken as the solution for the neutrino
energy, because the right hand side contains form factors which, in
general, are functions of the energy and other things.  But at least
it shows that in the limit ${\mathcal B}\to0$, the vacuum dispersion
relation is recovered.  If we retain the lowest order corrections in
${\mathcal B}$, we can treat the form-factors to be independent of
${\mathcal B}$ and can write~\cite{Bhattacharya:2002aj}
\begin{eqnarray}
\Omega^2 = \vec q^2 + (a_2^2 - a_3^2) {\mathcal B}^2 q_\perp^2 \,.
\label{disp_B}
\end{eqnarray}
Calculation of this self-energy was performed by Erdas and Feldman
\cite{Erdas:gy} using the Schwinger propagator, where they also
incorporated the modification of the $W$-propagator due to the
magnetic field.  Importantly, the $W$-propagator contains the same
phase factor as discussed in appendix \ref{s:LCP}. Therefore, the
phase factors from the charged lepton and the $W$-lines are of the
form $\kappa(x,x')\kappa(x',x)$. From appendix \ref{s:LCP} it is easy
to see that this is equal to unity, and therefore the phase factors do
not contribute to the final expression.  Detailed calculations show
that~\cite{Erdas:gy}
\begin{eqnarray}
a_2^2 - a_3^2 = {eg \overwithdelims() 2\pi M_W^2}^2
\left( \frac13 \ln {M_W \over m} + \frac18 \right) \,,
\label{a2sq-a3sq}
\end{eqnarray}
where $m$ is the mass of the charged lepton in the internal line.  The
above equation in conjunction with Eq.~(\ref{disp_B}) gives us
Eq.~(\ref{intro-nself}) appearing in the chapter \ref{chap1}.

Let us next concentrate on the terms which can occur only in a
magnetized medium.  In other words, we select out the terms which
cannot occur if the neutrino propagates in a background of pure
magnetic field without any material medium.  This means that, apart
from the term $\rlap/q$ which occurs also in the vacuum, we look for
the terms which contain both $u^\mu$ and $F_{\mu\nu}$.  Further, if
the background field is purely magnetic in the rest frame of the
medium, $u^\nu F_{\mu\nu}=0$ since $u$ has only the time component
whereas the only non-zero components of $F_{\mu\nu}$ are spatial.
Thus we are left with~\cite{D'Olivo:1989cr}:
\begin{eqnarray}
\Sigma_B(q) = 
\Big( a_1 q_\mu + b_1 u_\mu + b_3 u^\nu \widetilde F_{\mu\nu} 
\Big) \gamma^\mu L \,.
\label{Sigma_uB}
\end{eqnarray}
Once again, setting $a_1=0$ through a renormalization prescription, we
can find the dispersion relation of the neutrinos in the
form~\cite{D'Olivo:1989cr}: 
\begin{eqnarray}
\Omega = \Big| \vec q - b_3 \vec {\mathcal B} \Big| + b_1 \approx |\vec q| -
b_3 \hat q \cdot \vec {\mathcal B} + b_1 \,,
\label{magdisp}
\end{eqnarray}
where $\hat q$ is the unit vector along $\vec q$, and we have kept
only the linear correction in the magnetic field.  This form for the
dispersion relation was first arrived at by D'Olivo, Nieves and Pal
(DNP)~\cite{D'Olivo:1989cr} who essentially performed a calculation to
the first order in the external field.  As for the form factors, $b_1$
was known previously, obtained from the analysis of neutrino
propagation in isotropic matter, i.e., without any magnetic field.
The result was~\cite{Wolfenstein:1977ue, Notzold:1987ik, Pal:1989xs,
Nieves:ez}
\begin{eqnarray}
b_1 &=& \surd2 G_F (n_e - n_{\bar e}) \times \left(y_e +
\rho c_V \right) \,,  
\end{eqnarray}
where 
\begin{eqnarray}
\rho = {M_W^2 \over M_Z^2\cos^2\theta_W} \,,
\end{eqnarray}
$n_e$, $n_{\bar e}$ are the densities of electrons and positrons
in the medium, 
\begin{eqnarray}
y_e = \cases{1 & for $\nu_e$, \cr
0 & for $\nu\neq\nu_e$,}
\label{Ye}
\end{eqnarray}
and $c_V$ is defined through the coupling of the electron to the
$Z$-boson, whose Feynman rule is
\begin{eqnarray}
-{ig \over 2\cos\theta_W} \; \gamma_\mu (c_V - c_A\gamma_5) \,.
\end{eqnarray}
In other words, in the standard model
\begin{eqnarray}
c_V = - \frac12 + 2 \sin^2 \theta_W \,, \qquad c_A = - \frac12 \,.
\label{cvca}
\end{eqnarray}

The contribution to $b_3$ from background electrons and positrons was
calculated by DNP~\cite{D'Olivo:1989cr}.  They obtained\footnote{The
authors of \cite{D'Olivo:1989cr} used a convention in which
$e<0$.  Here we present the result in the convention $e>0$.}
\begin{eqnarray}
b_3 &=& -2\surd2eG_F 
\int {d^3p \over (2\pi)^3 2E} \; {d\over dE} (f_e - f_{\bar e})
\times \left(y_e + \rho c_A \right) \,, 
\label{mb3}
\end{eqnarray}
where $f_e$ and $f_{\bar e}$ are the Fermi distribution functions for
electrons and positrons, and
\begin{eqnarray}
E = \sqrt{\vec p^2 + m_e^2} \,.
\end{eqnarray}
Later authors have improved on this result in two different ways.
Some authors~\cite{D'Olivo:1997vi} have included the contributions
coming from nucleons in the background.  Some 
others~\cite{Elmfors:1996gy,Erdas:1998uu} have used the
Schwinger propagator and extended the results to all orders in the
magnetic field.
\section{The role of neutrino oscillations in a magnetized medium}
\label{noscillations}
Neutrino oscillation phenomenology in a medium has produced the
important MSW effect~\cite{Wolfenstein:1977ue, Mikheev:wj} which is now
widely applied in the solar neutrino problem and also in the case of
neutrinos coming from supernovas. But in most of the cases there is a
background magnetic field and so the following discussion is an
important part of neutrino oscillation phenomenology.

Calculation of neutrino self-energy has a direct consequence on
neutrino mixing and oscillations.  Of course neutrino oscillations
require neutrino mixing and therefore neutrino mass.  For the sake of
simplicity, we discuss mixing between two neutrinos which we will call
$\nu_e$ and $\nu_\mu$.  The eigenstates will in general be called
$\nu_1$ and $\nu_2$, which are given by
\begin{eqnarray}
{\nu_1 \choose \nu_2} = \left( \begin{array}{ccc}
\cos\theta && -\sin\theta \\
\sin\theta && \cos\theta
			\end{array} \right)
{\nu_e \choose \nu_\mu} \,.
\end{eqnarray}
We will denote the masses of the eigenstates by $m_1$ and $m_2$, and
assume that the neutrinos are ultra-relativistic.  Then in the vacuum,
the evolution equation for a beam of neutrinos will be given by
\begin{eqnarray}
i{d\over dt} {\nu_e \choose \nu_\mu} = {1\over 2\omega} M^2 
{\nu_e \choose \nu_\mu} \,.
\end{eqnarray}
where the matrix $M^2$ is given by
\begin{eqnarray}
M^2 = \left( \begin{array}{ccc}
-\frac12 \Delta m^2 \cos 2\theta && \frac12 \Delta m^2 \sin 2\theta \\
\frac12 \Delta m^2 \sin 2\theta && \frac12 \Delta m^2 \cos 2\theta
			\end{array} \right) \,,
\end{eqnarray}
where $\Delta m^2=m_2^2-m_1^2$.  In writing this matrix, we have
ignored all terms which are multiples of the unit matrix, which affect
the propagation only by a phase which is common for all the states.

In a non-trivial background, the dispersion relations of the neutrinos
change, as discussed in Sec.~\ref{vise}.  This adds new terms to the
diagonal elements of the effective Hamiltonian in the flavor basis,
which we denote by the symbol $A$.  As a result, the matrix $M^2$
should now be replaced by
\begin{eqnarray}
\widetilde M^2 = \left( \begin{array}{ccc}
-\frac12 \Delta m^2 \cos 2\theta + A_{\nu_e} 
&& \frac12 \Delta m^2 \sin 2\theta \\
\frac12 \Delta m^2 \sin 2\theta 
&& \frac12 \Delta m^2 \cos 2\theta + A_{\nu_\mu}
			\end{array} \right) \,,
\label{MM}
\end{eqnarray}
where the extra contributions are in general different for $\nu_e$ and
$\nu_\mu$.  The eigenstates and eigenvalues change because of these
new contributions.  For example, the mixing angle now becomes
$\widetilde\theta$, given by
\begin{eqnarray}
\tan 2\widetilde\theta = {\Delta m^2 \sin 2\theta \over \Delta m^2
\cos 2\theta + A_{\nu_\mu} - A_{\nu_e} } \,.
\label{RC}
\end{eqnarray}

In a pure magnetic field, the self-energies were shown in
\eqn{disp_B} and \eqn{a2sq-a3sq}.  The quantity $m$ appearing in
\eqn{a2sq-a3sq} is the mass of the charged lepton in the loop.  Thus,
for $\nu_e$, it is the electron mass whereas for $\nu_\mu$, it is the
muon mass.  Thus $A_{\nu_\mu} \neq A_{\nu_e}$.   However, the
difference appears in logarithmic form, and is presumably not very
significant. 

In a magnetized medium, however, the situation changes.  The reason is
that the medium contains electrons but not muons.  Accordingly, the
quantities $A_{\nu_e}$ and $A_{\nu_\mu}$ can be very different, as
seen by the presence of the term $y_e$ in \eqn{mb3}.  If we take
self-energy corrections only up to linear order in ${\mathcal B}$, as done in
\eqn{magdisp}, we obtain
\begin{eqnarray}
A_{\nu_\mu} - A_{\nu_e} = -\surd2 G_F (n_e - n_{\bar e}) - 2\surd2
eG_F \hat q \cdot \vec {\mathcal B} 
\int {d^3p \over (2\pi)^3 2E} \; {d\over dE} (f_e - f_{\bar e})
\,.
\label{A-A}
\end{eqnarray}
The first term on the right side comes just from the background
density of matter, and the second term is the magnetic field dependent
correction.  This quantity has been calculated for various
combinations of temperature and chemical potential of the background
electrons~\cite{Elmfors:1996gy, Esposito:1995db, D'Olivo:1995bq}.

If the denominator of the right side of \eqn{RC} becomes zero for some
value of $A_{\nu_\mu} - A_{\nu_e}$, the value of
$\tan2\widetilde\theta$ will become infinite.  This is the resonant
level crossing condition.  This was first discussed in the context of
neutrino oscillation in a matter background by Mikheev and 
Smirnov~\cite{Mikheev:wj}, where a particular value of density would ensure
resonance.  Presence of a magnetic field will modify this resonant
density, as seen from \eqn{A-A}.  The modification will be direction
dependent because of the factor $\hat q \cdot \vec {\mathcal B}$.
Some early authors~\cite{Esposito:1995db, D'Olivo:1995bq} contemplated
that, for large ${\mathcal B}$, the magnetic term might even drive the
resonance.  However, later it was shown~\cite{Nunokawa:1997dp} that
the magnetic correction would always be smaller than the other term.
So, if one considers values of ${\mathcal B}$ which are so large that
the last term in \eqn{A-A} is larger than the first term on the right
hand side, it means that one must take higher order corrections in
${\mathcal B}$ into account.

It should be noted that the type of corrections to the dispersion
relation discussed in Sec.~\ref{vise} appear from chiral neutrinos.
Thus, they produce chirality-preserving modifications to neutrino
oscillations. In addition, if the neutrino has a magnetic moment,
there will be chirality-flipping modifications as well. Many of these
modifications were analyzed in the context of the solar neutrino
problem, and we do not discuss them here.\footnote{A recent paper on
chirality-flipping oscillations is \cite{Egorov:1999ah}, where
one can obtain references to earlier literature. Some early references
are also found in \cite{Mohapatra:rq} and
\cite{Pal:1991pm}.} As pointed out in chapter \ref{chap1}, they are not
important for solar neutrinos, although may be important in other
stellar objects like the neutron star where the magnetic fields are
much larger.

An interesting possible consequence of neutrino oscillations have been
discussed~\cite{Kusenko:1996sr} in the context of high velocities of
neutron stars. 
In chapter
\ref{chap2} the puzzle about the high velocities of the pulsars was
addressed in terms of the asymmetric cross-section of the inverse beta
decay process which compels asymmetric emission of neutrinos from the
neutron stars. In the present situation the asymmetric emission of the
neutrinos is generated by resonant conversion of the neutrinos inside
the neutron star's body. In a material background containing electrons
but not any other charged leptons, the cross-section of $\nu_e$'s is
greater than that of any other flavor of neutrino.  If $\nu_e$'s can
oscillate resonantly to any other flavor, they can escape more easily
from a star.  In a proto-neutron star, the resonant density at an
angle $\theta$ with the magnetic field occurs at a distance
$R_0+\delta \cos\theta$ from the center, where $\delta$ is a function
of the magnetic field and specifies the deformation from a spherical
surface.  So this distance is direction dependent, as we have seen
from the previous discussion on the resonant level crossing condition
in a magnetized medium.  Therefore the escape of neutrinos is also
direction dependent, and the momentum carried away by them is not
isotropic.  The star would get a kick in the direction opposite to the
net momentum of escaped neutrinos.  This was suggested by Kusenko and
Segr\`e~\cite{Kusenko:1996sr}, who estimated that the momentum
imbalance is proportional to $\delta$ and can have a magnitude of
around 1\% for reasonable values of ${\mathcal B}$.  Later 
authors~\cite{Janka:1998kb} criticized their analysis and argued that the
effect was overestimated by them, because the kick momentum vanishes
in the lowest order in $\delta$.  A recent and detailed 
study~\cite{Barkovich:2002wh} indicates that these criticisms may not be
well-placed, and the kick momentum might indeed be proportional to the
surface deformation parameter $\delta$.
\section{Conclusion}\label{vios}
The main points discussed in this chapter are related to the
propagator of a charged particle and the neutrino self-energy in a
magnetic field and in a magnetized medium.  The Schwinger propagator
acts like the central point related to which all the discussion on
this chapter follows. This chapter also acts as a precursor for the
rest of the material in this thesis because what follows heavily
relies on the expression of the Schwinger propagator in a magnetized
medium.

While discussing about the Schwinger propagator due attention has been
paid to the phase factor accompanying it. It has been pointed out that
the origin and the purpose of the phase factor is related to the gauge
degree of freedom of the background gauge field.  Subsequently the
methods of statistical field theory has been utilized to write down
the expression of the Schwinger propagator in a thermal medium.

As a magnetic field interacts with the intermediate virtual charged
particles, so the neutrino self-energy also gets modified from its
vacuum value in presence of a magnetic field. In the section on
neutrino self-energy the expression of the self-energy is derived from
a purely general idea based on a form-factor analysis. From the
discussion on the neutrino self-energy it is evident that the general
form-factor analysis in a magnetized medium yields considerable
information.  The power of such an analysis is felt when it predicts
the dependence of the self-energy on the angle between neutrino
propagation direction and the external magnetic field in a magnetized
medium without the help of any actual loop calculation.

Section \ref{noscillations} discusses about neutrino
oscillations in a magnetized medium. The important question about the
effect of an external magnetic field on the resonant conversion of
neutrinos in a medium has been addressed, which is a very important
topic in astrophysics. In this section another attempt has been made
to explain the puzzle about the high velocities of the proto-neutron
stars by the mechanism of resonant conversion of neutrinos from one
flavour to the other. In literature this type of mechanism by which
the proto-neutron stars get a kick in one specific direction due to
asymmetric emission of neutrinos is often called the `neutrino rocket
mechanism'.

\chapter{Electromagnetic interactions in a magnetized medium}
\label{chap4}
\section{Introduction}
\label{nnp}
This chapter focuses on actual calculations done using the Schwinger
propagator. The following section will focus on the issue of the
effective electromagnetic vertex of the neutrinos in presence of a
magnetized medium. But before going into the techniques of calculation
a brief introduction to the topic of electromagnetic vertex of
neutrinos is presented here.

Previously a lot of work has been done concerning the
neutrino-neutrino-photon vertex in the presence of a background
magnetic field~\cite{Bhattacharya:2002aj}. The vertex arises from the
diagrams of \fig{f:magmom}, which contain internal $W$-lines. In
addition, there is a diagram mediated by the $Z$-boson, as shown in
\fig{f:Zvertex}.  For phenomenological purposes, we require the
electromagnetic vertex of neutrinos only in the leading order in Fermi
constant.  It should be realized that in this order, the diagram of
\fig{f:magmom}b does not contribute at all, since it has two
$W$-propagators.  The remaining diagrams, shown in \fig{f:magmom}a and 
\fig{f:Zvertex}, can both be represented in the form shown in 
\fig{f:4fermi}, where an effective 4-fermi vertex has been used.  
The effective 4-fermi interaction can be written as,
\begin{eqnarray}
\mathscr L_{\rm eff} = 
-\surd2 G_F \Big[\overline\psi_{(\nu)}\gamma^\lambda L\psi_{(\nu)} \Big]
\Big[\overline\psi_{(\ell)} \gamma_\lambda (g_V + g_A\gamma_5)
\psi_{(\ell)} \Big] \,,
\label{4fermi}
\end{eqnarray}
where $L = \frac12(1 - \gamma_5)$ is the left handed projection operator.
If the neutrino and the charged lepton belong to different generations
of fermions, this effective Lagrangian contains only the neutral
current interactions, and in that case $g_V$ and $g_A$ are equivalent
to $c_V$ and $-c_A$ defined in \eqn{cvca}.  On the other hand, if both
$\nu$ and $\ell$ belong to the same generation, we should add the
charged current contribution as well, and use
\begin{eqnarray}
g_V = c_V +1  \,, \qquad g_A = -(c_A + 1) \,.
\end{eqnarray}
%
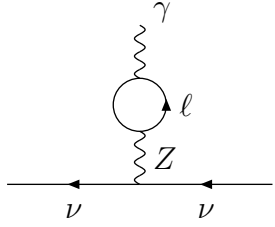
\begin{figure}[t]
\begin{center}
\begin{picture}(180,80)(-90,-20)
\ArrowLine(50,0)(0,0)
\Text(25,-10)[c]{$\nu$} 
\ArrowLine(0,0)(-50,0)
\Text(-25,-10)[c]{$\nu$} 
\Photon(0,0)(0,20)23
\Text(5,10)[l]{$Z$}
\ArrowArc(0,30)(10,-180,180)
\Text(15,30)[l]{$\ell$}
\Photon(0,40)(0,60)23
\Text(5,65)[l]{$\gamma$}
\end{picture}
\caption[$Z$-photon mixing diagram contributing to the neutrino
electromagnetic vertex.]{\sf $Z$-photon mixing diagram contributing to
the neutrino electromagnetic vertex.
\label{f:Zvertex}}
\end{center}
\end{figure}
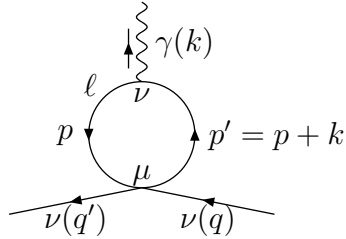
\begin{figure}[t]
\begin{center}
\begin{picture}(100,100)(-50,-20)
\ArrowLine(50,0)(0,10) 
\Text(25,-2)[c]{$\nu(q)$} 
\ArrowLine(0,10)(-50,0)
\Text(-25,-2)[c]{$\nu(q')$}
\Text(0,45)[c]{$\nu$} 
\ArrowArc(0,30)(20,90,270)
\Text(-29,30)[c]{$p$}
\ArrowArc(0,30)(20,-90,90)
\Text(0,15)[c]{$\mu$}
\Text(-22,49)[l]{$\ell$}
\Text(51,30)[c]{$p'=p+k$}
\Photon(0,50)(0,80)24
\ArrowLine(-5,55)(-5,70)
\Text(5,65)[l]{$\gamma(k)$}
\end{picture}
\caption[The neutrino electromagnetic vertex in the leading
order in the Fermi constant.]{\sf The neutrino electromagnetic vertex
in the leading order in the Fermi constant.
\label{f:4fermi}}
\end{center}
\end{figure}
Many processes involving neutrinos and photons have been calculated
using the 4-fermi Lagrangian of \eqn{4fermi}.  The calculations
simplify in this limit for various reasons.  First, we do not have to
use the momentum dependence of the gauge boson propagators.  Second,
since two charged lepton lines form a loop in \fig{f:4fermi}, the
phase factors of the propagators, as discussed in appendix \ref{s:LCP},
cancel each other. 

The background magnetic field, can give rise to many physical
processes which are forbidden in the vacuum. One such process is the
decay of a photon into a neutrino-antineutrino pair:
\begin{eqnarray}
\gamma \to \nu + \bar\nu \,.
\label{gammadk}
\end{eqnarray}
This was calculated using the Schwinger propagator in some very early
papers~\cite{GN72, DeRaad:1976kd}.  Assuming two generations of
fermions, the decay rate was found to be
\begin{eqnarray}
\Gamma = {\alpha^2 G_F^2 \over 48\pi^3 \omega} \Big|
\varepsilon^\mu k^\nu \widetilde F_{\mu\nu} \Big|^2  \Big| \mathscr M_e-
\mathscr M_\mu \Big|^2 \,,
\end{eqnarray}
where $\varepsilon^\mu$, $k^\mu$ and $\omega$ are the polarization
vector, the momentum 4-vector and the energy of the initial photon,
and the quantity $\mathscr M_\ell$ was evaluated in various limits by
these authors. $\widetilde F_{\mu\nu}$ is the dual of the
electromagnetic field strength tensor. For example, if $\omega\ll
m_\ell$, they found
\begin{eqnarray}
\mathscr M_\ell = {\omega^2 \over e{\mathcal B}} \sin^2\theta \times 
\cases{
{2\over 15} \left({e{\mathcal B} \over m_\ell^2} \right)^3 & for
$e{\mathcal B} \ll m_\ell^2$\,, 
\cr  
{1\over 3} \left( {e{\mathcal B} \over m_\ell^2} \right) & for
$e{\mathcal B} \gg m_\ell^2$\,,} 
\end{eqnarray}
where $\theta$ is the angle between the photon momentum and the
magnetic field.  No matter which neutrino pair the photon decays to,
both charged leptons appear in the decay rate because of the loop in
\fig{f:Zvertex}. In the calculation only the axial couplings of the
charged leptons contribute to the amplitude.

A related process is Cherenkov radiation from neutrinos:
\begin{eqnarray}
\nu \to \nu +\gamma \,.
\label{cheren}
\end{eqnarray}
Again, this is a process forbidden in the vacuum.  But a background
magnetic field modifies the photon dispersion relation, and so this
process becomes feasible.  The rate of this process has been
calculated by many authors~\cite{GN72, Sko76, Ioannisian:1996pn,
Gvozdev:1997mc}, and all of them do not get the same result.  
According to Ref.~\cite{Gvozdev:1997mc}, the rate for the process is
given by
\begin{eqnarray}
\Gamma = {\alpha G_F^2\over 8\pi^2} (g_V^2 + g_A^2) (e{\mathcal B})^2
\Omega\sin^2\theta F(\Omega^2\sin^2\theta/e{\mathcal B}) \,,
\end{eqnarray}
where $\Omega$ is the initial neutrino energy and $\theta$ is the
angle between the initial neutrino momentum and the background
field.  For large magnetic fields satisfying the condition $e{\mathcal B} \gg
\omega^2\sin^2\theta$, the function $F$ is given by
\begin{eqnarray}
F(x) = 1 - {x \over 2} + {x^2 \over 3} - {5x^3 \over 24} + {7x^4 \over
60} + \cdots \,.
\end{eqnarray}
The modification of this process in the presence of background matter
has also been calculated~\cite{Chistyakov:1999ii}.

Another process that has been discussed is the radiative neutrino
decay
\begin{eqnarray}
\nu_a \to \nu_b + \gamma \,.
\end{eqnarray}
Unlike the previous processes, this can occur in the vacuum as well
when the neutrinos have mass and mixing.  However, a background
magnetic field adds new contributions to the amplitude, and the rate
can be enhanced.  Gvozdev, Mikheev and
Vasilevskaya~\cite{Gvozdev:1996kx} 
calculated the rate of this decay in a variety
of situations depending on the field strength and the energy of the
initial neutrino.  For a strong magnetic field (${\mathcal B}\gg
{\mathcal B}_e$), they 
found the decay rate of an ultra-relativistic neutrino of energy
$\Omega$ to be
\begin{eqnarray}
\Gamma = {2\alpha G_F^2 \over \pi^4} {m^6 \over \Omega} {{\mathcal B}
\overwithdelims() {\mathcal B}_e}^2 |U_{ae} U_{be}^*|^2
J(\Omega\sin\theta/2m_e) \,, 
\end{eqnarray}
where $U$ is the leptonic mixing matrix, $\theta$ is the angle between
the magnetic field and the neutrino momentum, $m$ is the electronic
mass, and
\begin{eqnarray}
J(z) = \int_0^z dy \; (z-y) \left( {1 \over y\sqrt{1-y^2}} \tan^{-1}
{y \over \sqrt{1-y^2}} - 1 \right)^2 \,.
\end{eqnarray}
The curious feature of this result is that this is independent of the
initial and the final neutrino masses.

The form for the 4-fermi interaction in \eqn{4fermi} suggests that the
neutrino electromagnetic vertex function $\Gamma_\mu$ can be
written as~\cite{Nieves:1993er}:
\begin{eqnarray}
\Gamma_\mu = - \, {\surd2 G_F \over e} \gamma^\nu L \Big( g_V
\Pi_{\mu\nu} + g_A \Pi^5_{\mu\nu} \Big) \,.
\label{nvf}
\end{eqnarray}
Here, the term $\Pi_{\mu\nu}$ is exactly the expression for the
vacuum polarization of the photon, and appears from the vector
interaction in the effective Lagrangian.  The other term,
$\Pi^5_{\mu\nu}$, differs from $\Pi_{\mu\nu}$ in that it
contains an axial coupling from the effective Lagrangian.
This equality is valid even when one has a magnetic field and a
material medium as the background. Thus, the calculation of the
photon self-energy in a background magnetic field in matter can give
us information about the neutrino electromagnetic vertex in the same
situation.  

Here we concentrate on calculations done in presence of a medium with
a uniform background magnetic field. We consider various cases where
the charged particles affected by the magnetic fields remain as
virtual particles in the Feynman diagrams of the respective
processes. The techniques of the calculation involve the use of the
propagators introduced in chapter \ref{chap3}. This chapter discusses
the electromagnetic vertex function of neutrinos and consequently the
structure of the two second rank tensors $\Pi_{\mu\nu}$ and
$\Pi^5_{\mu\nu}$ which are necessary to calculate it, as is evident
from Eq.~(\ref{nvf}).  $\Pi_{\mu\nu}$ has been calculated in a
magnetic field and in a magnetized medium by various
authors~\cite{Tsai:1, D'Olivo:2002sp, Ganguly:1999ts}. In the
following section only the relevant portions of the results are
presented briefly. The main discussion is around the calculation of
$\Pi^5_{\mu\nu}$ in presence of a magnetized medium. For the sake of
comparison the results of the calculations has been given for various
backgrounds as vacuum, medium, magnetic field and obviously a
magnetized medium. Some comments on neutrino-photon scattering in a
background magnetic field, which is highly suppressed in vacuum, is
provided in subsection \ref{nnps}.  Section \ref{efftcharge} discusses
the concept of the effective electric charge of a neutrino from a
quantum field theoretical point of view. As neutrinos do not couple to
photons in the tree level in the standard model so naturally if the
neutrinos acquire some electric charge it must be an effective
one. This effective charge of neutrinos in a magnetized medium has
been calculated to odd orders in the background magnetic field in
section \ref{efftcharge}.  Section \ref{s:infty} describes the methods
for handling ultraviolet divergences in calculations involving
background magnetic fields.  This chapter concludes with a general
discussion on the various ideas exposed in the following sections.
\section{Neutrino-photon scattering and the electromagnetic vertex of 
neutrinos}
\label{npsev}
In this section we will discuss effective neutrino photon interactions
in a magnetized medium~\cite{Bhattacharya:2001nm, Bhattacharya:2003hq},
and the constituents of the neutrino photon electromagnetic vertex.
But before taking up the issue of the $\nu\nu\gamma$ vertex some
comments on the neutrino-photon scattering in a background magnetic
field follows. 
\subsection{Some comments on neutrino-photon scattering}
\label{nnps}
The cross-section of neutrino-photon scattering is highly
suppressed in the standard model due to Yang's theorem~\cite{Yang},
which makes the scattering cross-section vanish to the first order of
the Fermi coupling $G_F$.  But in presence of a magnetic field,
neutrino-photon scattering can occur and to orders of $G_F$ the
cross-section has been calculated~\cite{Shaisultanov:1997bc}. In the
following paragraphs of this subsection a brief overview on neutrino
photon scattering in a magnetic field background is supplied.

Gell-Mann showed~\cite{Gell-Mann} that the amplitude of the reaction
\begin{eqnarray}
\gamma + \nu \to \gamma + \nu
\label{gnugnu}
\end{eqnarray}
is exactly zero to order $G_F$ because by Yang's 
theorem~\cite{Yang, Landau} two photons cannot couple to a $J = 1$ state.  In
the standard model, therefore, amplitude of the above process appears
only at the level of $1/M_W^4$ and as a result the cross-section is
exceedingly small~\cite{Dicus:iy}.

\begin{figure}[btp]
\begin{center}
%
%
\begin{picture}(180,120)(-90,-35)
\Text(0,-30)[ct]{\large\bf (a)}
\ArrowLine(80,0)(40,0)
\Text(60,-10)[c]{$\nu$}
\Photon(40,0)(-40,0)37
\Text(0,-10)[c]{$W$}
\ArrowLine(-40,0)(-80,0)
\Text(-60,-10)[c]{$\nu$}
\ArrowArc(0,0)(40,0,45)
\Text(-15,47)[l]{$e$}
\ArrowArc(0,0)(40,45,90)
\Text(-40,20)[r]{$e$}
\ArrowArc(0,0)(40,90,135)
\Text(15,47)[r]{$e$}
\ArrowArc(0,0)(40,135,180)
\Text(45,20)[r]{$e$}
\Photon(-30,27)(-60,60){2}{4}
\Text(-45,60)[r]{$\gamma$}
\Photon(0,40)(0,80){2}{4}
\Text(10,65)[r]{$\gamma$}
\Photon(30,27)(60,60){2}{4}
\Text(49,57)[r]{$\gamma$}
\end{picture}
%
%
\qquad
\begin{picture}(180,120)(-90,-35)
\ArrowLine(40,0)(0,0) 
\Text(25,-10)[c]{$\nu$} 
\ArrowLine(0,0)(-40,0)
\Text(-25,-10)[c]{$\nu$} 
\Photon(0,0)(0,20)24
\Text(5,10)[l]{$Z$}
\ArrowArc(0,40)(20,0,90)
\Text(20,60)[l]{$e$}
\Photon(0,60)(0,90)24
\Text(5,75)[l]{$\gamma$}
\ArrowArc(0,40)(20,90,180)
\Text(-22,60)[l]{$e$}
\Photon(-20,40)(-50,40)24
\Text(-40,30)[l]{$\gamma$}
\ArrowArc(0,40)(20,180,270)
\Text(-22,20)[l]{$e$}
\Photon(20,40)(50,40)24
\Text(35,30)[l]{$\gamma$}
\ArrowArc(0,40)(20,270,360)
\Text(18,20)[l]{$e$}
\Text(0,-30)[ct]{\large\bf (b)}
\end{picture}
\caption[The 1-loop effective vertex for two neutrinos and three
photons.]{\sf The 1-loop effective vertex for two neutrinos and three
photons. 
\label{f:nugamma}}
\end{center}
\end{figure}
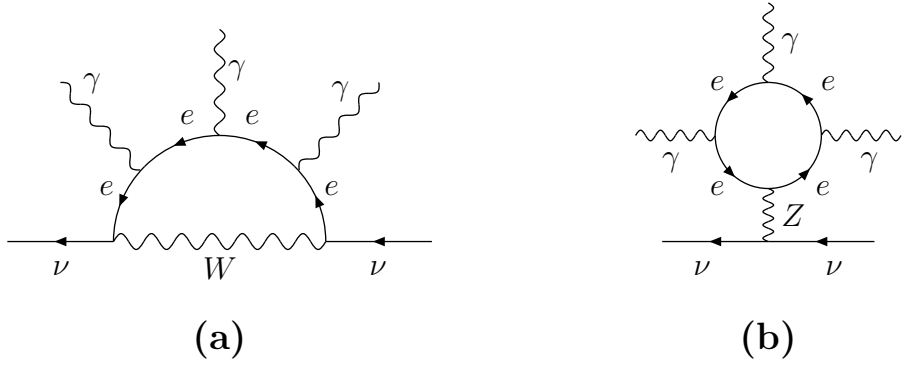
But there is no such restriction on the coupling of three photons with
neutrinos as,
\begin{eqnarray}
\gamma + \nu \to \gamma + \gamma + \nu.
\end{eqnarray}
The cross-section of the above process can be calculated from the
effective Lagrangian proposed by Dicus and Repko~\cite{Dicus:1997rw}. 
The diagrams for the two neutrino three photon
interaction are shown in \fig{f:nugamma} where
\fig{f:nugamma}a shows the contribution from the $W$ exchange
diagram and \fig{f:nugamma}b shows the contribution from $Z$
exchange. Denoting the photon field tensor as ${\bar F}_{\mu \nu}$ and the
neutrino fields by $\psi_{(\nu)}$, and integrating out the particles in the
loop the effective Lagrangian comes out as
\begin{eqnarray}
\mathscr L_{\rm eff} = {G_F \over \surd2}
\frac{e^3 (c_V+1)}{360 \pi^2 m^4}
\left[ 5 (N_{\mu \nu} {\bar F}^{\mu \nu})({\bar F}_{\lambda \rho}
{\bar F}^{\lambda \rho}) - 14 N_{\mu \nu} {\bar F}^{\nu \lambda}
{\bar F}_{\lambda \rho} {\bar F}^{\rho \mu}\right],
\label{EL}
\end{eqnarray}
where $c_V$ was defined in \eqn{cvca}, $m$ is the electronic mass, and
\begin{eqnarray}
N_{\mu \nu} = \partial_\mu(\bar{\psi}_{(\nu)} \gamma_\nu L \psi_{(\nu)}) - 
\partial_\nu(\bar{\psi}_{(\nu)} \gamma_\mu L \psi_{(\nu)}) \,.
\end{eqnarray}
For energies much smaller than the electron mass, this can be used as
an effective Lagrangian to calculate various processes involving
photons and neutrinos in the presence of a background magnetic field
$F_{\mu\nu}$.  For this, we simply have to write
\begin{eqnarray}
{\bar F}_{\mu\nu} = f_{\mu\nu} + F_{\mu\nu} \,,
\end{eqnarray}
where now $f_{\mu\nu}$ is the dynamical photon field, and look for the
terms involving $F_{\mu\nu}$.  For example, 
Shaisultanov~\cite{Shaisultanov:1997bc} calculated the rate of $\gamma\gamma\to
\nu\bar\nu$ in a background field.  \eqn{EL} shows that in the lowest
order, the amplitude for involving $\nu_e$'s would be proportional to
\begin{eqnarray}
{G_FB \over m^4} \sim {{\mathcal B} \over M_W^2 m^2 {\mathcal B}_e} \,,
\end{eqnarray}
where ${\mathcal B}_e$ is the value of the magnetic field defined in
\eqn{Bc}.  Since the amplitude without any magnetic
field~\cite{Dicus:iy} 
is of order $1/M_W^4$, it follows that the background
field increases the amplitude by a factor of order $(M_W/m)^2
{\mathcal B}/{\mathcal B}_e$, or the rate by a factor $(M_W/m_e)^4
({\mathcal B}/{\mathcal B}_e)^2$.  Later
calculations~\cite{Dicus:2000cz} 
have extended these results by including other
processes obtained by crossing, like $\nu\bar\nu\to\gamma\gamma$ and
$\nu\gamma\to\nu\gamma$.  To obtain higher ${\mathcal B}$ terms in
these cross-sections, one needs the effective Lagrangian containing
higher order terms in the electromagnetic field strength.  Such an
effective Lagrangian has been derived by Gies and
Shaisultanov~\cite{Gies:2000tc}.

Alternatively, the amplitudes can be calculated using the Schwinger
propagator for charged leptons.  Such calculations for 
$\gamma\gamma\to \nu\bar\nu$  were done some time 
ago~\cite{Chyi:1999wy, Chyi:1999fc}.  One of the important features of
this calculation is that in the 4-fermi limit, the diagram contains
three electron propagators.  In such situations, the phase factor
$\Psi(x,x')$ appearing in the Schwinger propagator of
\eqn{schwingprop} cannot be disregarded\footnote{A detailed 
discussion on the phase factor of the Schwinger propagator is
presented in subsection \ref{ss:sp} of chapter \ref{chap3} and appendix
\ref{s:LCP}.}. In the calculation, only the linear term in 
${\mathcal B}$ was retained in the amplitude so that the results are
valid only for small magnetic fields.  However, since no effective
Lagrangian was used, the results are valid even when the energies of
the neutrinos and/or the photons are comparable to, or greater than,
the electron mass.  Later authors~\cite{Dicus:2000cz} reported some
mistakes in this calculation and corrected them.

The previous discussions about neutrino-photon scattering relied on
effective Lagrangians and the use of Schwinger propagator to calculate
a 3-point function, but aside these scattering phenomena there are a
lot of phenomena, discussed in the introduction of this chapter, which
involves the $\nu\nu\gamma$ vertex in a background magnetic field.  To
calculate the rates of these processes one has to calculate the
effective electromagnetic vertex of neutrinos in a background magnetic
field and in a magnetized medium.  This section is primarily concerned
about the electromagnetic vertex of the neutrinos in a magnetized
medium. 

For simplicity we consider the background temperature and
neutrino momenta to be small compared to the masses of the W and Z
bosons. We can, therefore, neglect the momentum dependence in the W
and Z propagators, which is justified if we are performing a
calculation to the leading order in the Fermi constant, $G_F$. In this
limit the 4-fermi interaction is given by Eq.~(\ref{4fermi}) written
in the beginning of this chapter. 
Moreover if we restrict the temperature of the system such that muons
and taons are not produced in the medium then in the leptonic part of
the Lagrangian in Eq.~(\ref{4fermi}), $\psi_{(\ell)}$ stands for
electrons alone. For electron neutrinos,
\begin{eqnarray}
g_{\rm V} &=& \frac{1}{2} + 2 \sin^2 \theta_{\rm W},\nonumber \\
g_{\rm A} &=& - \frac{1}{2}\nonumber\,.
\end{eqnarray}
For muon and tau neutrinos,
\begin{eqnarray}
g_{\rm V} &=& -\frac{1}{2} + 2 \sin^2 \theta_{\rm W},\nonumber \\
g_{\rm A} &=&  \frac{1}{2}\,.\nonumber
\end{eqnarray}
With the interaction Lagrangian as given in Eq.~(\ref{4fermi}) the
$\nu \nu \gamma$ vertex, as shown in \fig{f:4fermi}, can be written in
terms of two tensors. The vector-vector amplitude $\Pi_{\mu \nu}(k)$
and the axialvector-vector amplitude $\Pi^5_{\mu \nu}(k)$. This fact
is symbolically written down in Eq.~(\ref{nvf}).  The vector-vector
amplitude term $\Pi_{\mu \nu}(k)$ turns out to be:
\begin{eqnarray}
i\Pi_{\mu \nu}(k)=(-i e)^2(-1) \int {{d^4 p}\over {(2\pi)^4}}\Tr \left
[ \gamma_\mu iS(p) \gamma_\nu iS(p')\right]\,.
\label{npimunu}
\end{eqnarray}
Here and henceforth $p'= p + k$. From the above equation it can be
seen that $\Pi_{\mu \nu}(k)$ is the photon vacuum polarization tensor.
The axialvector-vector
amplitude $\Pi^5_{\mu \nu}(k)$ comes out as:
\begin{eqnarray}
i\Pi^5_{\mu \nu}(k)=(-i e)^2(-1) \int {{d^4 p}\over {(2\pi)^4}}\Tr \left
[\gamma_\mu \gamma_5 iS(p) \gamma_\nu iS(p')\right].
\label{npimunu5}
\end{eqnarray}
Both tensors are obtained by calculating the Feynman diagram given in
\fig{f:4fermi}. The following subsections discusses about the second
rank tensors $\Pi_{\mu \nu}(k)$ and $\Pi^5_{\mu \nu}(k)$.
\subsection{The photon vacuum polarization tensor $\Pi_{\mu \nu}(k)$
in different backgrounds}
\label{pimunumb}
The calculation of the photon vacuum polarization tensor in a magnetized
medium has been done by previous authors~\cite{D'Olivo:2002sp,
Ganguly:1999ts}. In this subsection only a brief summary
of the previously known results is reproduced. The form of
$\Pi_{\mu\nu}(k)$ to one-loop is given in Eq.~(\ref{npimunu}) and it
is also pointed out that $\Pi_{\mu\nu}(k)$ is a necessary quantity for
understanding the neutrino electromagnetic vertex.

The form of $\Pi_{\mu\nu}(k)$ is restricted by two symmetries.
Owing to gauge invariance, $\Pi_{\mu\nu}(k)$ satisfies the conditions
\begin{eqnarray}
k^\mu \Pi_{\mu\nu} (k) = 0 \,, \quad 
k^\nu \Pi_{\mu\nu} (k) = 0 \,.
\label{kpi=0}
\end{eqnarray}
In addition, Bose symmetry implies
\begin{eqnarray}
\Pi_{\mu\nu} (k) = \Pi_{\nu\mu} (-k) \,.
\label{bose}
\end{eqnarray}
These conditions restrict the form of the possible tensor structure of
the vacuum polarization tensor. The general form of
$\Pi_{\mu \nu}(k)$ in various backgrounds is dictated by
Eq.~(\ref{kpi=0}) and Eq.~(\ref{bose}) as discussed below.

\subsubsection{In vacuum}
In the vacuum, the tensor $\Pi_{\mu\nu}(k)$ depends only on the
momentum vector $k$. Thus, the most general form for
$\Pi_{\mu\nu}(k)$ is given as,
\begin{eqnarray}
\Pi_{\mu\nu}(k) = \Pi(k^2) \left[k^2 g_{\mu\nu} - k_\mu
k_\nu \right] \,. 
\label{vacpi}
\end{eqnarray}
This form satisfies both the conditions given in Eq.~(\ref{kpi=0}) and 
Eq.~(\ref{bose}).
\subsubsection{In a medium}
In a medium the expression of $\Pi_{\mu\nu}(k)$ differs from that in
the vacuum.  Although $\Pi_{\mu\nu}(k)$ still has to satisfy
Eq.~(\ref{kpi=0}) and Eq.~(\ref{bose}), the form given in
Eq.~(\ref{vacpi}) does not follow. This is because $\Pi_{\mu\nu}(k)$
can now depend, apart from the momentum vector $k^\mu$, on various
vectors or tensors which characterize the background medium. Even for
a homogeneous and isotropic medium, there is an extra vector in the
form of the velocity of its center of mass, $u^\mu$.  In general in
place of $k^2$ the Lorentz scalars will now depend upon
\begin{eqnarray}
\omega &=& k\cdot u\,,
\label{medomega}\\
     K &=& \sqrt{\omega^2 - k^2}\,.
\label{medK}
\end{eqnarray}
The above expressions of $\omega$ and $K$ reduce to the energy and
momentum of the photon in the rest frame of the thermal medium where 
$u^\mu$ is given as in Eq.~(\ref{urf}). 
	\begin{figure}
\begin{center}
\begin{picture}(150,50)(0,-25)
\Photon(0,0)(40,0){2}{4}
\Text(20,5)[b]{$k\rightarrow$}
\Photon(110,0)(150,0){2}{4}
\Text(130,5)[b]{$k\rightarrow$}
\Text(75,30)[b]{$p+k\equiv p'$}
\Text(75,-30)[t]{$p$}
\SetWidth{1.2}
\Oval(75,0)(25,35)(0)
\ArrowLine(74,25)(76,25)
\ArrowLine(76,-25)(74,-25)
\end{picture}
\end{center}
\caption[One-loop diagram for vacuum polarization.]{One-loop diagram 
for vacuum polarization.}\label{f:1loop}
\end{figure}
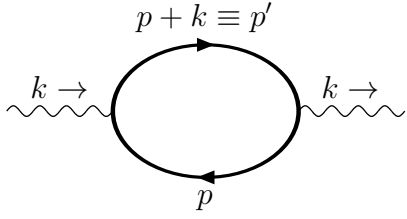

In presence of a medium without any magnetic field the most general
expression of the polarization tensor can be written in terms of form
factors along with the new tensors constructed out of $u^{\mu}$ and
the ones we already had in absence of a medium as~\cite{Nieves:1988qz},
\begin{eqnarray}
\Pi_{\mu\nu}(K,\omega)=\Pi_T\,T_{\mu\nu} + \Pi_L\,L_{\mu\nu} +
\Pi_P\,P_{\mu \nu}.  
\label{mvacpol}
\end{eqnarray}
Here
\begin{eqnarray}
T_{\mu\nu}&=& {\widetilde g}_{\mu\nu} - L_{\mu\nu}\,,\\
L_{\mu\nu}&=&\frac{{\widetilde u}_{\mu}{\widetilde
u}_{\nu}}{{\widetilde u}^2}\,,\\
P_{\mu \nu}&=& \frac{i}{K} \varepsilon_{\mu \nu \alpha \beta} k^\alpha 
u^\beta\,,
\label{t}
\end{eqnarray}
with
\begin{eqnarray}
{\widetilde u}_{\mu}={\widetilde g}_{\mu\rho} u^{\rho}\,,
\end{eqnarray}
and
\begin{eqnarray}
{\widetilde g}_{\mu\rho} = g_{\mu\rho} - \frac{k_\mu k_\rho}{k^2}\,.
\end{eqnarray}
In Eq.~(\ref{mvacpol}) all form-factors 
are functions of $\omega$, $K$ and other parameters as chemical
potential $\mu$ and inverse temperature $\beta$.  
In Eq.~(\ref{mvacpol}), 
$\Pi_P$ is nonzero only if the
background medium or the interaction Lagrangian of the photon break
both $\bf{P}$ and $\bf{CP}$ symmetries~\cite{Nieves:1988qz}.
\subsubsection{In presence of a background magnetic field}
The vacuum polarization of the photon in a magnetic field has also
been calculated~\cite{Tsai:1} up to one loop but to all orders in the
external field and the expression of $\Pi_{\mu \nu}(k)$ in our
convention, as discussed in section \ref{conv}, is given by 
\begin{eqnarray}
\Pi_{\mu \nu}(k) = \frac{e^3 {\mathcal B}}{(4\pi)^2}\left[(g_{\mu \nu}k^2 -
k_{\mu} k_{\nu}) N_0 - (g_{\mu \nu}^{\para}k^2_{\para} -
k_{\mu}^{\para}  k_{\nu}^{\para}) N_{\para} - (g_{\mu
\nu}^{\pr} k^2_{\pr} + k_{\mu}^{\pr}  k_{\nu}^{\pr}) N_{\pr}
\right]\,.
\label{tsai1}
\end{eqnarray}
Here $N_0$, $N_\para$ and $N_\pr$ are the form-factors, which are
functions of ${\mathcal B}$, $k^2_{\para}$, and $k^2_{\pr}$.
\subsubsection{In presence of a magnetized medium}
The discussion on $\Pi_{\mu \nu}(k)$ in presence of a magnetized
medium is presented with some more details than the previous
ones. This is because the form of $\Pi_{\mu \nu}(k)$
is important for the expression of the effective charge of the
neutrinos in a magnetized medium. It also sets the stage for the
calculation of $\Pi^5_{\mu \nu}(k)$ in a magnetized medium.

At the 1-loop level, the vacuum polarization tensor arises from the
diagram in \fig{f:1loop}. To evaluate this diagram, one needs to use
the electron propagator within a thermal medium in the presence of a
background electromagnetic field. Rather than working with the
complicated expression for a general background field, we will
specialize to the case of a uniform magnetic field. Once this is
assumed, the field can be taken in the $z$-direction without any
further loss of generality.

The amplitude of the 1-loop diagram of \fig{f:1loop} is given by the
expression in Eq.~(\ref{npimunu}). Only in this case the
propagators appearing in Eq.~(\ref{npimunu}) like $S(p)$ and $S(p')$
must be replaced by the appropriate ones suitable in a magnetized
medium, $S_B(p)$ and $S_B(p')$ as introduced in chapter \ref{chap3},
Eq.~(\ref{sbp}).  As the loop is a 2-point function the phase factor
drops out of the calculation, as is discussed in appendix \ref{s:LCP}.
From Eq.~(\ref{sbp}), we see that there are two terms in the
propagators $S_B(p)$ in a magnetized medium, the vacuum part and the
other part which involves background matter distribution. In the trace
appearing in the right hand side of Eq.~(\ref{npimunu}) there
appears a product of two propagators $S_B(p)$ and $S_B(p')$ with
other quantities in between. This product will give rise to four
quantities of which one term will contain two factors of $S_B^V$s from
the two propagators. It has no importance as far as the background
effects are concerned. There will be one term containing two factors
of $S^\eta_B$s, which will contribute only to the absorptive part of
the vacuum polarization.  There will be two terms in the trace each
containing a factor of $S_B^V$ and a factor of $S^\eta_B$.  These terms
contribute to the present calculation, as they are important for
calculating the effective charge of the neutrinos in a magnetized
medium. The relevant part of $\Pi_{\mu\nu}(k)$ is therefore made up of
the two terms inside the trace each containing one vacuum part and one
thermal part of the total electron propagator. With this introduction 
the expression of $\Pi_{\mu\nu}(k)$ is written as
\begin{eqnarray}
\Pi_{\mu\nu}(k) = -ie^2 \int \frac{d^4p}{(2\pi)^4}
\; \Tr\, \Big[\gamma_\mu S_B^\eta(p) \gamma_\nu iS_B^V(p') +
\gamma_\mu \, iS_B^V(p) \gamma_\nu \, S_B^\eta (p') \Big] \,.
\label{SS'terms}
\end{eqnarray}
%
%
%
Using the form of the propagators from Eq.~(\ref{SB}) and
Eq.~(\ref{Seta}) and manipulating under the integral
signs~\cite{Ganguly:1999ts}, the 1-1-component\footnote{The 1-1-component
of the propagator is discussed in appendix \ref{rtf}, where
quantum statistical field theory is discussed.} of the polarization
tensor to odd powers in the external magnetic field can be written
as~\cite{Ganguly:1999ts},
\begin{eqnarray}
\Pi_{\mu\nu}(k) 
&=& 4ie^2 \varepsilon_{\mu\nu\alpha_\parallel\beta} k^\beta
\int \frac{d^4p}{(2\pi)^4} \eta_-(p)
\int_{-\infty}^\infty ds \; e^{\Phi(p,s)}
\int_0^\infty ds' \; e^{\Phi(p',s')} \nonumber\\*
&\times & \Bigg[ 
p^{\widetilde\alpha_\parallel} \tan (e{\mathcal B}s) + 
p'^{\widetilde\alpha_\parallel} \tan (e{\mathcal B}s') 
- {\tan (e{\mathcal B}s) \; \tan (e{\mathcal B}s') \over 
\tan (e{\mathcal B}(s+s'))} \;
(p+p')^{\widetilde\alpha_\parallel} \Bigg] \,.
\label{Ofinal}
\end{eqnarray}
The above form of the vacuum polarization tensor is valid in the rest
frame of the background medium. $\eta_-(p_0)$ contains the information
about the distribution of electrons and positrons in the background
and is given by 
\begin{eqnarray}
\eta_-(p_0) = \eta_F(p_0) - \eta_F(-p_0)\,,
\end{eqnarray}
where $\eta_F(p_0)$ is specified in Eq.~(\ref{eta}). In the above
equation the 
argument of the $\eta_F$ function is not 
$p\cdot u$ but only the energy component $p_0$, because in
the rest frame of the medium $p\cdot u = p_0$ due to the form of $u$
in the rest frame, given in Eq.~(\ref{urf}). From the form of
$\Pi_{\mu\nu}(k)$ as given above it is seen that it is antisymmetric
in its tensor indices. 
\subsection{The axialvector-vector amplitude $\Pi^5_{\mu \nu}(k)$ in
different backgrounds}
\label{form}
The form of the axialvector-vector amplitude $\Pi^5_{\mu \nu}(k)$ to
one-loop is supplied in Eq.~(\ref{npimunu5}). In this regard it must
be noted that the importance of this second rank tensor arises from
the fact that it is an important constituent of the neutrino
electromagnetic vertex as given in Eq.~(\ref{nvf}). The neutrino
electromagnetic vertex to one-loop is depicted in \fig{f:4fermi}.
While discussing about $\Pi^5_{\mu \nu}(k)$ it should be remembered
that for the electromagnetic vertex i.e., the vertex in the Feynman
diagram where the external photon couples to the fermion loop in
\fig{f:4fermi}, we have the gauge invariance condition,
\begin{eqnarray}
k^{\nu} \Pi^5_{\mu \nu}(k) = 0\,.
\label{ngi}
\end{eqnarray}
In the following paragraphs the form of $\Pi^5_{\mu \nu}(k)$ in
various backgrounds is discussed.
\subsubsection{In vacuum}
$\Pi^5_{\mu \nu}(k)$ violates parity due to the presence of $\gamma_5$
in its expression as given in Eq.~(\ref{npimunu5}).  In accordance
with this fact the available vectors and tensors at hand are the
following,
\begin{eqnarray}
k_{\mu},~g_{\mu\nu} \mbox{~and~} \epsilon_{\mu\nu\lambda\sigma}.
\label{vt}
\end{eqnarray}
The two point axialvector-vector amplitude $\Pi^5_{\mu\nu}(k)$ can be
expanded in a basis constructed out of the above tensors. Given the
axial-tensor like parity structure of $\Pi^5_{\mu\nu}(k)$ the relevant
tensors at hand are $\epsilon_{\mu\nu\lambda\sigma}
k^{\lambda}k^{\sigma}$ and $\epsilon_{\mu\nu\lambda\sigma} g^{\lambda
\sigma}$ both of which are zero. So in vacuum $\Pi^5_{\mu\nu}(k)$
vanishes.
\subsubsection{In a medium}
On the other hand, in a medium in absence of any magnetic field, we
have an additional vector $u^{\mu}$, i.e the velocity of the centre of
mass of the medium. Therefore the axialvector-vector amplitude can be
expanded in terms of form-factors along with the new tensors
constructed out of $u^{\mu}$ and the ones we already had in absence of
a medium. A second rank tensor constructed out of them can be
$\varepsilon_{\mu \nu \alpha \beta}u^{\alpha}k^{\beta}$, \cite{pal1,
Mohapatra:rq} which would verify the current conservation condition
for the two point function.
\subsubsection{In presence of a background magnetic field}
In this case neither {\bf C} or {\bf P} but {\bf CP} is a symmetry,
first we look at the {\bf CP} transformation properties of the
axialvector-vector amplitude\footnote{This is a preliminary attempt to
get some idea about the tensorial basis of $\Pi^5_{\mu\nu}(k)$.
Strictly speaking this attempt to use {\bf CP} transformation
properties of the axialvector-vector amplitude must be incomplete
because we know that the {\bf CP} symmetry is also broken in
nature.}. As the $Z_\mu-A_\nu$ mixing Lagrangian is of the form,
\begin{eqnarray}
\mathscr L_{\rm Z-A} \propto Z^\mu(k) 
\Pi^5_{\mu \nu}(k) A^\nu(-k)\,,   
\end{eqnarray}
we can specify the ${\bf CP}$ transformation property of $\Pi^5_{\mu
\nu}(k)$ knowing the ${\bf CP}$ transformation properties of the
fields $Z_\mu$ and $A_\nu$. Under a {\bf CP} transformation the
various components of the tensor transforms as,
\begin{eqnarray}
{\rm{CP}}:\,\Pi^5_{0 0} &\to& \Pi^5_{0 0}\,,\nonumber\\
{\rm{CP}}:\,\Pi^5_{i j} &\to& \Pi^5_{i j}\,,
\label{pi5cpij}\\
{\rm{CP}}:\,\Pi^5_{0 i} &\to& -\Pi^5_{0 i}\,.\nonumber
\end{eqnarray}
At first, out of all the tensors available in the
present situation, a set of mutually orthogonal 4-vectors are
constructed with specific {\bf CP} transformation properties. In the
next step these different 4-vectors are combined in  pairs to
produce a second rank tensor which has similar {\bf CP} transformation
properties to that of $\Pi^5_{\mu\nu}(k)$. As the 4-vectors whose
combinations give the second rank tensors were mutually orthogonal to
start with, the second rank tensors are also mutually orthogonal.
The basic task in this section is to list all the second rank tensors
which have similar {\bf CP} transformation properties to that of
$\Pi^5_{\mu\nu}(k)$.

Before starting the classification of the various second rank tensors
at hand which have similar {\bf CP} transformation properties as that
of $\Pi^5_{\mu\nu}(k)$, some preliminary remarks about the form-factor
analysis is in order. In the beginning of this section we have seen
that $\Pi^5_{\mu\nu}(k)$ is zero in the vacuum. As a consequence it can be
inferred that $\Pi^5_{\mu\nu}(k)$ calculated in the presence of a
background uniform magnetic field vanishes as the magnetic field goes
to zero.  In a uniform background magnetic field, the vectors and
tensors at hand are
\begin{eqnarray}
F_{\mu\nu},\qquad {\widetilde{F}_{\mu\nu}},\qquad k^{\mu}_{\para},
\qquad k^{\mu}_{\pr}.   
\label{lt}
\end{eqnarray}
%
$F_{\mu \nu}$ is the field strength tensor for the uniform background 
magnetic field and ${\widetilde F}^{\mu \nu}$ is its dual given as 
${\widetilde F}^{\mu \nu} = \frac{1}{2}{\varepsilon^{\mu \nu \rho
\sigma}} F_{\rho \sigma}$. For a magnetic field directed along the 
$z$-axis the expressions of $F_{12}$, $F_{21}$, $F_{03}$, $F_{30}$ 
are given in Eq.~(\ref{edual}) in chapter \ref{chap3}.  
%
%
%
All other components of $F_{\mu\nu}$ zero. The sign of the
4-dimensional totally antisymmetric tensor is specified in
Eq.~(\ref{leviciv}). Now due to Lorentz invariance the form-factors
multiplying the basis tensors can be functions of $k^2$, $k^2_\para$,
$k^2_\pr$ and even functions of the background magnetic field as
$F_{\mu \nu}F^{\mu \nu}$, $F_{\mu \nu}F^{\nu \rho}F_{\rho
\sigma}F^{\sigma \mu}$ and other higher powers. Terms containing an
odd number of $F_{\mu \nu}$s such as,
\begin{eqnarray}
F_{\lambda \mu} F^{\mu \rho} F_\rho^{\,\,\,\lambda}\,,
\end{eqnarray}
will yield zero because only $F_{12}$ and $F_{21}$ are non zero.
The first thing to notice from
the above discussion is that the form-factors themselves are {\bf CP}
even and so the {\bf CP} transformation property of $\Pi^5_{\mu
\nu}(k)$ is solely dependent on the basis tensors of
$\Pi^5_{\mu\nu}(k)$.  Secondly, from Eq.~(\ref{SB}) of chapter
\ref{chap3} it is evident that in the ${\cal B}\to 0$ limit the
propagator in a magnetic field transforms into the vacuum propagator
and it is shown~\cite{Chyi:1999fc} that the Schwinger propagator
can be expanded in a series of powers of the external field magnitude
i.e.${\cal B}$. From this observation it is assumed that calculations
based on the Schwinger propagator are perturbative in the the external
field i.e. can be expanded in a series of ${\cal B}$. In the present
circumstance the form-factors can also be expanded in a power series
of the the even powers of ${\cal B}$ where the first term of the
series is ${\cal B}^2$ raised to the power zero i.e. a term
independent of ${\cal B}$.  So if the external magnetic field vanishes
the form-factors themselves in general do not vanish. As a result the
second rank tensors which multiply the form-factors must tend to zero
as the external field is put off because $\Pi^5_{\mu\nu}(k)$ itself
vanishes when the external field is absent.

Now all the possible 4-vectors that can be made
from the quantities enlisted in Eq.~(\ref{lt}) are given. They are,
\begin{eqnarray}
\begin{array}{cclc}
b^{\mu}_1 & = & (Fk)^{\mu}\,,\\
b^{\mu}_2 &=& (\widetilde{F}k)^{\mu}\,,\\          
b^{\mu}_3 &=& k^{\mu}_{\para}\,\\
b^{\mu}_4 &=& k^{\mu}_{\pr}.
\label{b4}
\end{array}
\end{eqnarray}
The expressions $(Fk)_{\mu}$, and $({\widetilde F}k)_{\mu}$ stands for
\begin{eqnarray}
\begin{array}{cclc}
(Fk)_{\mu}&=& F_{\mu \nu} k^{\nu}\,,\\
({\widetilde F}k)_{\mu}&=&{\widetilde F}_{\mu \nu} k^{\nu}\,.
\end{array}
\end{eqnarray}
The set of 4-vectors $b^{\mu}_i$s, $i=1,2,3,4$ are mutually
orthogonal to each other and can serve as the basis vectors to build
up the tensor basis of $\Pi^5_{\mu \nu}(k)$.

Next the {\bf CP} transformation properties of these vectors are
summarized.
\begin{eqnarray}
\begin{array}{cccc}
{\rm{CP}}:&\,b^0_1 &\to& b^0_1\,,\\
{\rm{CP}}:&\,\vec{b}_1 &\to& \vec{b}_1.
\label{cpb1}
\end{array}
\end{eqnarray}
The other three vectors have similar transformation properties as
\begin{eqnarray}
\begin{array}{cccc}
{\rm{CP}}:&\,b^0_{2,3,4} &\to&   b^0_{2,3,4}\,\\
{\rm{CP}}:&\,\vec{b}_{2,3,4} &\to& - \vec{b}_{2,3,4}.
\end{array}
\end{eqnarray}
From Eq.~(\ref{pi5cpij}) we can see that a suitable tensor basis can
be built up from vectors $b^{\mu}_i$ where $i = 2,3,4$. The {\bf CP}
transformation of the axialvector-vector amplitude compels us to
disregard $b^{\mu}_1$ as a basis vector.

Now we can list all the possible second rank tensors, having similar
{\bf CP} transformation properties to that of $\Pi^5_{\mu \nu}(k)$, made
by combining in pairs the mutually orthogonal 4-vectors
$b^{\mu}_2$, $b^{\mu}_3$, $b^{\mu}_4$. There are nine of them. For
later usage we denote them as follows:
\begin{eqnarray}
\begin{array}{r@{=}c@{=}l@{\quad}r@{=}c@{=}l@{\quad}r@{=}c@{=}l}
B^{\mu \nu}_1 & b^{\mu}_2 b^{\nu}_2
                    & (\widetilde{F}k)^{\mu}(\widetilde{F}k)^{\nu}\,,
&\quad B^{\mu \nu}_2 & b^{\mu}_3 b^{\nu}_3
                    & k^{\mu}_{\para} k^{\nu}_{\para}\,,\quad&
B^{\mu \nu}_3 & b^{\mu}_4 b^{\nu}_4 
                    & k^{\mu}_{\pr} k^{\nu}_{\pr}\,,\\
B^{\mu \nu}_4 & b^{\mu}_2 b^{\nu}_3 
                    & (\widetilde{F}k)^{\mu} k^{\nu}_{\para}\,,\quad
&\quad B^{\mu \nu}_5 & b^{\mu}_3 b^{\nu}_2 
                    & (\widetilde{F}k)^{\nu} k^{\mu}_{\para}\,,\quad& 
B^{\mu \nu}_6 & b^{\mu}_2 b^{\nu}_4 
                    & (\widetilde{F}k)^{\mu} k^{\nu}_{\pr}\,,\\
B^{\mu \nu}_7 & b^{\mu}_4 b^{\nu}_2
                    & (\widetilde{F}k)^{\nu} k^{\mu}_{\pr}\,, 
&\quad B^{\mu \nu}_8 & b^{\mu}_3 b^{\nu}_4
                    & k^{\mu}_{\para} k^{\nu}_{\pr}\,,\quad&
B^{\mu \nu}_9 & b^{\mu}_4 b^{\nu}_3
                    & k^{\nu}_{\para} k^{\mu}_{\pr}\,. 
\end{array}
\end{eqnarray}
This basis gives nine second rank mutually orthogonal tensors. The
orthogonal condition for the second rank tensors is given by,  
\begin{eqnarray}
B_i^{\mu \nu} B_{j\,\mu \nu} = 0\,.
\end{eqnarray}
In the above equation $i,j=1,..,9$ and $i\neq j$.  Any second rank
tensor containing higher field dependence can be represented by
suitable combinations of these tensors.

Out of these nine basis tensors,
$B^{\mu \nu}_2$, $B^{\mu \nu}_3$, $B^{\mu \nu}_8$ and 
$B^{\mu \nu}_9$
do not vanish in the ${\mathcal B} \to 0$ limit and are therefore
redundant.  The remaining five second rank tensors qualify
successfully as probable building blocks of the axialvector-vector
amplitude in the light of {\bf CP} transformation properties. 
But till now the gauge invariance condition as given
in Eq.~(\ref{ngi}) has not been used. Regarding gauge invariance the
first thing to note is that $B^{\mu \nu}_1$ can also be written as,
\begin{eqnarray}
B^{\mu \nu}_1 &=& (\widetilde{F}k)^{\mu}(\widetilde{F}k)^{\nu}\,\nonumber\\
              &=& -{\cal B}^2 (g^{\mu \nu}_{\para}k^2_{\para} -
                   k^{\mu}_{\para}  k^{\nu}_{\para})\,,
\label{bmunu1}
\end{eqnarray}
and as a result
\begin{eqnarray}
k_\mu B^{\mu \nu}_1 = B^{\mu \nu}_1 k_\nu = 0\,.
\end{eqnarray}
From the above equation it is seen that $B^{\mu \nu}_1$ satisfies the
gauge invariance condition for the photon vacuum polarization tensor.
From Eq.~(\ref{bmunu1}) and the expression of the vacuum polarization
given in Eq.~(\ref{tsai1}) it can be seen that, except a numerical
constant which can be absorbed in $N_{\para}$, $B^{\mu \nu}_1$ actually
occurs in the expression of the polarization tensor. From these
observations it can be inferred that $B^{\mu \nu}_1$ is not a suitable
tensor to be in the set of the probable second rank tensors which can
act as the basis of axialvector-vector amplitude. The basis tensors of
the axialvector-vector amplitude can be made by suitable combinations
of the other four second rank tensors listed above, satisfying the
{\bf CP} transformation properties as given in Eq.~(\ref{pi5cpij}) and
the gauge invariance condition. They are,
\begin{eqnarray}
P^{\mu \nu}_1 = (\widetilde{F}k)^{\nu} k^{\mu}_{\para}\,, 
\end{eqnarray}
and
\begin{eqnarray}
P^{\mu \nu}_2 = (\widetilde{F}k)^{\mu} k^{\nu}_{\pr} + (\widetilde{F}k)^{\nu}
k^{\mu}_{\pr} + \frac{k^2_\pr}{k^2_\para}[(\widetilde{F}k)^{\mu}
k^{\nu}_{\para} - (\widetilde{F}k)^{\nu} k^{\mu}_{\pr}]\,.
\end{eqnarray}
$P^{\mu \nu}_2$ can be simply written as:
\begin{eqnarray}
P^{\mu \nu}_2 = (\widetilde{F}k)^{\mu} k^{\nu}_{\pr} + (\widetilde{F}k)^{\nu}
k^{\mu}_{\pr} + k^2_\pr {\widetilde{F}}^{\mu \nu}\,, 
\end{eqnarray}
utilizing the identity,
\begin{eqnarray}
{\widetilde F}^{\mu \nu} = \frac{1}{k^2_{\para}}\left[(\widetilde{F}k)^{\mu}
k^{\nu}_{\para} - (\widetilde{F}k)^{\nu} k^{\mu}_{\para} \right]\,.
\end{eqnarray}
The result as given in the papers by Hari Dass and Raffelt
\footnote{However the metric used by the authors in references
mentioned is different from that of ours, and in their calculation the 
$\mu$ vertex is the vector type vertex. Due to this change the sign of
$k^2_\pr{\widetilde{F}}^{\mu \nu}$ appearing in $P^{\mu \nu}_2$ is
different from the one they obtained. The metric used in this thesis
is specified in section \ref{conv}}
verifies this choice~\cite{DeRaad:1976kd,Ioannisian:1996pn},
\begin{eqnarray}
& &\Pi^{5}_{\mu\nu}(k) = \frac{e^3}{(4\pi)^2 m^2}\left[C_\para P_{1\,\mu \nu}
 + C_\pr P_{2\,\mu \nu}\right]\,,
\label{harid}
\end{eqnarray}
where $C_\para$ and $C_\pr$ are the form-factors.
\subsubsection{In presence of a magnetized medium}
In presence of a magnetized medium the situation complicates. In this
analysis we are not going into an in depth study of the tensorial basis
as is done in the case where there was no medium. No new techniques
are used in this discussion, it is only a continuation of the previous
analysis in a new situation.
   
To start with we again emphasize on the {\bf CP} transformation
properties of the axialvector-vector amplitude $\Pi^5_{\mu
\nu}(k)$. Unlike the vacuum case now the theory may not be {\bf CP}
invariant. If the background contains more number of electrons than
positrons and/or if the electrons momenta are not random but along some
specific direction then {\bf CP} is violated by the background.  For
simplicity here we will discuss only those cases where the background
does not break {\bf CP}.

Now the form-factors can be functions of odd powers of
the magnetic field as well, as now new scalars as:
\begin{eqnarray}
\begin{array}{ccc}
(Fk)u & = & (Fk)_\mu u^\mu\,,\\
({\widetilde F}k)u & = & ({\widetilde F}k)_\mu u^\mu\,,
\end{array}
\end{eqnarray}
are also available. These scalars change sign under {\bf CP}
transformation. Some of the form-factors may contain odd powers of the
scalars listed above multiplied by equal odd powers of chemical
potential of the background charged fermions. These form-factors will
then not change sign under a {\bf CP} transformation. So in a
magnetized medium there can be two kinds of form-factors, one which
does not change sign and the other which does change sign under a {\bf
CP} transformation. This is unlike the previous case where there was
no medium. So in this case the basis tensors can have different {\bf
CP} transformation properties as the form-factors which multiply them
can also have different transformation properties. The vector
$b^{\mu}_1$ in Eq.~(\ref{b4}) had a {\bf CP} transformation property
which was different from the other three vectors $b^{\mu}_{2,3,4}$, as
a result it was dropped from the set of vectors which could combine in
pairs to give a set of second rank tensors having similar {\bf CP}
transformation property as that of $\Pi^5_{\mu \nu}(k)$. In the
present circumstances $b^{\mu}_1$ is not excluded as in the vacuum
because the {\bf CP} transformation property of the basis tensors have
changed.  

In presence of a medium we can have two different sets of
mutually orthogonal vectors. The first set is as supplied in
Eq.~(\ref{b4}).  The other set of orthogonal vectors useful in a
medium, are
\begin{eqnarray}
\begin{array}{cclc}
b'^{\mu}_1 &=& (\widetilde{F}u)^{\mu}\,,\\
b'^{\mu}_2 &=& u^{\mu}_{\para}\,,
\label{bp3}
\end{array}
\end{eqnarray}
In this list we have omitted two vectors. One is $(F u)^{\mu}$ and the
other one is $u^{\mu}_\pr$. The reason is that ultimately we are
interested in the rest frame of the medium. In the medium rest frame
there is no electric field. Also $u^{\mu}_\pr = 0$ in this frame.

This above set of vectors as given in Eq.~(\ref{bp3}) has similar {\bf
CP} transformation properties to those of $b^{\mu}_2$ and
$b^{\mu}_3$. But the vectors given in Eq.~(\ref{b4}) are not
orthogonal to the vectors specified in Eq.~(\ref{bp3}). So although
the number of vectors at hand has increased, four from Eq.~(\ref{b4})
and two from Eq.~(\ref{bp3}), the number of mutually orthogonal
vectors have not increased. So this six vectors as such cannot serve
as the basis vectors to build up the second rank tensor basis of
$\Pi^5_{\mu \nu}(k)$. Only a linear combination of the two sets of
vectors can make a orthogonal vector basis. Now we list the set of
orthogonal basis vectors which can be made from the two set of
vectors, they are:
\begin{eqnarray}
\begin{array}{cclc}
b''^{\mu}_1 &=& (Fk)^{\mu}\,,\\
b''^{\mu}_2 &=& ({\widetilde F}u)^{\mu} + ({\widetilde F}k)^{\mu}\,,\\ 
b''^{\mu}_3 &=& k^{\mu}_{\pr}\,,\\
b''^{\mu}_4 &=& k^{\mu}_{\para} + u^{\mu}_{\para}\,.
\label{bpp4}
\end{array}
\end{eqnarray}
In a magnetized medium we have these four basis vectors which serves
as the building blocks of the axialvector-vector amplitude. The basis
tensors in this case will be the direct product of these basis
vectors. There will be sixteen of them but all of them will not be
useful.  The electromagnetic current conservation condition
will reduce the number of admissible basis tensors. 
\subsection{One  loop calculation of the  axialvector-vector amplitude
$\Pi^5_{\mu \nu}(k)$ in a magnetized medium}
\label{capt}
Since we investigate the case with a uniform background magnetic
field, without any loss of generality it can be taken to be in the
$z$-direction. The relevant Feynman diagram of the neutrino
electromagnetic vertex to one loop appears in \fig{f:4fermi}.
Following that diagram the axialvector-vector amplitude $\Pi^5_{\mu
\nu}(k)$ is given as in Eq.~(\ref{npimunu5}). For calculating 
$\Pi^5_{\mu\nu}(k)$ in presence of a magnetized medium the propagators
appearing in Eq.~(\ref{npimunu5}) like $S(p)$ and $S(p')$ must be 
replaced by the appropriate ones suitable, $S_B(p)$ and $S_B(p')$ as
introduced in chapter \ref{chap3}, Eq.~(\ref{sbp}). 

Similar to the case of $\Pi_{\mu\nu}(k)$ here also the phase factor
does not appear because the quantity in question is a one loop 2-point
function.  From Eq.~(\ref{sbp}), we see that there are two terms in
the propagator $S_B(p)$ in a magnetized medium, the vacuum part and
the other part which involves background matter distribution. In the
trace appearing in the right hand side of Eq.~(\ref{npimunu5}) there
appears a product of two propagators $S_B(p)$ and $S_B(p')$ with other
quantities in between. This product will give rise to four quantities
of which one term will contain two factors of $S_B^V$s from the two
propagators. It has no importance as far as the background effects are
concerned. There will be one term containing two factors of
$S^\eta_B$s, which will contribute only to the absorptive part of the
axialvector-vector amplitude.  There will be two terms in the trace
each containing a factor of $S_B^V$ and a factor of $S^\eta_B$.  These
terms contribute to the present calculation, as they are important for
calculating the effective charge of the neutrinos in a magnetized
medium.

With this introduction the relevant expression of $\Pi^5_{\mu\nu}(k)$
can be written as,
\begin{eqnarray}
i\Pi^5_{\mu\nu}(k)= e^2 \int {{d^4 p}\over {(2\pi)^4}}\Tr \left
[ \gamma_\mu \gamma_5 S^\eta_B(p) \gamma_\nu
iS^V_B(p')
+\gamma_\mu
\gamma_5 iS^V_B(p) \gamma_\nu S^\eta_B(p')\right]\,.
\label{pi-ini}
\end{eqnarray}
Using the form of the fermion propagators in presence of a magnetic
field and a magnetized medium, given by Eq.~(\ref{SB}) and
Eq.~(\ref{Seta}) in sections \ref{propv} and \ref{smed} of chapter
\ref{chap3} we get
\pagebreak
\begin{eqnarray}
i\Pi^5_{\mu\nu}(k)&=& -e^2 \int {{d^4 p}\over {(2\pi)^4}}
\int_{-\infty}^\infty ds\, e^{\Phi(p,s)}\nonumber\\
&\times &\int_0^\infty
ds'e^{\Phi(p',s')}\big[ \Tr \left[\gamma_\mu\gamma_5 G(p,s)
\gamma_\nu G(p',s')\right]\eta_F(p\cdot u)\big. \nonumber\\
& &\big.\hskip 1cm + \Tr \left[\gamma_\mu
\gamma_5 G(-p',s') \gamma_\nu G(-p,s)\right]\eta_F(-p\cdot u)
 \big]\nonumber\\
&=& -e^2 \int {{d^4 p}\over {(2\pi)^4}}
\int_{-\infty}^\infty ds\, e^{\Phi(p,s)}
\int_0^\infty ds'\,e^{\Phi(p',s')}\mbox{R}_{\mu\nu}(p,p',s,s')\,,
\label{compl}
\end{eqnarray}
where $\mbox{R}_{\mu\nu}(p,p',s,s')$ contains the traces. 
\subsubsection{$\mbox{R}_{\mu\nu}$ to even and odd orders in magnetic field}
We calculate $\mbox{R}_{\mu\nu}(p,p',s,s')$ to even and odd orders in the
external magnetic field and call them $\mbox{R}^{(e)}_{\mu\nu}$
and $\mbox{R}^{(o)}_{\mu\nu}$. The reason for doing this is 
that the two contributions have different properties as far as their
dependence on medium is concerned, and the contributions are
\begin{eqnarray}
\mbox{R}^{(e)}_{\mu\nu}&=&
4i\eta_{-}(p\cdot u)\varepsilon_{\mu \nu \alpha \beta}
\left[p^{\alpha_\para} p'^{\beta_\para}(1 + \tan(e{\mathcal B}s)\tan(e{\mathcal
B}s'))\right.\nonumber\\ 
&+&\left. p^{\alpha_\para} p'^{\beta\pr} \sec^2 (e{\mathcal
B}s') 
+ p^{\alpha\pr} 
p'^{\beta_\para} \sec^2 (e{\mathcal B}s)\right.\nonumber\\
&+&\left. p^{\alpha\pr} p'^{\beta\pr}\sec^2 (e{\mathcal
B}s)\sec^2 (e{\mathcal B}s')\right]\,,
\label{reven}
\end{eqnarray}
and
\begin{eqnarray}
\mbox{R}^{(o)}_{\mu\nu}&=& 4i\eta_{+}(p\cdot u)\left[ m^2\varepsilon_{\mu \nu
1 2}(\tan(e{\mathcal B}s) + \tan(e{\mathcal B}s'))\right.\nonumber\\
&+&\left.\left\{(g_{\mu \alpha_\para} p^{\widetilde{\alpha_\para}}
p'_{\nu_\para} - g_{\mu \nu} p'_{\alpha_\para}
p^{\widetilde{\alpha_\para}} +g_{\nu \alpha_\para} p^{\widetilde{\alpha_\para}}
p'_{\mu_\para} )\right.\right.\nonumber\\
&+&\left.\left.(g_{\mu \alpha_\para} p^{\widetilde{\alpha_\para}}
p'_{\nu\pr} + g_{\nu \alpha_\para} p^{\widetilde{\alpha_\para}}
p'_{\mu\pr}) \sec^2(e{\mathcal B}s')\right\} \tan(e{\cal
B}s)\right.\nonumber\\
&+& \left.\left\{(g_{\mu \alpha_\para} p'^{\widetilde{\alpha_\para}}
p_{\nu_\para} - g_{\mu \nu} p_{\alpha_\para}
p'^{\widetilde{\alpha_\para}} +g_{\nu \alpha_\para} 
p'^{\widetilde{\alpha_\para}}
p_{\mu_\para} )\right.\right.\nonumber\\
&+&\left.\left.(g_{\mu \alpha_\para} p'^{\widetilde{\alpha_\para}}
p_{\nu\pr} + g_{\nu \alpha_\para} p'^{\widetilde{\alpha_\para}}
p_{\mu\pr}) \sec^2(e{\mathcal B}s)\right\} \tan(e{\mathcal
B}s') \right]. 
\label{rodd}
\end{eqnarray}
Here
\begin{eqnarray}
\eta_+(p\cdot u)&=&\eta_F(p\cdot u) + \eta_F(-p\cdot u)\,, \label{etaplus}\\
\eta_-(p\cdot u)&=&\eta_F(p\cdot u) - \eta_F(-p\cdot u)\,,
\label{etaminus}
\end{eqnarray}
which contain the information about the distribution functions.
Also it should be noted that, in our convention
\begin{eqnarray}
a_{\mu} b^{{\widetilde \mu}_\para}=a_0 b^3 + a_3 b^0\,.
\end{eqnarray}
As stated we have split the contributions to $\Pi^5_{\mu\nu}(k)$ to
odd and even orders in the external constant magnetic field. The main
reason for doing so is the fact that, $\Pi^{5(o)}_{\mu\nu}(k)$ and
$\Pi^{5(e)}_{\mu\nu}(k)$, the axialvector-vector amplitude to odd and
even powers in $e{\mathcal B}$, have different dependence on the
background matter. Pieces proportional to even powers in ${\mathcal
B}$ are proportional to $\eta_{-}(p\cdot u)$, an odd function of the
chemical potential. On the other hand pieces proportional to odd
powers in ${\mathcal B}$ depend on $\eta_{+}(p\cdot u)$, and are even
in $\mu$ and as a result it survives in the limit $\mu \to 0$. 

From Eq.~(\ref{reven}) we notice that $\Pi^5_{\mu \nu}(k)$ to even
orders in the magnetic field satisfies the current conservation
condition in both the vertices. In Eq.~(\ref{rodd}) we see that all
the terms in the right hand side are symmetric in the $\mu$ and $\nu$
indices except the first term. This term differentiates between the
two vertices in this case and as $\Pi^5_{\mu \nu}(k)$ to odd orders in
magnetic field is gauge invariant in the $\nu$ vertex we do not get
the same condition for the axialvector vertex. If in Eq.~(\ref{rodd})
we put $m=0$ then all the terms in the right will be symmetric in both
the tensor indices, and as a result current conservation condition
will hold for both vertices. If the mass of the looping fermion is not
zero then from the above analysis we can say that only Eq.~(\ref{ngi})
will hold.

If we concentrate on the rest frame of the medium, then $p\cdot
u=p_0$. Thus, the distribution function does not depend on the spatial
components of $p$. In this circumstance we can manipulate under the
momentum integral signs, and modify the above equations. The
techniques used to convert expressions inside the momentum integrals
is briefly discussed in appendix {\ref{muis}}, while other related
calculations showing the derivations of the equations given below are
supplied in appendix \ref{somecal}. Using the relations discussed in 
the appendix we can rewrite,
\begin{eqnarray}
\mbox{R}^{(e)}_{\mu \nu}&\stackrel{\circ}{=}&
4i\eta_{-}(p_0)\left[\varepsilon_{\mu \nu \alpha_\para \beta_\para}
p^{\alpha_\para} p'^{\beta_\para}(1 + \tan(e{\mathcal
B}s)\tan(e{\mathcal B}s'))
+ \varepsilon_{\mu \nu \alpha_\para
\beta\pr} p^{\alpha_\para} p'^{\beta\pr} \sec^2 (e{\cal
B}s')\right.\nonumber\\
&+&\left.\varepsilon_{\mu \nu \alpha\pr \beta_\para} p^{\alpha\pr} 
p'^{\beta_\para} \sec^2 (e{\mathcal B}s)\right]\,,
\label{reven1}
\end{eqnarray}
and 
\begin{eqnarray}
\mbox{R}^{(o)}_{\mu \nu}&\stackrel{\circ}{=}&
4i\eta_+(p_0)\left[-\varepsilon_{\mu \nu 1 2} 
\left\{ \frac{\sec^2(e{\mathcal B}s)\tan^2(e{\mathcal
B}s')}{\tan(e{\mathcal B}s) 
+ \tan(e{\mathcal B}s')}
k_{\pr}^2 
+ (k\cdot p)_\para (\tan(e{\mathcal B}s) +
\tan(e{\mathcal B}s'))\right\}\right.\nonumber\\ 
&+&\left. 2\varepsilon_{\mu 1 2 \alpha_\para}(p'_{\nu_\para}
p^{\alpha_\para}\tan(e{\mathcal B}s) +
p_{\nu_\para}p'^{\alpha_\para}\tan(e{\mathcal B}s'))
+ g_{\mu\alpha_\para} k_{\nu\pr}\left\{p^{\widetilde
\alpha_\para}(\tan(e{\mathcal B}s)
 - \tan(e{\mathcal B}s'))\right.\right.\nonumber\\
&-&\left.\left. k^{\widetilde \alpha_\para}\,
{\sec^2(e{\mathcal B}s)\tan^2(e{\mathcal B}s')\over{\tan(e{\mathcal B}s) +
\tan(e{\mathcal B}s')}}\right\}
+\{g_{\mu\nu}(p\cdot \widetilde k)_\para + g_{\nu \alpha_\para}
p^{\widetilde \alpha_\para} k_{\mu\pr}\}
(\tan(e{\mathcal B}s) - \tan(e{\mathcal B}s'))\right.\nonumber\\
&+&\left. g_{\nu \alpha_\para}
k^{\widetilde\alpha_\para}p_{\mu\pr}\sec^2(e{\mathcal B}s)
\tan(e{\mathcal B}s')\right].
\label{crmunu}
\end{eqnarray}
The sign `$\stackrel\circ=$' means that the expressions on both sides
of the above equations, though not necessarily equal algebraically,
yield the same momentum integral. To be precise if there are two
functions of 4-momenta $p$, $p'$ and parameters $s$, $s'$, as 
$A(p,p',s,s')$ and $B(p,p',s,s')$ and
\begin{eqnarray}
A(p,p',s,s') \stackrel\circ= B(p,p',s,s')\,, 
\end{eqnarray}
then in general,
\begin{eqnarray}
A(p,p',s,s') \neq B(p,p's,s')\,,
\end{eqnarray}
but 
\begin{eqnarray}
\int d^4p\, A(p,p',s,s')\,e^{\Phi(p,s)+\Phi(p',s')} = 
\int d^4p\, B(p,p',s,s')\,e^{\Phi(p,s)+\Phi(p',s')}\,. 
\label{symb}
\end{eqnarray}
As the momentum integrals appearing in the right hand side of 
Eq.~(\ref{compl}) are of the same form as those in Eq.~(\ref{symb})  
we can safely use the right hand sides of
Eq.~(\ref{reven1}) and Eq.~(\ref{crmunu}) as representatives of
$\mbox{R}^{(e)}_{\mu \nu}$ and $\mbox{R}^{(o)}_{\mu \nu}$.

Before going into the next section we comment on the nature of the
integral appearing in Eq.~(\ref{compl}). The first point to make is
that from the form of $\mbox{R}^{(e)}_{\mu \nu}$ in Eq.~(\ref{reven1})
we note the axialvector-vector amplitude in a magnetized medium to even
orders in the magnetic field is antisymmetric, as it was in a medium
without any magnetic field. Contrary to this $\mbox{R}^{(o)}_{\mu
\nu}$ does not have any well defined symmetry property.

Secondly, as the integrals are not done explicitly something must be
said about possible ultraviolet divergence which may creep up in
evaluating them. It is a complicated topic but fortunately the
integrals we are dealing with do not have any divergences as we are
working in presence of a medium which offers a natural ultraviolet
cut-off as the temperature of the system. The penultimate section in
this chapter discusses the points about ultraviolet divergences of the
two cases we have considered in this chapter.
 
The integral expression for $\Pi^5_{\mu
\nu}(k)$ in our case does not have any singularities. We can now
write the full expression of the axialvector-vector amplitude as
\begin{eqnarray}
i\Pi^5_{\mu\nu}(k)&=& -e^2 \int {{d^4 p}\over {(2\pi)^4}}
\int_{-\infty}^\infty ds\, e^{\Phi(p,s)}
\int_0^\infty ds'\,e^{\Phi(p',s')}\left[\mbox{R}^{(o)}_{\mu\nu} 
+ \mbox{R}^{(e)}_{\mu\nu}\right]\,,
\label{alord}
\end{eqnarray}
where $\mbox{R}^{(o)}_{\mu\nu}$ and
$\mbox{R}^{(e)}_{\mu\nu}$ are given by
Eqs.(\ref{crmunu}) and (\ref{reven1}) in the rest frame of the
medium. A thorough proof of the gauge invariance of  $\Pi^5_{\mu\nu}(k)$ is
given in appendix \ref{gauin}.
\section{Neutrino effective charge in various backgrounds}
\label{efftcharge}
\subsection{Definition of effective charge}
Intuitively when a neutrino moves inside a thermal medium composed of
electrons and positrons, they interact with these background
particles. The background electrons and positrons themselves have
interaction with the electromagnetic fields, and this fact gives rise
to an effective coupling of the neutrinos to the photons. Under these
circumstances the neutrinos may acquire an ``effective electric
charge'' through which they interact with the ambient plasma. The
proper definition of this charge follows.

The off-shell electromagnetic vertex function $\Gamma_{\nu}$ is
defined in such a way that, for on-shell neutrinos, the $\nu \nu
\gamma$ amplitude is given by:
\begin{eqnarray}
{\cal M} = - i \bar{u}(q') \Gamma_{\nu} u(q) A^{\nu}(k),
\label{chargedef}
\end{eqnarray}
where, $k$ is the photon momentum. Here, $u(q)$ is the the neutrino
spinor and $A^{\nu}$ stands for the electromagnetic vector
potential. In general $\Gamma_{\nu}$ would depend on $k$ and the
characteristics of the medium. With our effective Lagrangian in
Eq.~(\ref{4fermi}), the form of $\Gamma_{\nu}$ is as given in 
Eq.~(\ref{nvf}).
%
%
The effective charge of the neutrinos is defined in terms of the
vertex function by the following relation~\cite{pal1, Mohapatra:rq}:
\begin{eqnarray}
e_{\rm eff} = {1\over{2 q_0}} \, \bar{u}(q) \, \Gamma_0(k_0=0, {\vec k}
\rightarrow 0) \, u(q) \,.
\label{chargedef1}
\end{eqnarray}
For left handed massless Weyl spinors this definition can be rendered 
into the form:
\begin{eqnarray}
e_{\rm eff} = {1\over{2 q_0}} \, \Tr  \left[\Gamma_0(k_0=0, {\vec
k}\rightarrow 0) \, (1 - \gamma^5) \, \rlap/q \right]\,.
\label{nec1}
\end{eqnarray}
%

Both the vector-vector amplitude and the axialvector-vector amplitude
contribute to the effective charge of neutrinos. In the following
subsections the the effective charge of neutrinos in different 
backgrounds are discussed. 
\subsection{In vacuum}
In vacuum
$\Pi_{\mu\nu}(k)$ vanishes in the limit $k_0=0,~{\vec k}\to 0$ and
$\Pi^5_{\mu\nu}(k)$ is zero. As a result there can be no effective
charge of neutrinos in vacuum.

\subsection{In a medium} 
In a medium the axialvector-vector amplitude is of the form
$\varepsilon_{\mu \nu \alpha \beta}u^{\alpha}k^{\beta}$, and so it
does not contribute for the effective electric charge of the neutrinos
since for charge calculation we have to put the index $\nu = 0$. In
the rest frame only the time component of the 4-vector $u$ exists,
that forces the totally antisymmetric tensor to vanish. In a medium
the vector-vector amplitude is same as the vacuum polarization as
given in Eq.~(\ref{mvacpol}) except one point. Eq.~(\ref{mvacpol})
contains $\Pi_P$ which is an effect of $\bf{P}$ and $\bf{CP}$
violation, either by the Lagrangian or by the background
medium. Phenomenologically it will come in orders of $G_F$ and so its
inclusion will make the effective charge of neutrinos to be higher in
orders in the Fermi coupling. So the effect of $\Pi_P$ is neglected in
the effective charge calculations of neutrinos.  The longitudinal
projector $L_{\mu\nu}$ is not zero in the limit
$k_0=0,\vec{k}\rightarrow 0$ and $\Pi_L$ is also not zero in the above
mentioned limit~\cite{pal1,Mohapatra:rq}. This fact is responsible for
giving nonzero contribution to the effective charge of neutrinos in a
medium. The effective charge of neutrinos in a medium has been
calculated previously by many authors~\cite{alth, pal1, Mohapatra:rq}.
\subsection{In presence of a background magnetic field} 
From the Eq.~(\ref{harid}) we see that the axialvector-vector
amplitude in a background magnetic field without any medium does not
survive when the momentum of the external photon vanishes. Also from
Eq.~(\ref{tsai1}) it is evident that $\Pi^{\mu \nu}(k)$ also vanishes
when the momentum of the external photon vanishes.  As a result there
cannot be any effective electric charge of the neutrinos in a constant
background magnetic field.  Actually this formal statement could have
been spoilt by the presence of possible infrared divergence in the
loop; i.e to say in $C_\para$ and $C_\pr$ \cite{Ioannisian:1996pn} and
$N_0$, $N_\para$ and $N_\pr$~\cite{Tsai:1}. Since the particle inside
the loop is massive so there is no scope of having infrared
divergences.

\subsection{In presence of a magnetized medium}
To odd orders in the external magnetic field the expression of
$\Pi_{\mu \nu}(k)$ is given by Eq.~(\ref{Ofinal}) and it shows that
the result is dependent on $k^\beta$. In the limit when
the momentum of the external photon goes to zero $\Pi_{\mu \nu}(k)$
will vanish unless the parametric integrals can produce terms
which cancels $k^\beta$. In the present circumstance this cancellation
does not occur and as a result $\Pi_{\mu \nu}(k)$ vanishes. 
On the other hand all the components of $\Pi_{\mu\nu}^5(k)$ does not
vanish in the specified momentum limit. The relevant components of
$\Pi_{\mu\nu}^{5(e)}(k)$ which are required for the effective charge
calculation, vanish completely when the external momentum goes to zero.
But all the components of $\Pi_{\mu \nu}^{5(o)}(k)$ does not vanish
when the external photon momentum goes to zero.  As a result this can
give rise to an effective neutrino charge in a magnetized medium. The
calculations showing the validity of the above statements appear in
appendix \ref{somecal}.  

Now we concentrate on the zero momentum limit of that part of the
axial polarization tensor which is going to contribute for the
neutrino effective charge in the rest frame of a magnetized medium.
\subsubsection{Effective charge to odd orders in external field}
Denoting $\Pi^{5}_{\mu\nu}(k_0=0, {\vec k } \to 0)$ by
$\Pi^{5}_{\mu\nu}$, we obtain
\begin{eqnarray}
\Pi^5_{\mu 0}&=&\lim_{k_0=0
\vec{k}\rightarrow 0}4 e^2 \int{d^4p\over{(2\pi)^4}} 
\int^{\infty}_{-\infty} ds\, e^{\Phi(p,s)}
\int^{\infty}_0 ds'
e^{\Phi(p',s')}(\tan(e{\mathcal B}s) + \tan(e{\mathcal B}s')) \nonumber\\
&\times &\eta_+(p_0)\left[2 p^2_0 -  (k\cdot p)_\para
\right]\varepsilon_{\mu 0 1 2} 
\label{pi5k0}
\end{eqnarray}
the other terms turns out to be zero in this limit.  Although it looks
that the term $(k\cdot p)_\para$ will vanish in the limit when $k_0=0$
and $\vec{k}\rightarrow 0$, but it does not. This is because the
parametric integrals yield terms which contain $(k\cdot p)_\para$ in
the denominator of the momentum integrals and there remains a scope
for cancellation. The calculation in appendix \ref{somecal} clarifies the
point. The above equation shows that, except the exponential
functions, the integrand is free of the perpendicular components of
momenta. This is a peculiarity of this case that the perpendicular
excitations of the loop momenta are only present in the phase like
part of the integrals and in effect decouples from the scene once they
are integrated out. Its presence is felt only through a linear
dependence of the external field ${\mathcal B}$ when the perpendicular
components of $k$ vanish. Upon performing the gaussian integration
over the perpendicular components and taking the limit $k_{\pr} \to
0$, we obtain,
\begin{eqnarray}
\Pi^5_{3 0}&=&\lim_{k_0=0,\vec{k}\rightarrow 0} \frac{(4i e^3
{\mathcal B})}{4\pi} \int{d^2 p_\para \over{(2\pi)^2}} 
\int^{\infty}_{-\infty} ds 
\, e^{is(p^2_\para - m^2) - \varepsilon|s|}
\int^{\infty}_0 ds'
e^{is'(p'^{\,2}_\para - m^2) - \varepsilon|s'|} \nonumber\\
&\times &\eta_+(p_0)\left[2 p^2_0 -  (k\cdot p)_\para
\right].
\label{charge1}
\end{eqnarray}
The details of the calculation for the effective charge is given in
appendix \ref{somecal}.  For a classical gas background 
the expression of the effective charge, to odd orders in the external
magnetic field, is given by,
\begin{eqnarray}
e^{\nu}_{\rm eff} &=& - 2 \sqrt{2} g_{A} m \beta G_F
\frac{e^2 {\mathcal B}}{\pi^2} 
\cos\theta
\cosh(\beta\mu)  
K_1(m \beta)\,.
\label{nucharge}
\end{eqnarray}
Here $\theta$ is the angle between the neutrino three momentum and the
background magnetic field. The superscript $\nu$ in $e^{\nu}_{\rm
eff}$ denotes that we are calculating the effective charge of the 
neutrino. $K_1(m \beta)$ is the modified Bessel function (of the
second kind) of order one which sharply falls off as we move away from
the origin in the positive direction.  As temperature tends to zero,
$K_1(m\beta)$ grows as $e^{-m\beta}$, as a result from
Eq.~(\ref{nucharge}) it is seen that the effective charge vanishes.
In \cite{Nieves:2003kw} a more general treatment of the effective
charge, relinquishing the assumption of a classical gas, has been
presented.

Before ending this section a general discussion can me made on the
form-factors of $\Pi^5_{\mu \nu}(k)$ in the zero momentum limit.  In a
background magnetic field
the form-factors can be functions of scalars containing the 
magnetic field. The scalars can be of the following form: 
%
\begin{eqnarray}
k^{\mu}F_{\mu\nu}F^{\nu\lambda}k_{\lambda} 
\mbox{~~~and~~~}
F_{\mu\nu}F^{\mu\nu}\,,  
\end{eqnarray}
or
\begin{eqnarray}
({\widetilde F}u)^{\mu}({\widetilde F}u)_{\mu} \mbox{~~~and~~~}
({\widetilde F}u)^{\mu}k_\mu\,.
\end{eqnarray}
The scalars listed above are the most elementary ones, there can be
many more possible scalars where more number of $k^{\mu}$s,
$F^{\mu\nu}$s and $u_\mu$s are mutually fully contracted.
The thing which must be noted is when $k$ tends to zero only terms
that can survive in the form-factors must be even functions of
${\mathcal B}$.

Of all possible tensorial structures for the axialvector-vector amplitude
in a magnetized plasma, there exists one which is independent of the
external momentum $k$ given by, 
\begin{eqnarray}
{\widetilde F}_{\mu \alpha} u^{\alpha} 
u^{_\para}_{\nu}\,.  
\label{ww}
\end{eqnarray}
It is seen from the above expression, which is odd in the external
field, that it survives in the zero external momentum limit in the
rest frame of the medium. We have earlier noted that the form-factors
which exist in the rest frame of the medium and in the zero momentum
limit are even in powers of the external field. This tells us directly
that the axial polarization tensor must be odd in the external field
in the zero external momentum limit, a result which is verified by
actual calculations. Some similar calculations regarding the
axialvector-vector coupling is also done by Konar and Das~\cite{Konar:2002gy}.
\section{A short note on divergences in a background magnetic field}
\label{s:infty}
While calculating closed loops in a Feynman diagram in vacuum we
encounter ultraviolet divergences in many places. For renormalizable
quantum field theories there are standard methods of handling this
divergences. First the theory is regulated and then some of the bare
parameters in the theory are rescaled to give a meaningful
theory. There are various regularization schemes as the cut-off
regularization, the Pauli-Villars regularization and the most
prominently used dimensional regularization. Not going into the full
details of the renormalization programme in vacuum, which has been 
studied extensively, here we focus on the methods by which ultraviolet
divergences are handled in those calculations where we have a
background magnetic field.

The first thing to note here is that magnetic fields do not bring in
any new kind of divergences, they only modify the properties of the
divergences present in the vacuum structure of the theory. Vacuum
polarization of photon contains ultraviolet divergence and
consequently the divergence remains when we calculate it in a
background magnetic field. The expression of $\Pi_{\mu \nu}(k)$
calculated in a magnetic field includes the $\Be=0$ vacuum result
implicitly, which we denote by $\lim_{\Be\to 0}\Pi_{\mu \nu}(k)$.  The
divergent part in $\Pi_{\mu
\nu}(k)$ calculated in a background magnetic field is actually present
in $\lim_{\Be\to 0}\Pi_{\mu\nu}(k)$. The usual procedure is to fix the
divergent part in $\lim_{\Be\to 0}\Pi_{\mu\nu}(k)$ and subtract it
from the the final expression of $\Pi_{\mu \nu}(k)$ calculated in presence
of a magnetic field, to get rid of the divergence.


The ultraviolet nature of $\Pi^5_{\mu \nu}(k)$ is interesting. In
vacuum we have seen from subsection \ref{form} that $\Pi^5_{\mu
\nu}(k)$ is zero and so it cannot have any divergences when we
calculate it in presence of a background magnetic field. In this
connection it can be said that in absence of the medium but in
presence of the background magnetic field another divergent structure
could arise, the Adler anomaly~\cite{Adler:gk}, due to the presence of
the axial-vector vertex in the Feynman diagram of $\Pi^5_{\mu
\nu}(k)$.
A brief discussion of this topic is given in the work by Raffelt
and Ioannisian~\cite{Ioannisian:1996pn}.

As stated previously a medium affects the ultraviolet nature of the
calculations. The finite temperature of the system automatically act
as an ultraviolet cut-off of the system and so momentum integrals
cannot run to infinity to damage the logical consistency of the
theory. As a result the calculations in a magnetized medium are also
expected to be free from ultraviolet divergences. 
\section{Conclusion}
This chapter was about the electromagnetic interactions of neutrinos
in a magnetized medium. In the introduction various decay processes as
the photon splitting into two neutrinos, the Cherenkov process and
radiative decays of heavy neutrinos were discussed.  Section
\ref{npsev} deals with neutrino-photon scattering and the
electromagnetic vertex of neutrinos. The topic of neutrino-photon
scattering, which is highly suppressed in the standard model, is
briefly discussed in subsection \ref{nnps}.  The interesting thing
about this scattering process is that in presence of an external
magnetic field the neutrino-photon cross-section increases which can
be an important ingredient for astrophysics.  

In the standard model of particle physics neutrinos do not interact
with photons in the tree level so the neutrino electromagnetic vertex
is an effective one mediated by the charged fermions in the
loop. Section \ref{npsev} deals with the two important second rank
tensors, $\Pi_{\mu \nu}(k)$ and $\Pi^5_{\mu \nu}(k)$, which are the
building blocks of the electromagnetic vertex function of the
neutrinos. The structure of $\Pi_{\mu \nu}(k)$ in various possible
backgrounds has been discussed in subsection \ref{pimunumb}. 
Here for the first time methods of statistical field theory is
used to find out the form of $\Pi_{\mu\nu}(k)$, to odd orders in the
external field, in a magnetized medium. Subsection \ref{form} of the
present chapter deals with the axialvector-vector amplitude
$\Pi^5_{\mu\nu}(k)$ in a magnetized medium and its possible tensor
structures in various possible backgrounds.  Section \ref{efftcharge}
deals with the issue of the effective charge of the neutrino in a
magnetized medium. The formula of the effective charge calculated
shows that it depends upon the angle between the neutrino 3-momentum
and the external magnetic field directions. Also the effective charge
vanishes as temperature tends to zero.  

As no discourse in quantum field theory can be complete without
solving the thorny issue of divergences popping out from the loops, 
in section \ref{s:infty} this issue is discussed. It is pointed out
that the temperature of the medium can act as an ultraviolet regulator
for the various processes which we have discussed and we can safely
assume that the calculations are free of ultraviolet divergences.

\chapter{Conclusion}\label{chap5}
It is now an established fact that magnetic fields of various
magnitudes and various shapes pervade our universe. They are present
in the intergalactic medium, in the galactic region and also in
planetary and stellar atmospheres. Order of magnitude estimates of
these fields have been presented in chapter \ref{chap1}. 

Two different
ways have been followed in this thesis to incorporate the magnetic
field effects. The first way is straight forward and involves the
solution of the Dirac equation in presence of a classical uniform
background magnetic field. In presence of a magnetic field the Dirac
equation yields an exact solution.  The exact wave functions in a
uniform magnetic field can be used to produce a consistent quantum
field theory and various scattering cross-sections involving charged
fermions can be calculated, as elaborately discussed in the chapter
\ref{chap2}.  Magnetic fields found in astrophysical objects like
neutron stars, active galactic nuclei vary both in magnitude and
direction but the typical length scale $D$ in which this variation
occurs is much much higher than the Compton wavelengths of the
elementary particles and as a result this variation can be neglected
in a quantum field theoretical calculation\footnote{For neutron stars
$D \sim 10$Km.}.

Magnetic fields also enter into the elementary particle regime
indirectly through the effect of virtual charged particles.  To
evaluate quantities like neutrino self-energy and vacuum polarization
of photon intermediate virtual charged particles like electrons,
W-bozons are required whose propagators are modified in a magnetic
field. Although ultimately the intermediate particles are integrated
out, the magnetic field effects remains frozen in and affect the final
results. In chapters \ref{chap3} and \ref{chap4} we have encountered
some of the calculations involving the Schwinger propagator and seen
how magnetic fields creep in to modify the physical properties of
those particles which does not have any interaction with a magnetic
field normally.

Historically the URCA processes, as specified in section \ref{intneu},
were the first to be studied in an external field. Closely related
with the URCA processes is the inverse beta decay process. In chapter
\ref{chap2} the inverse beta decay cross-section of arbitrarily
polarized neutrons in an external magnetic field is calculated. The
magnitude of the external magnetic field assumed is much less than
$m_p^2/e$ and so the magnetic field modification of the proton wave
function is neglected. Using the solutions of the Dirac equation for
electrons in a background of a magnetic field a quantum field
theoretical formalism is built up to tackle various scattering
processes consistently.  From Eq.~(\ref{b0inv}) of chapter \ref{chap2}
it is seen that the inverse beta decay scattering cross-section in a
background magnetic field has a smooth ${\mathcal B}\to0$ limit. In
Eq.~(\ref{sigma}) the cross-section of the inverse beta decay process
is given and is seen to be dependent on the angle between the incoming
neutrino 3-momentum and the magnetic field vector.  From \fig{f:sigma}
it is apparent that the cross-section of the inverse beta decay in an
external magnetic field becomes considerably enhanced compared to its
corresponding value in vacuum when the magnetic field magnitude is
greater than ${\mathcal B}_e$, the critical field defined in
Eq.~(\ref{Bc}). Moreover when the initial neutrinos are taken to be
monochromatic \fig{f:sigma} shows that the cross-section of the
inverse beta decay process is plagued with spikes which goes all the
way to infinity. This disease is cured and the spikes smear out, as
shown in \fig{f:smooth}, when a flat probability distribution of
initial neutrino energy is considered. Using the fact that the beta
decay cross-section is sensitive to the angle between the neutrino
3-momentum and the magnetic field direction attempt has been made to
explain the high velocities of pulsars, of the order of
$450\pm90\;{\rm Km\,s}^{-1}$ in section \ref{asymnem}.

The charged fermion propagator in a background magnetic field as
derived by Schwinger is given in chapter \ref{chap3}, subsection
\ref{ss:sp}.  The propagator is defined as an integral over a
parameter called `proper time'. Using form-factor analysis the general
structure of the neutrino self-energy in a magnetized medium has been
found out in section \ref{vise}. From Eq.~(\ref{magdisp}) it is seen
that in a magnetized medium the neutrino self-energy becomes sensitive
to the angle between the neutrino propagation direction and the
magnetic field direction. It is also sensitive to the nature of the
background i.e., whether the background is made up of electrons or
muons etc., as is evident from Eq.~(\ref{Ye}).  Both of these facts
affect the resonant level crossing condition in context of neutrino
oscillations discussed in section \ref{noscillations}.

The topic of electromagnetic interaction of neutrinos is discussed in
chapter \ref{chap4}. All the calculations assumes an effective 4-fermi
vertex. There are various processes which are either forbidden or
highly restrained in vacuum, like a photon decaying into a
neutrino-antineutrino pair, Cherenkov radiation of neutrinos, 
neutrino-photon scattering, which can become possible in presence of a
background magnetic field. These processes, briefly discussed in
sections \ref{nnp} and \ref{npsev}, opens up new channels for energy
emission from newly born neutron stars and so are astrophysically
important. The neutrino electromagnetic vertex function 
$\Gamma_\mu$ can be written in terms of two second rank tensors 
$\Pi_{\mu\nu}(k)$ and $\Pi^5_{\mu\nu}(k)$ as mentioned in chapter
\ref{chap1} and explicitly shown in Eq.~(\ref{nvf}). 
$\Pi_{\mu\nu}(k)$ is exactly the vacuum polarization of the photon and
its form in a thermal medium and in presence of a magnetized medium is
presented in subsection \ref{pimunumb}. The general form of the
axialvector-vector amplitude $\Pi^5_{\mu \nu}(k)$, as defined in
Eq.~(\ref{npimunu5}), is calculated in a magnetized medium in section
\ref{form}. 
A general tensor analysis based on the discrete symmetries shows that
$\Pi^5_{\mu \nu}(k)$ must vanish in vacuum. A similar analysis gives
the tensor structure of $\Pi^5_{\mu \nu}(k)$ in a thermal
medium. Based upon {\bf CP} transformation property of $\Pi^5_{\mu
\nu}(k)$ the tensor basis of the axialvector-vector amplitude is
analyzed in a background magnetic field. Some comments on the tensor
basis of $\Pi^5_{\mu\nu}(k)$ in a magnetized medium is supplied at the
end of subsection \ref{form}. Subsequently $\Pi^5_{\mu \nu}(k)$ has been
calculated in a specific limit i.e., when $k_0=0$ and ${\vec k} \to
0$, which is related to the effective charge of the neutrinos. In this
limit it is observed that $\Pi^5_{\mu \nu}(k)$ does exist to odd
orders in the external magnetic field but to even orders it does
not. It is seen that when $k_0=0$ and ${\vec k} \to 0$, $\Pi^5_{\mu
\nu}(k)$ becomes linear in the external field. Using a classical gas
approximation the expression of the effective charge to odd orders in
the external field has been written down. The formula of the effective
charge calculated shows that it depends upon the angle between the
neutrino 3-momentum and the external magnetic field directions. Also
the effective charge vanishes as temperature tends to zero.

To conclude it must be said that most of the effects discussed in this
thesis depends upon large magnetic fields, where ${\mathcal
B}>{\mathcal B}_e$. These fields are obtained only at astronomical
distances, presumably in neutron stars and magnetars.  For such
distant objects, observational data are not clean enough to resolve
the effects of the magnetic field. Perhaps more accurate measurements
of indirect evidences like the velocities of the pulsars or rate of
cooling of newly born neutron stars can shine some light on the
calculations done in this thesis in near future.

%
%

\appendix      
\chapter{Spin Sum in a uniform background magnetic field}
\label{assum}
From Eq.~(\ref{Usoln}) the spin sum
$\sum_s U_s (y,n,\vec p\omit y) \overline U_s (y_\star ,n,\vec p\omit
y)$ can be written as:
\begin{eqnarray}
\sum_s U_s (y,n,\vec p\omit y) \overline U_s (y_\star ,n,\vec p\omit
y) = \frac{1}{E_n+m} \sum_{i,j = n-1}^n I_i(\xi) I_j(\xi_*)\,T_{i,j} 
\label{ssum1}
\end{eqnarray}
where the $T_{i,j}$\,s are $4\times4$ matrices. The $T_{i,j}$ matrices
are obtained by performing the matrix multiplication in the left hand
side of the above equation.

Using the dispersion relation $E_n^2 = p_z^2 + m^2 + 2ne{\mathcal B}$,
$T_{n,n}$ can be written as,
\begin{eqnarray}
T_{n,n}
 = 
\left( \begin{array}{ccccccc}
0 & & 0 & & 0 & & 0\\
0 & & (E_n+m) & & 0 & & p_z\\
0 & & 0 & & 0 & & 0 \\
0 & & -p_z & & 0 & & -(E_n-m)
\end{array} \right)\,.
\label{tnn}
\end{eqnarray}
In the $2\times2$ notation the above matrix can be written as,
\begin{eqnarray}
T_{n,n}
 &=& 
E_n\left( \begin{array}{ccc}
\frac12(1-\sigma_3) & & 0\\
0 & & -\frac12(1-\sigma_3)\\
\end{array} \right)
+
p_z\left( \begin{array}{ccc}
0 & & \frac12(1-\sigma_3)\\
-\frac12(1-\sigma_3)& & 0 \\
\end{array} \right)\nonumber\\
&+&
m\left( \begin{array}{ccc}
\frac12(1-\sigma_3) & & 0\\
0 & & \frac12(1-\sigma_3)\\
\end{array} \right)\,,
\label{ssum2}
\end{eqnarray}
where $\sigma_3$ is the third Pauli matrix.
%
%
In the $4\times4$ notation Eq.~(\ref{ssum2}) can be written as,
\begin{eqnarray}
T_{n,n} &=& \frac12[m(1 - \sigma_z) + E_n(\gamma^0 + \gamma^5\gamma^3) -
p_z(\gamma^5\gamma^0 + \gamma^3)]\,,\nonumber\\
&=& \frac12[m(1 - \sigma_z) + \rlap/p_\parallel + \widetilde{\rlap/p}_\parallel \gamma_5]\,,
\label{n1}
\end{eqnarray}
where $\sigma_z=i\gamma^1\gamma^2$.

In a similar way $T_{n-1,n-1}$ can be written as:
\begin{eqnarray}
T_{n-1,n-1}
 = 
\left( \begin{array}{ccccccc}
(E_n+m) & & 0 & & -p_z & & 0\\
0 & & 0 & & 0 & & 0\\
p_z & & 0 & & -(E_n-m) & & 0 \\
0 & & 0 & & 0 & & 0
\end{array} \right)\,.
\end{eqnarray}
In the $2\times2$ notation the above equation looks like,
\begin{eqnarray}
T_{n-1,n-1}
 &=& 
E_n\left( \begin{array}{ccc}
\frac12(1+\sigma_3) & & 0\\
0 & & -\frac12(1+\sigma_3)\\
\end{array} \right)
+
p_z\left( \begin{array}{ccc}
0 & & -\frac12(1+\sigma_3)\\
\frac12(1+\sigma_3)& & 0 \\
\end{array} \right)\nonumber\\
&+&
m\left( \begin{array}{ccc}
\frac12(1+\sigma_3) & & 0\\
0 & & \frac12(1+\sigma_3)\\
\end{array} \right)\,.
\end{eqnarray}
In the $4\times4$ notation the above equation becomes,
\begin{eqnarray}
T_{n-1,n-1} &=& \frac12[m(1 + \sigma_z) + E_n(\gamma^0 - \gamma^5\gamma^3) +
p_z(\gamma^5\gamma^0 - \gamma^3)]\,,\nonumber\\
&=& \frac12[m(1 + \sigma_z) + \rlap/p_\parallel - 
\widetilde{\rlap/p}_\parallel \gamma_5]\,.
\label{n2}
\end{eqnarray}

From the matrix multiplication in the left hand side of Eq.~(\ref{ssum1})
it can be seen that $T_{n-1,n}$ is given as,
\begin{eqnarray}
T_{n-1,n}
 = 
\sqrt{2ne{\mathcal B}}\left( \begin{array}{ccccccc}
0 & & 0 & & 0 & & 1\\
0 & & 0 & & 0 & & 0\\
0 & & -1 & & 0 & & 0 \\
0 & & 0 & & 0 & & 0
\end{array} \right)\,.
\end{eqnarray}
In the $2\times2$ notation the above equation looks like,
\begin{eqnarray}
T_{n-1,n} =
\sqrt{2ne{\mathcal B}}\left( \begin{array}{ccc}
0 & & \frac12(\sigma_1 + i\sigma_2)\\
-\frac12(\sigma_1 + i\sigma_2) & & 0
\end{array} \right)\,.
\end{eqnarray}
Here $\sigma_1$ and $\sigma_2$ are the first two Pauli matrices.
When converted back to the $4\times4$ notation the above equation
becomes,
\begin{eqnarray}
T_{n-1,n} =
-\frac12\sqrt{2ne{\mathcal B}}(\gamma_1 + i\gamma_2)\,.
\label{n3}
\end{eqnarray}

Similarly $T_{n,n-1}$ is given by,
\begin{eqnarray}
T_{n,n-1}
 = 
\sqrt{2ne{\mathcal B}}\left( \begin{array}{ccccccc}
0 & & 0 & & 0 & & 0\\
0 & & 0 & & 1 & & 0\\
0 & & 0 & & 0 & & 0 \\
-1 & & 0 & & 0 & & 0
\end{array} \right)\,.
\end{eqnarray}
In the $2\times2$ notation the above equation looks like,
\begin{eqnarray}
T_{n,n-1} =
\sqrt{2ne{\mathcal B}}\left( \begin{array}{ccc}
0 & & \frac12(\sigma_1 - i\sigma_2)\\
-\frac12(\sigma_1 - i\sigma_2) & & 0
\end{array} \right)\,,
\end{eqnarray}
which when converted back to the $4\times4$ notation becomes,
\begin{eqnarray}
T_{n-1,n} =
-\frac12\sqrt{2ne{\mathcal B}}(\gamma_1 - i\gamma_2)\,.
\label{n4}
\end{eqnarray}
Supplying the values of $T_{i,j}$s from Eq.~(\ref{n1}),
Eq.~(\ref{n2}), Eq.~(\ref{n3}) and Eq.~(\ref{n4}) to Eq.~(\ref{ssum1})
we get the result given in Eq.~(\ref{PU}). A similar procedure is
followed to obtain Eq.~(\ref{PV}).

\chapter{Phase factor of the Schwinger propagator}
\section{Loops of charged particles}\label{s:LCP}
As we have discussed previously in chapter 3, the phase
factor in the Schwinger propagator appears because the fermion
propagator attaches two points with different gauge transformation
properties. The Schwinger mechanism is best suited for processes where
we do not have external charged particles and all the charged
particles are inside the loop. For example one of the figures that
contribute to the neutrino self-energy and another one which
contributes for two neutrino three photon scattering as shown in
\fig{f:magmom} are perfect examples for these kind of diagrams.
One important fact that can be noticed from the form of the
propagators of charged fermions or gauge bosons~\cite{Erdas:1998uu} is
that the form of the phase factor\footnote{In the reference cited the
authors did not explicitly take $\lambda(\xi)$ as they were not
interested in the general properties of the phase factor.} of the
different propagators are the same as given in
Eq.~(\ref{compactIPF}). To understand its importance we take concrete
examples.

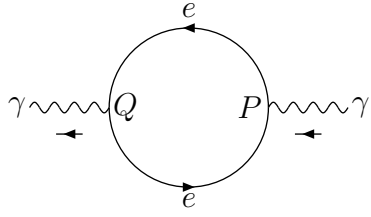
\begin{figure}[btp]
\begin{center}
\begin{picture}(180,120)(-90,-35)
\Photon(-60,0)(-30,0)24
\ArrowLine(-40,-10)(-50,-10)
\Text(-65,0)[c]{$\gamma$}
\ArrowArc(0,0)(30,0,180)
\Text(23,0)[c]{$P$}
\Text(0,37)[c]{$e$}
\ArrowArc(0,0)(30,180,360)
\Text(-24,0)[c]{$Q$}
\Text(0,-35)[c]{$e$}
\Photon(30,0)(60,0)24
\ArrowLine(50,-10)(40,-10)
\Text(65,0)[c]{$\gamma$}
\end{picture}
\caption[One loop two point function containing virtual charged
fermion internal lines.]{\sf One loop two point function containing
virtual charged fermion internal lines, Vacuum polarization diagram of
photon.
\label{f:vacpol}}
\end{center}
\end{figure}

First we take a one loop two point function containing virtual charged
fermion internal lines. The one loop photon vacuum polarization
diagram as shown in \fig{f:vacpol} is a good example. The electron
propagators connect points $P$ and $Q$ in space time.
If we are interested in finding out the overall phase factor
accompanying the vacuum polarization tensor
then we will have to use the Schwinger propagator for the
electrons. The contribution from the phase factors $\kappa(Q,P)$ and
$\kappa(P,Q)$ which we denote as $\Psi(P,Q)$ will be according to
Eq.~(\ref{PF}) and Eq.~(\ref{compactIPF})
\begin{eqnarray}
\Psi(P,Q) &=& \kappa(Q,P) \kappa(P,Q)\nonumber\\
           &=& \exp\left\{ ie\frac12 \left[P^\mu F_{\mu \nu} Q^\nu +
           Q^\mu F_{\mu \nu} P^\nu \right]\right\}\,,
\label{VP}
\end{eqnarray}
which reduces to unity because of the antisymmetry of $F_{\mu
\nu}$. From Eq.~(\ref{VP}) it is seen that the phase factor's
contribution in the one loop calculation is trivial and obviously gauge
invariant. In any kind of a loop where we have two charged particles
the phase factor contribution is unimportant as all the charged
particles have the same form of phase factor with the propagators. 

\begin{figure}[b]
\begin{center}
\begin{picture}(180,120)(-90,-35)
\ArrowLine(-60,0)(80,20)
\Text(-65,-5)[c]{$P$}
\ArrowLine(80,20)(10,90)
\Text(85,15)[c]{$Q$}
\ArrowLine(10,90)(-60,0)
\Text(15,95)[l]{$R$}
\end{picture}
\caption[A loop consisting of three space-time points joined by
charged particle propagators.]{\sf A loop consisting of three
space-time points joined by charged particle propagators. The arrow
heads indicate the direction of charge flow.
\label{f:3Ptloop}}
\end{center}
\end{figure}
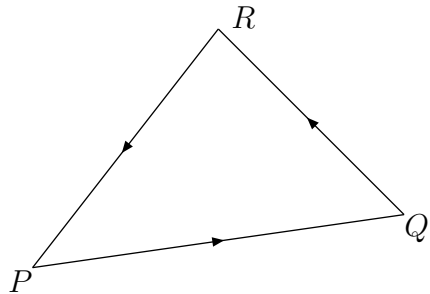

Next we take a loop with three charged particle propagators
connecting three space-time points $P,~Q,~R$. We have shown that
the phase factor between two points do not depend upon the path which
joins them. Consequently to find the contribution of the phase factors in
the calculation for the loop we join the three points by straight lines
as shown in \fig{f:3Ptloop}. The overall phase contribution
can then be calculated using 
Eq.~(\ref{PF}) and Eq.~(\ref{compactIPF}) and is given by
\begin{eqnarray}
\Psi(P,Q,R) &=& \kappa(Q,P) \kappa(R,Q) \kappa(P,R)\nonumber\\
          &=& \exp\left\{ ie\frac12 \left[P^\mu F_{\mu \nu} Q^\nu +
          Q^\mu F_{\mu \nu} R^\nu + R^\mu F_{\mu \nu} P^\nu\right]\right\}\,.
\label{TPF}
\end{eqnarray}
As all the phase factors are of the same form the first point to
notice is that the contribution from the function $\lambda(\xi)$
cancel out in the overall factor, showing that the contribution is
explicitly gauge invariant.

The next point which requires to be discussed is about the path
independence of the phase factor. From subsection \ref{ss:sp} we know
that $\kappa(Q,P)$ is independent of the path which joins them,
but here the  
path independence of $\kappa(Q,P)$ does not imply  
\begin{eqnarray}
\Psi(P,Q,R) = \kappa(Q,P) \kappa(R,Q) \kappa(P,R) = 1\,,
\end{eqnarray}
or from Eq.~(\ref{compactIPF})
\begin{eqnarray}
I(P,P) = I(Q,P) + I(R,Q) + I(P,R) = 0.
\label{AdIPF}
\end{eqnarray}
Instead of the above expectation we get a finite contribution from
Eq.~(\ref{TPF}). As $I(Q,P)$ consists of the product of the two end
points instead of their difference, the different phase factors from
the different paths connecting two intermediate points of the loop
when multiplied does not reduce to unity.

It can be shown in general that if we include loops where we have more
than three charged particle propagators joining more than three points
in space time then also the phase factor is going to contribute. To
understand where we can actually neglect the phase factor and where
not we have to calculate the flux of the uniform magnetic field
attached to the area of the loop.
\section{The flux rule}\label{s:FR}
The phase factor contribution to the one loop calculations of various
diagrams as shown in \fig{f:vacpol} and \fig{f:3Ptloop} show
that they are explicitly gauge invariant. In the case of
\fig{f:vacpol} it was shown that it contributes nothing for the
phase factor. The contribution from \fig{f:3Ptloop} can be understood
in another way also. As the magnetic field is directed along the
$z$-direction the component of area enclosed by the looping particles
along the $x-y$ plane is only active for its flux.  From the path
independence of $\kappa$ we know that $\Psi(P,Q,R)$ can always be
calculated from the triangular path connecting the three points as
shown in \fig{f:3Ptloop} and so the flux attached with this loop must
be the magnitude of the magnetic field ${\mathcal B}$, times the area
of this triangle in the $x-y$ plane denoted by
$\Delta_\perp(P,Q,R)$. Calculating the flux we get
\begin{eqnarray}
{\cal F}(P,Q,R) &=& {\mathcal B} \Delta_\perp(P,Q,R) \nonumber\\ 
                &=& \frac12 {\mathcal B} \left[(\vec{Q} - \vec{P}) \times 
                    (\vec{R} - \vec{Q})\right]_z\,,
\label{FR}
\end{eqnarray}
where $[\vec{M} \times \vec{N}]_z$ represents $z$-component of the
antisymmetric cross-product between two arbitrary 3-vectors
$\vec{M}$ and $\vec{N}$. Eq.~(\ref{FR})
can also be written as
\begin{eqnarray}
{\cal F}(P,Q,R) = \frac12 (Q - P)^\mu F_{\mu \nu} (R - Q)^\nu\,,
\label{RFR}
\end{eqnarray}

and in this form we see from Eq.~(\ref{TPF}) that it can be written as
\begin{eqnarray}
\Psi(P,Q,R) = \exp\left\{ie{\cal F}(P,Q,R)\right\}\,.
\label{RPF}
\end{eqnarray}

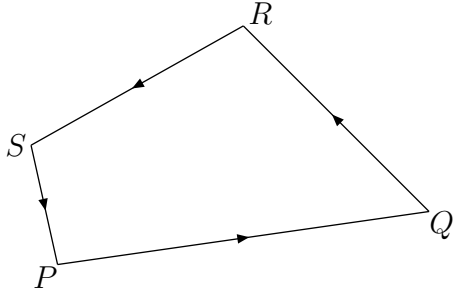
\begin{figure}[btp]
\begin{center}
\begin{picture}(180,120)(-90,-35)
\ArrowLine(-60,0)(80,20)
\Text(-65,-5)[c]{$P$}
\ArrowLine(80,20)(10,90)
\Text(85,15)[c]{$Q$}
\ArrowLine(10,90)(-70,45)
\Text(12,95)[l]{$R$}
\ArrowLine(-70,45)(-60,0)
\Text(-80,45)[l]{$S$}
\end{picture}
\caption[A loop consisting of four space-time points joined by
charged particle propagators.]{\sf A loop consisting of four
space-time points joined by charged particle propagators. The arrow
heads indicate the direction of charge flow.
\label{f:4Ptloop}}
\end{center}
\end{figure}

Next we consider a loop containing four charged particle
propagators. The four photon interaction in QED is a process where
this loop occurs naturally.  Here the the internal virtual particles
may be electrons. In coordinate space the four points where the
photons attach to the loop may be taken as $P$, $Q$, $R$ and $S$. As
far as the contribution from the overall phase is concerned we can
join them by straight lines forming a general quadrilateral as shown
in \fig{f:4Ptloop}.
Using the same technique for finding the overall phase as discussed in 
section \ref{s:LCP} we find here 
\begin{eqnarray}
\Psi(P,Q,R,S) &=& \kappa(Q,P) \kappa(R,Q) \kappa(S,R) \kappa(P,S)\nonumber\\
              &=& \exp\left\{ ie\frac12 \left[P^\mu F_{\mu \nu} Q^\nu 
                  + Q^\mu F_{\mu \nu} R^\nu + R^\mu F_{\mu \nu} S^\nu
                  + S^\mu F_{\mu \nu} P^\nu\right]\right\}\,,
\end{eqnarray}
which can also be written as
\begin{eqnarray}
\Psi(P,Q,R,S) = \exp\left\{ie{\cal F}(P,Q,R,S)\right\}\,,
\label{4F}
\end{eqnarray}
where
\begin{eqnarray}
{\cal F}(P,Q,R,S) = \frac12 \left[(Q - P)^\mu F_{\mu \nu} (R - Q)^\nu 
                    + (S - R)^\mu F_{\mu \nu} (P - S)^\nu \right]\,.
\label{4FE}
\end{eqnarray}
Working similarly as was done in the case of Eq.~(\ref{RFR}) it can be
shown that the above flux is just the sum of the areas of the
triangles $PQR$ and $SRP$ along the $x-y$ plane times the magnitude of
the magnetic field.

From Eq.~(\ref{RPF}) and  Eq.~(\ref{4F}) we can generalize that the
overall phase depends on the flux of the magnetic field attached with
the area of the loop. Given an arbitrary diagram we can initially
calculate this flux to find out the overall phase. Also this
connection with magnetic flux explains another important fact that
this phase is absent for a loop consisting of just two points. If we
join this two points with straight lines then in no way will those
straight lines enclose any area and as a result no resultant flux will
emerge.

The Schwinger propagator as given in Eq.~(\ref{schwingprop}) is
specific for the case of those gauges which gives a background
magnetic field. In the general derivation by
Schwinger~\cite{Schwinger:nm} the form of the fermionic propagator was
derived for general constant gauge fields. If we choose a pure gauge
field which does not give rise to any electric or magnetic field as
\begin{eqnarray}
A_\mu(\xi) = K \delta_\mu^{\,\,\,\alpha} \xi_\alpha\,,
\label{PG}
\end{eqnarray}
where $K$ is a constant, then also we can find the Schwinger
propagator and it differs from the vacuum propagator only by the phase
factor. Denoting this propagator in presence of a pure gauge as given in 
Eq.~(\ref{PG}) as $iS_A(x,x')$ we have
\begin{eqnarray}
iS_A(x,x') = \exp (ie I(x,x'))
             \int \frac{d^4 p}{ 2 \pi^4} \frac{i(\rlap/p + m)}{(p^2 - m^2)}\,,
\label{PGP}
\end{eqnarray}
where $I(x,x')$ is given by Eq.~(\ref{AI}). From generalized Stokes
theorem, which is  
\begin{eqnarray}
\oint d\xi^\mu A_\mu(\xi) = \int d\sigma^{\mu \nu} F_{\mu \nu}(\xi)\,,
\end{eqnarray}
where $d\sigma^{\mu \nu}$ is the infinitesimal surface area in the
$\xi_\mu - \xi_\nu$ plane and $F_{\mu \nu}$ is the field strength
tensor, we can say that $I(x,x')$ is path independent as here $F_{\mu
\nu} = 0$.  The path independence of $I(x,x')$ allows us to join the
two points by a straight line and proceeding similarly as done in
chapter \ref{chap3} while discussing about the phase factor, we get
\begin{eqnarray}
I(x,x') &=& \int_{x'}^x d\xi^\mu A_\mu(\xi) \nonumber\\
        &=& K(x^\alpha x_\alpha - x'^\alpha x'_\alpha)\,.
\label{PGI}
\end{eqnarray}
The above result implies that the overall phase factor for diagrams
like those in \fig{f:3Ptloop} and \fig{f:4Ptloop} will
always be unity. As a result rates of those processes will remain
similar to the vacuum results.

\chapter{A short note on `Real Time Formulation'}
\label{rtf}
\section{A brief introduction to thermal propagators} 
There are more than one way in which we can proceed in real time
formalism, out of them here we take the canonical
approach\cite{Nieves:1990ne} as is prevalent in the zero temperature
case. The main purpose of this section is to give a glimpse of the
real time formalism, to show how a genuinely temperature dependent term
and a temperature independent term appears in the propagators. Also
some of the relevant formulas will be cited in this section.

In this section, the essence of the real time formulation is presented
in a heuristic manner using the real scalar fields. Suppose that we
have a Lagrangian,
\begin{eqnarray}
{\mathscr L}=\frac{1}{2}\partial_\mu\phi\partial^\mu\phi-\frac{m^2}{2}\phi^2\,,
\end{eqnarray}
which describes a real scalar field. The Fourier mode expansion of the
free field is given as,
\begin{eqnarray}
\phi(x)=\int\frac{d^3p}{\sqrt{(2\pi)^32E_p}}\left[a(p)e^{-ip\cdot x}
+a^{\dag}(p) e^{ip\cdot x}\right]\,,
\label{fffe}      
\end{eqnarray}
where the $a(p)$ and $a^{\dag}(p)$ are the annihilation and the
creation operators of the bosons with momentum $p$. Formally the
propagator is represented as
\begin{eqnarray}
i\Delta_F(x-y)=\left<0\mid T\phi(x)\phi(y)\mid0\right>\,,
\label{c.1}
\end{eqnarray}
where $T$ is the time ordered product defined as
\begin{eqnarray}
T\phi(x)\phi(y)=\Theta(x_0-y_0)\phi(x)\phi(y) +
\Theta(y_0-x_0)\phi(y)\phi(x)\,.
\end{eqnarray}
For obtaining the expression of the propagator from Eq.~(\ref{c.1}) we
use the fact that $\left<0\mid a^{\dag} (p)a(p)\mid 0\right>=0$,
always due to the basic definition of the operators
$a(p)$ and $a^{\dag}(p)$, and get
\begin{eqnarray}
\Delta_F(x-y)=\int\frac{d^4p}{(2\pi)^4}\frac{e^{-ip\cdot(x-y)}}
{p^2-m^2+i\epsilon}\,.
\end{eqnarray}
In a medium there are real particles and so the expectation value of
the number operator $a^{\dag}a$ is not zero
$\left<a^{\dag}(p)a(p)\right>_\beta\neq 0$.  So now we take
\begin{eqnarray}
\left<a^{\dag}(p)a(p')\right>_\beta=\delta^3(\vec{p}-\vec{p'})f_B(p)\,, 
\end{eqnarray}
where 
\begin{eqnarray}
f_B(p)=\frac{1}{e^{\beta p\cdot u}-1}\,,\nonumber
\end{eqnarray}
is the Bose-Einstein distribution function. Here $u$ is the four-velocity
of the centre of mass of the heat bath. By the commutation relations
naturally we have
\begin{eqnarray}
\left<a(p)a^{\dag}(p')\right>_\beta=\delta^3(\vec{p}-\vec{p'})
\left[f_B(p)+1\right]\,.
\end{eqnarray}
In a thermal medium the following relations
\begin{eqnarray}
\left<a^{\dag}(p)a^{\dag}(p')\right>_\beta =0\;\mbox{and}\;
\left<a(p)a(p')\right>_\beta =0\,, \nonumber 
\end{eqnarray}
remain the same as those in vacuum.  Utilizing these facts and doing
the calculation as is done in the zero temperature case we find the
expression of the thermal propagator as
\begin{eqnarray}
\Delta_F(p)=\frac{1}{p^2-m^2+i\epsilon}-2\pi i\delta(p^2-m^2)
\frac{1}{e^{\beta p\cdot u}-1}\,,
\label{c.2}
\end{eqnarray}
which has a genuine temperature dependent term as the second term in
the right hand side. Here the temperature dependent term comes in with
a delta-function showing this propagator corresponds to exchange of
real particles unlike the exchange of virtual particles present in
the first term. The expression becomes realistic because the thermal
bath always has some real particles, so in any process these particles
can be exchanged. Eq.~(\ref{c.2}) is really not the whole story, there
are other propagators also accompanying it. In the next section some
technical details is given which will elucidate the complex structure
of this formalism.
\section{A bit of formal theory}
\subsection{The n-point Green's function in vacuum}
The Fourier expansion of the scalar field is given in
Eq.~(\ref{fffe}). When we are talking about interacting field theory
the free field expansion as given in Eq.~(\ref{fffe}) is often called
$\phi_{\mbox{in}}(x)$. To do interacting field theory we can always find
an operator $U$ which transforms the incoming free fields
$\phi_{in}$ to the interacting Heisenberg fields $\phi(x)$, and
the transformation is given by
\begin{eqnarray}
\phi(x)=U^{-1}\phi_{in}(x)\,U,\hskip 1cm t=x^0.
\label{ff}
\end{eqnarray}
For the above transformation to hold, we must have 
\begin{eqnarray}
U(t)=T \,\mbox{exp}\left[i \int^t_{-\infty} \int d^3 x
\,{\mathscr L}_{int}[\phi_{in}(x)]\right]\,,
\label{evo}
\end{eqnarray}
which is the evolution operator. It must satisfy the following property,
\begin{eqnarray}
\lim_{t\rightarrow -\infty}U(t) = 1.
\label{evo1}
\end{eqnarray}
Generally an n-point Green's function in the interacting theory is defined as
\begin{eqnarray}
G(x_1,\cdot,\cdot,\cdot,x_n)=\langle0|T\{\phi(x_1) \phi(x_2)\cdot \cdot
\cdot \phi(x_n)\}|0\rangle.
\label{green}
\end{eqnarray}
Introducing a very large time $\tau$ such that
\begin{eqnarray}
-\tau < t_1,t_2,\cdot,\cdot,\cdot,t_n < \tau
\end{eqnarray}
and choosing the convention 
\begin{eqnarray}
U(\tau,-\tau)=U(\tau) U^{-1}(-\tau)\,,
\end{eqnarray}
we can manipulate the expression inside the time ordering. Using
Eq.~(\ref{ff}), 
\begin{eqnarray}
& &T\{\phi(x_1) \phi(x_2)\cdot\cdot\cdot \phi(x_n)\} \nonumber\\
&=& T\{U^{-1}(t_1)\phi_{in}(x_1)\,U(t_1)
U^{-1}(t_2)\phi_{in}(x_2) U(t_2)\cdot \cdot\cdot
U^{-1}(t_n)\phi(x_n) U(t_n)\}\nonumber\\
 &=& U^{-1}(\tau)\left[ T\, U(\tau)U^{-1}(t_1)
\phi_{in}(x_1)\,U(t_1)\cdot \cdot \cdot U^{-1}(t_n)
\phi_{in}(x_n) U(t_n) U^{-1}(-\tau)\right]U(-\tau)\nonumber\\
 &=& U^{-1}(\tau)\left[ T\,\{\phi_{in}(x_1)
\phi_{in}(x_2)\cdot \cdot \cdot \phi_{in}(x_n)\} U^{-1}(\tau,-\tau)\right]U(-\tau).
\label{tord}
\end{eqnarray}
Using the above equation we can write
\begin{eqnarray}
G(x_1,\cdot,\cdot,\cdot,x_n)=\langle0|U^{-1}(\tau)\left[T\{\phi_{in}(x_1)\cdot
\cdot \cdot \phi_{in}(x_n)\,U(\tau,-\tau)\}\right]U(-\tau)|0\rangle.
\label{green1}
\end{eqnarray}
In vacuum
\begin{eqnarray}
\lim_{\tau \to \infty}U(\tau)|0\rangle  &=& e^{i\delta}|0\rangle\,,
\nonumber\\
\lim_{\tau \to \infty}U(-\tau)|0\rangle &=& |0\rangle\,,\nonumber
\label{vphase}
\end{eqnarray}
where $\delta$ is some real number. To take account of the phase the
Green's function in vacuum is defined as
\begin{eqnarray}
G^{(0)}(x_1,\cdot,\cdot,\cdot,x_n)=\frac{\langle0|
\left[T\{\phi_{in}(x_1)\cdot 
\cdot \cdot \phi_{in}(x_n)\,U(\infty)\}\right]
|0\rangle}{\langle0|U(\infty)|0\rangle}\,,
\label{green2}
\end{eqnarray}
where
\begin{eqnarray}
U(\infty)=U(\tau,-\tau)\,,
\label{infu}
\end{eqnarray}
when $\tau$ tends to infinity.
The perturbation expansion is obtained by expanding $U(\infty)$ 
and reducing the products of field operators through Wick's theorem.
\subsection{Interacting Scalar Fields in Presence of a Thermal Bath.}
When there is a thermal bath present, then
\begin{eqnarray}
\lim_{\tau \to \infty}U(-\tau)|0,\beta\rangle &=& |0,\beta\rangle\,,\nonumber
\end{eqnarray}
(where $|0,\beta\rangle$ is the thermal vacuum)but 
\begin{eqnarray}
\lim_{\tau \to \infty}U(\tau)|0,\beta\rangle  &\neq& 
e^{i\delta}|0,\beta\rangle\,,\nonumber
\end{eqnarray}
as there are real particles inside the thermal bath (thermal vacuum),
they can interact and the state may change physically. As a result
$\langle 0,\beta | U^{-1}(\tau)$ cannot be just replaced by
$e^{-i\delta}\langle 0,\beta |$ as is done in Eq.~(\ref{green2}). Here
\begin{eqnarray}
G(x_1,\cdot,\cdot,\cdot,x_n)=\langle
U^{-1}(\infty)\left[T\{\phi_{in}(x_1)\cdot  
\cdot \cdot \phi_{in}(x_n)\,U(\infty)\}\right] \rangle.
\label{thgreen}
\end{eqnarray}

The expectation value of some operator $O$ is defined as
\begin{eqnarray}
\langle O \rangle = \frac{\Tr Z \,
O}{\Tr\, Z}\,,
\label{expt}
\end{eqnarray}
where $\Tr$ indicates a trace has to be taken (with respect to any
basis) and $Z$ is the partition function given by
\begin{eqnarray}
Z=\mbox{exp} \left[-\beta H + \sum_A \alpha_A Q_A \right]\,,
\label{partition}
\end{eqnarray}
where $Q_A$ are the conserved charges that commute with the Lagrangian
and $\alpha_A$ are the chemical potentials that parameterize the
composition of the medium. Here we are working in presence of a
thermal bath, and it can have a velocity. For the next part of the
discussions we will work in the rest frame of the heat bath.

The perturbation expansion is obtained by expanding the formal formula
for $U(\infty)$, but in this case we have to take into account the
factor $U^{-1}(\infty)$, which is given by
\begin{eqnarray}
U^{-1}(\infty)=\bar{T}\mbox{exp} \left[ -i \int d^4 x
{\mathscr L}_{int}[\phi_{in}(x)]\right]\,,  
\label{uinv}
\end{eqnarray}
and expand it also. $\bar{T}$ indicates anti-time ordered product.
\subsection{Calculation of the Thermal Green's Function.}
Writing ${\mathscr L}_{int}[\phi_{in}(x)]$ as ${\mathscr L}_{int}(x)$
for brevity, the whole expression of the Green's function can be
written as,
\begin{eqnarray}
G(x_1,\cdot,\cdot,\cdot,x_n) &=& \sum_{s=0}^{\infty}\frac{(-i)^s}{s!}
\sum_{p=0}^{\infty} \frac{(i)^p}{p!}\langle \bar{T}\,\left[\int d^4 z_1
\int d^4 z_2 \cdot \cdot \int d^4 z_s {\mathscr L}_{int}(z_1)
{\mathscr L}_{int}(z_2)\cdot \cdot
{\mathscr L}_{int}(z_s)\right]\nonumber\\
&\times& T \,\left[\phi_{in}(x_1) \cdot \cdot
\phi_{in}(x_n) \int d^4 y_1 \cdot \cdot d^4 y_p
{\mathscr L}_{int}(y_1)\cdot \cdot {\mathscr L}_{int}(y_p)\right]\rangle.
\label{thgreen1}
\end{eqnarray}
Concentrating on $\phi^4$ theory where the interaction term is given
by
\begin{eqnarray}
{\mathscr L}_{int}(x) =-\frac{\lambda}{4!}\,\phi^4_{in}(x)\,,\nonumber
\end{eqnarray}
we can write the Green's function as 
\begin{eqnarray}
G(x_1,\cdot,\cdot,\cdot,x_n)&=& \sum_{s=0}^{\infty}\frac{(i\lambda)^s}{s!}
\sum_{p=0}^{\infty} \frac{(-i\lambda)^p}{p!}\left\langle
\bar{T}\,\left[\int d^4 z_1 
\int d^4 z_2 \cdot \cdot \int d^4 z_s
\frac{\phi^4_{in}(z_1)}{4!} \cdot
\cdot\frac{\phi^4_{in}(z_s)}{4!}\right]\right.\nonumber\\
&\times& \left. T \,\left[\phi_{in}(x_1) \cdot \cdot
\phi_{in}(x_n) \int d^4 y_1 \cdot \cdot d^4 y_p
\frac{\phi^4_{in}(y_1)}{4!}\cdot \cdot
\frac{\phi^4_{in}(y_p)}{4!}\right]\right\rangle \nonumber\\
&=&\sum_{s,p=0}^{\infty} \frac{(i\lambda)^s(-i\lambda)^p}{s! p!} \int
d^4 z_1 \cdot \cdot \cdot \int d^4 z_s \int d^4 y_1 \cdot \cdot \cdot
\int d^4 y_p\nonumber\\
&\times& \left\langle\left[\bar{T}\{\frac{\phi^4_{in}(z_1)}{4!}
\cdot \cdot \cdot
\frac{\phi^4_{in}(z_s)}{4!}\}\,T\{\phi_{in}(x_1)\cdot \cdot \cdot
\phi_{in}(x_n)\frac{\phi^4_{in}(y_1)}{4!} \cdot \cdot \cdot
\frac{\phi^4_{in}(y_p)}{4!} \} \right] \right\rangle\,.\nonumber\\ 
\label{thgreen3}
\end{eqnarray}
This expansion shows that the general expectation value term will be
of the form 
\begin{eqnarray}
\langle \bar{T} (\bar{A}\,\bar{B}\cdot \cdot \cdot \bar{F})\,T (A\,B \cdot
\cdot \cdot F)\rangle
\label{wick}
\end{eqnarray}
where the bars over $A,B,...$ indicate they are operators under the
$\bar{T}$ sign, or they are anti-time ordered operators. The objects
$A,B,..$ are the fields that appear in the last line of the
Eq.~(\ref{thgreen3}). The vertices that come under $T$ are called
the type-1 vertices, and those that come under $\bar{T}$ are called
the type-2 vertices. A propagator which joins two vertices can now be
of four kinds, 
\begin{itemize}
\item 1-1 type, which joins two vertices under the $T$ sign.
\item 2-2 type, which joins two vertices under the $\bar{T}$ sign.
\item 1-2 type, which joins a type-1 vertex to a type-2 vertex.
\item 2-1 type, which joins a type-2 vertex to a type-1 vertex.
\end{itemize}
Another important thing to be noticed is, any external vertex where
real particles come in or go out must be of type-1 variety, as those
fields $\phi(x)$s are under the $T$ operation in
Eq.~(\ref{thgreen3}). If now we calculate the two point function
where both the fields are of type-1 by simplifying the
Eq.~(\ref{wick}) properly then we will arrive at Eq.~(\ref{c.2}). But
there can be other options as well, where the fields are both of type-2,
or are mixed. Not going into further details here the other
propagators are supplied for the scalar theory.  $\Delta_{F 1 1}(k)$
is given by Eq.~(\ref{c.2}), the others are
\begin{eqnarray}
\Delta_{F 2 2}(p)&=& \frac{-1}{p^2-m^2+i\varepsilon}-2\pi i\delta(p^2-m^2)
\eta(p\cdot u)\,,\\
\Delta_{F 2 1}(p)&=& -2\pi i\delta(p^2-m^2)[\eta(p\cdot u) +
\theta(-p\cdot u)]\,,\\
\Delta_{F 1 2}(p)&=& -2\pi i\delta(p^2-m^2)[\eta(p\cdot u) +
\theta(p\cdot u)]\,,
\end{eqnarray}
where 
\begin{eqnarray}
\eta(p\cdot u) = \theta(p\cdot u)n_B(x) + \theta(-p\cdot u)n_B(-x)\,,
\end{eqnarray}
with
\begin{eqnarray}
n_B(x) = \frac{1}{e^x - 1}\,,
\end{eqnarray}
and $x$ is defined as,
\begin{eqnarray}
x = \beta p\cdot u\,.
\end{eqnarray}

For fermions the procedure is same, except here we have
anti-commutation relations in place of commutation relations, and the
expressions for the four propagators are,
\begin{eqnarray}
S_{F 1 1}(p)&=&(\rlap/p + m)\left[\frac{1}{p^2 - m^2 + i\varepsilon} 
+ 2\pi i\delta(p^2 - m^2)\eta_F(p\cdot u) \right]\,,\\
S_{F 2 2}(p)&=&(\rlap/p + m)\left[\frac{-1}{p^2 - m^2 + i\varepsilon} 
+ 2\pi i\delta(p^2 - m^2)\eta_F(p\cdot u) \right]\,,\\
S_{F 1 2}(p)&=&(\rlap/p + m) 2\pi i\delta(p^2 - m^2)
[\eta_F(p\cdot u) - \theta(-p\cdot u)]\,,\\
S_{F 2 1}(p)&=&(\rlap/p + m) 2\pi i\delta(p^2 - m^2)
[\eta_F(p\cdot u) - \theta(p\cdot u)]\,.
\end{eqnarray}
In the case of fermions $\eta_F(p\cdot u)$ is given by,
\begin{eqnarray}
\eta_F(p\cdot u) = \Theta(p\cdot u) f_F(p,\mu,\beta) 
+ \Theta(-p\cdot u) f_F(-p,-\mu,\beta) \,,
\end{eqnarray}
where,
\begin{eqnarray}
f_F(p,\mu,\beta) = {1\over e^{\beta(p\cdot u - \mu)} + 1} \,,
\end{eqnarray}
is the Fermi-Dirac distribution function.

With this brief exposition of statistical field theory in the real
time formalism we understand why instead of one propagator we have
four of them. But usually most of the time we only work with the 1-1
component as is done in this thesis. But formal calculations in
statistical field theory requires all four propagators as specified
above.

\chapter{Manipulations under the integral sign}
\label{muis}
All the results appearing in this appendix have been derived
previously in \cite{Ganguly:1999ts}. They are explicitly written down here
because the following techniques are very useful in manipulating the
integrand inside the integral sign in an equation like
Eq.~(\ref{compl}). If we concentrate on the rest frame of the medium,
then $p\cdot u=p_0$. Thus, the distribution function does not depend
on the spatial components of $p$.  From the form of Eq.~(\ref{reven})
and Eq.~(\ref{rodd}) we find that, the integral
over the transverse components of $p$ has the following generic
structure:
\begin{eqnarray}
\int d^2 p_\perp \; e^{\Phi(p,s)} e^{\Phi(p',s')} \times
\mbox{($p^{\beta_\perp}$ or $p'^{\beta_\perp}$)} \,.
\end{eqnarray}
Notice now that
\begin{eqnarray}
{\partial \over \partial p_{\beta_\perp}} 
\Big[ \; e^{\Phi(p,s)} e^{\Phi(p',s')} \Big] = 
{2i\over e{\mathcal B}} \Big( \tan (e{\mathcal B}s) \; p^{\beta_\perp} + \tan
(e{\mathcal B}s') \; p'^{\beta_\perp} \Big)
e^{\Phi(p,s)} e^{\Phi(p',s')} \,.
\label{single_derivative}
\end{eqnarray}
However, this expression, being a total derivative, should integrate
to zero. Thus we obtain
\begin{eqnarray}
\int d^2 p_\perp \; \tan(e{\mathcal B}s)p^{\beta_\perp}
e^{\Phi(p,s)} e^{\Phi(p',s')} = 
- \int d^2 p_\perp \; \tan(e{\mathcal B}s') p'^{\beta_\perp}
e^{\Phi(p,s)} e^{\Phi(p',s')}\,.
\end{eqnarray}
This above equation is symbolically written as,
\begin{eqnarray}
\tan (e{\mathcal B}s) \; p^{\beta_\perp} \stackrel\circ= - \tan
(e{\mathcal B}s') \; p'^{\beta_\perp} \,,
\end{eqnarray}
where the sign `$\stackrel\circ=$' means that the expressions on both
sides of it, though not necessarily equal algebraically, yield the
same integral. This gives
\begin{eqnarray}
p^{\beta_\perp} &\stackrel\circ=& - \, {\tan (e{\mathcal B}s') \over \tan
(e{\mathcal B}s) + \tan (e{\mathcal B}s')} \; k^{\beta_\perp}\,,
\label{pperpint}\\
p'^{\beta_\perp} &\stackrel\circ=&  {\tan (e{\mathcal B}s) \over \tan
(e{\mathcal B}s) + \tan (e{\mathcal B}s')} \; k^{\beta_\perp} \,.
\label{primeperpint}
\end{eqnarray}
Similarly we can derive some other relations which can be used under
the momentum integral signs. To write them in a useful form, we turn
to Eq.\ (\ref{single_derivative}) and take another derivative with
respect to $p^{\alpha_\perp}$. From the fact that this derivative
should also vanish on $p$ integration, we find
\begin{eqnarray}
p_\perp^\alpha p_\perp^\beta \stackrel\circ= {1\over \tan (e{\cal
B}s) + \tan (e{\mathcal B}s')} \Bigg[{ie{\mathcal B} \over 2}
g_\perp^{\alpha\beta} +  {\tan^2 (e{\mathcal B}s') \over \tan (e{\cal
B}s) + \tan (e{\mathcal B}s')} \;
k_\perp^\alpha k_\perp^\beta \Bigg] \,.
\label{perppapb}
\end{eqnarray}
In particular, then,
\begin{eqnarray}
p_\perp^2 \stackrel\circ= {1\over \tan (e{\cal
B}s) + \tan (e{\mathcal B}s')} \Bigg[ -ie{\mathcal B} +
{\tan^2 (e{\mathcal B}s') \over \tan (e{\mathcal B}s) + \tan (e{\mathcal B}s')} \;
k_\perp^2 \Bigg] \,.
\label{psq}
\end{eqnarray}
It then simply follows that
\begin{eqnarray}
p_\perp^{\prime2} \stackrel\circ= {1\over \tan (e{\cal
B}s) + \tan (e{\mathcal B}s')} \Bigg[ -ie{\mathcal B} +
{\tan^2 (e{\mathcal B}s) \over \tan (e{\mathcal B}s) + \tan
(e{\mathcal B}s')} \; 
k_\perp^2 \Bigg] \,.
\label{p'sq}
\end{eqnarray}
And finally using the definition of the exponential factor in
Eq.~(\ref{Phi}) we can write
\begin{eqnarray}
m^2 &\stackrel{\circ}{=}&\left(i{d\over
ds} + (p^2_\para -\sec^2(e{\mathcal B}s) p^2_\pr)\right)\,.
\label{msqr}
\end{eqnarray}
Here the `$\stackrel\circ=$' means that both sides of the above
equation, when multiplied by $e^{\Phi(p,s)} e^{\Phi(p',s')}$,
are equivalent inside the momentum integrals.

This set of relations help us considerably to handle the terms present
inside the integrals appearing in Eq.~(\ref{rodd}) and
Eq.~(\ref{reven}).

\chapter{A list of relevant integrals}
\label{relint}
These kinds of integrals often crop up when we are calculating with the
Schwinger propagator.  
%
\begin{eqnarray}
\int_{-\infty}^{\infty} \frac{dx}{2\pi} e^{iax^2}
&=& \frac{e^{{i\pi}/{4}}}{2\sqrt{\pi a}}\,,\\
\int_{-\infty}^{\infty} \frac{dx}{2\pi} e^{-iax^2}
&=& \frac{e^{{-i\pi}/{4}}}{2\sqrt{\pi a}}\,.
\end{eqnarray}
Here $a$ is a real number. Obviously the second integral can be
obtained from complex conjugating the first one. 

Next we consider integrals containing vector indices.
%
\begin{eqnarray}
\int \frac{d^4 p}{(2\pi)^4} e^{-ia(p - B)^2}&=&
\frac{i}{16\pi^2 a^2}\,,\\
\int \frac{d^4 p}{(2\pi)^4} p_\mu e^{-ia(p - B)^2}&=&
\frac{i}{16\pi^2 a^2} B_\mu\,,\\
\end{eqnarray}
Here, as before  $a$ is a real number, $p$ and $B$ are 4-vectors.
For each one of the above integrals there exists one corresponding
complex conjugate integral. The set of complex conjugated integrals
is not written here.

Finally comes those integrals which involve the parallel and
perpendicular components of the vectors.
%
\begin{eqnarray}
\int \frac{d^2 p_\para}{(2\pi)^2} e^{-ia(p - 
B)^2_\para}&=&\frac{1}{4 \pi a}\,,\\
\int \frac{d^2 p_\pr}{(2\pi)^2} e^{-ia(p - B)^2_\pr}
&=&-\frac{i}{4 \pi a}\,,\\
\int \frac{d^2 p_\para}{(2\pi)^2} p_{\mu_\para}e^{-ia(p - 
B)^2_\para}&=&\frac{1}{4 \pi a} B_{\mu_\para}\,,\\
\int \frac{d^2 p_\pr}{(2\pi)^2} p_{\mu_\pr} e^{-ia(p - 
B)^2_\pr}&=&-\frac{i}{4 \pi a} B_{\mu_\pr}\,,\\
\end{eqnarray}
Here also we have one corresponding complex conjugated integral for
each one of the above integrals which are not written.

\chapter{Some calculations in a magnetized medium}
\label{somecal}
\section{Derivation of $\mbox{R}^{(e)}_{\mu \nu}$ and 
$\mbox{R}^{(o)}_{\mu \nu}$}
In this appendix those calculations are done which were side stepped
in chapter \ref{chap4}. First it is shown how we get Eq.~(\ref{reven1}) from
Eq.~(\ref{reven}) and how Eq.~(\ref{crmunu}) is derived.  From
Eq.~(\ref{reven}) it is clear that if we apply Eq.~(\ref{perppapb})
and then Eq.~(\ref{pperpint}) to the last term in the right hand side
then it drops out and we get Eq.~(\ref{reven1}).

An expression of $\mbox{R}^{(o)}_{\mu\nu}$ is given in Eq.~(\ref{rodd}),
leaving it aside we derive Eq.~(\ref{crmunu}) in a different way.
Now from Eq.~(\ref{compl}) it is known that in the rest frame of the medium,
\begin{eqnarray}
\mbox{R}_{\mu\nu}(p,p',s,s') = \Tr[\gamma_\mu \gamma_5 G(p,s)
\gamma_\nu G(p',s')] \eta_F(p_0) + \Tr[\gamma_\mu \gamma_5 G(-p',s')
\gamma_\nu G(-p,s)] \eta_F(-p_0)\,.\nonumber\\
\label{firstrmunu}
\end{eqnarray}
$\mbox{R}_{\mu\nu}(p,p',s,s')$ contains the sum of two traces
multiplied by the functions $\eta_F(\pm p_0)$. Except the $\eta$
functions, the two traces are equal, so for the part which is odd in
orders of the external field, Eq.~(\ref{firstrmunu}) is equivalent to,
\begin{eqnarray}
\mbox{R}^{(o)}_{\mu\nu} = \eta_+(p_0)\Tr[T^{(o)}_{\mu \nu}]\,
\label{deftmunu}
\end{eqnarray}
where for brevity the arguments of $\mbox{R}^{(o)}_{\mu\nu}$ and
$T^{(o)}_{\mu \nu}$ are suppressed. Here $\eta_+(p_0) = \eta(p_0) +
\eta(-p_0)$ and $T^{(o)}_{\mu \nu}$ is given by,
\begin{eqnarray}
T^{(o)}_{\mu \nu}&=& i\tan(e \Be s) \left[ m^2 \gamma_\mu \gamma_5
\sigma_z \gamma_\nu + \gamma_\mu \gamma_5 \sigma_z \rlap/p_\para
\gamma_\nu \rlap/p'_\para + \gamma_\mu \gamma_5 \sigma_z \rlap/p_\para
\gamma_\nu \rlap/p'_\pr \sec^2(e \Be s')\right]\nonumber\\
&+& i\tan(e \Be s')\left[m^2 \gamma_\mu \gamma_5 \gamma_\nu \sigma_z + 
\gamma_\mu \gamma_5 \rlap/p_\para \gamma_\nu \sigma_z \rlap/p'_\para 
+ \gamma_\mu \gamma_5 \rlap/p_\pr \gamma_\nu \sigma_z \rlap/p'_\para 
\sec^2(e\Be s)\right]\,.\nonumber\\ 
\label{to1}
\end{eqnarray}
Writing out the terms in $T^{(o)}_{\mu \nu}$, we get
\begin{eqnarray}
\Tr[\gamma_\mu \gamma_5 \sigma_z \rlap/p_\para \gamma_\nu \rlap/p'_\para] &=&
\Tr \left[2 p'_{\nu_\para}\gamma_\mu \gamma_5 \sigma_z \rlap/p_\para -
p^2_\para \gamma_\nu \gamma_\mu \gamma_5 \sigma_z  - (p\cdot k)_\para 
\gamma_\nu \gamma_\mu \gamma_5 \sigma_z \right.\nonumber\\
&+& \left.(p^0 k^3 - p^3 k^0)\gamma_\nu \gamma_\mu\right]\,,
\label{1st term}
\end{eqnarray}
%
%
\begin{eqnarray}
\Tr[\gamma_\mu \gamma_5 \rlap/p_\para \gamma_\nu \sigma_z \rlap/p'_\para] &=&
\Tr\left[2 p_{\nu_\para} \gamma_\mu \gamma_5 \sigma_z \rlap/p'_\para + 
p^2_\para 
\gamma_\mu \gamma_\nu \gamma_5 \sigma_z + (p \cdot k)_\para \gamma_\mu
\gamma_\nu  
\gamma_5 \sigma_z\right.\nonumber\\
&-& \left.(p^0 k^3 - p^3 k^0) \gamma_\mu \gamma_\nu\right]\,.
\label{2nd term}
\end{eqnarray}
Using Eq.~(\ref{1st term}) and Eq.~(\ref{2nd term}) we can write
Eq.~(\ref{to1}) as, 
\begin{eqnarray}
\Tr[T^{(o)}_{\mu \nu}] &=& i\tan(e\Be s)\Tr\left[\{m^2 - p^2_\para - (p\cdot k)_\para\}
\gamma_\nu\gamma_\mu \gamma_5 \sigma_z + (p^0 k^3 - p^3 k^0)\gamma_\nu \gamma_
\mu\right.\nonumber\\
&+&\left. 2 p'_{\nu_\para} \gamma_\mu \gamma_5 \sigma_z \rlap/p_\para 
+ \gamma_\mu \gamma_5 \sigma_z \rlap/p_\para \gamma_\nu \rlap/p'_\pr 
\sec^2(e\Be s')\right]\nonumber\\
&+& i\tan(e\Be s')\Tr\left[-\{m^2 - p^2_\para - (p\cdot k)_\para\}\gamma_\mu
\gamma_\nu \gamma_5 \sigma_z - (p^0 k^3 - p^3 k^0)\gamma_\mu \gamma_\nu
\right.\nonumber\\
&+&\left. 2 p_{\nu_\para} \gamma_\mu \gamma_5 \sigma_z \rlap/p'_\para 
+ \gamma_\mu \gamma_5 \rlap/p_\pr \gamma_\nu \sigma_z \rlap/p'_\para 
\sec^2(e\Be s)\right]\,.
\label{to2}
\end{eqnarray}
As we are interested in taking the trace of $T^{(o)}_{\mu \nu}$, so
now we write down the traces of the components of $T^{(o)}_{\mu
\nu}$. The easier ones comes first, and they are
\begin{eqnarray}
\Tr[\gamma_\mu \gamma_5 \sigma_z \rlap/p_\para] &=& 4 \varepsilon_{\mu 1 2 \alpha
_\para} p^{\alpha_\para}\,,\\
\label{3rd step}
\Tr[\gamma_\mu \gamma_5 \sigma_z \rlap/p'_\para] &=& 4 \varepsilon_{\mu 1 2 \alpha
_\para} p'^{\alpha_\para}\,,\\
\label{4th step}
\Tr[\gamma_\mu \gamma_\nu \gamma_5 \sigma_z]&=&- \Tr[\gamma_\nu \gamma_\mu \gamma_5 
\sigma_z]\,,\nonumber\\
&=& - 4\varepsilon_{\mu \nu 1 2}\,.
\label{5th step}
\end{eqnarray}
Next comes the trace of a part of $T^{(o)}_{\mu \nu}$, and it is 
\begin{eqnarray}
& &\Tr\left[\gamma_\mu \gamma_5 \sigma_z \rlap/p_\para \gamma_\nu \rlap/p'_\pr 
\tan(e\Be s)\sec^2(e\Be s') + \gamma_\mu \gamma_5 \rlap/p_\pr \gamma_\nu 
\sigma_z \rlap/p'_\para \tan(e\Be s')\sec^2(e\Be s) \right]\nonumber\\
&=& 4g_{\mu \alpha_\para}\left[p^{\widetilde \alpha_\para}
p'_{\nu_\pr} \sec^2(e\Be s')\tan(e\Be s) + p'^{\widetilde \alpha_\para} 
p_{\nu_\pr} \sec^2(e\Be s)\tan(e\Be s')\right]\nonumber\\
&+& 4g_{\nu \alpha_\para} \left[p^{\widetilde \alpha_\para}
p'_{\mu_\pr} \sec^2(e\Be s')\tan(e\Be s) + p^{\widetilde \alpha_\para} 
p_{\mu_\pr} \sec^2(e\Be s)\tan(e\Be s')\right]\nonumber\\
&+& 4g_{\nu \alpha_\para} k^{\widetilde \alpha_\para}p_{\mu_\pr}
\sec^2(e\Be s)\tan(e\Be s')\,.
\label{6th step}
\end{eqnarray}
Now using Eq.~(\ref{pperpint}) and Eq.~(\ref{primeperpint}) from appendix
\ref{muis}, the above trace can be rewritten as,
\begin{eqnarray}
& &\Tr\left[\gamma_\mu \gamma_5 \sigma_z \rlap/p_\para \gamma_\nu \rlap/p'_\pr 
\tan(e\Be s)\sec^2(e\Be s') + \gamma_\mu \gamma_5 \rlap/p_\pr \gamma_\nu 
\sigma_z \rlap/p'_\para \tan(e\Be s')\sec^2(e\Be s) \right]\nonumber\\
&=& 4(g_{\mu \alpha_\para}p^{\widetilde \alpha_\para}k_{\nu_\pr}
+ g_{\nu \alpha_\para}p^{\widetilde \alpha_\para}k_{\mu_\pr})
\{\tan(e\Be s) - \tan(e\Be s')\}\nonumber\\
&-& 4g_{\mu \alpha_\para}k_{\nu_\pr}
k^{\widetilde \alpha_\para}
\frac{\sec^2(e\Be s)\tan^2(e\Be s')}
{\tan(e\Be s) + \tan(e \Be s')}
+ 4g_{\nu \alpha_\para}  k^{\widetilde \alpha_\para}p_{\mu_\pr}
\sec^2(e\Be s)\tan(e\Be s')\,.
\label{7th step}
\end{eqnarray}
Combining the previous steps, the trace of $T^{(o)}_{\mu \nu}$ can be
written as,
\begin{eqnarray}
\Tr[T^{(o)}_{\mu \nu}] &=& 4i\left[\varepsilon_{\mu \nu 1 2}
\{m^2 - p^2_\para - (p\cdot k)_\para\}\{\tan(e\Be s) + 
\tan(e\Be s')\}\right.\nonumber\\
&+&\left. 2\varepsilon_{\mu 1 2 \alpha_\para}\{p'_{\nu_\para}
p^{\alpha_\para} \tan(e\Be s) + p_{\nu_\para}p'^{\alpha_\para}
\tan(e\Be s')\}\right.\nonumber\\
&+&\left. g_{\mu \nu} (p^0 k^3 - p^3 k^0)\{\tan(e\Be s) - 
\tan(e\Be s')\}\right.\nonumber\\
&+&\left.p^{\widetilde \alpha_\para}(g_{\mu \alpha_\para} 
k_{\nu_\pr} + g_{\nu \alpha_\para} k_{\mu_\pr})
\{\tan(e\Be s) - \tan(e\Be s')\}\right.\nonumber\\
&+&\left. g_{\mu \alpha_\para} k^{\widetilde \alpha_\para}
k_{\nu_\pr}\frac{\sec^2(e\Be s)\tan^2(e\Be s')}
{\tan(e\Be s) + \tan(e \Be s')} + g_{\nu \alpha_\para}  k^{\widetilde
\alpha_\para}p_{\mu_\pr} 
\sec^2(e\Be s)\tan(e\Be s')\right]\,.\nonumber\\ 
\label{8th step}
\end{eqnarray}
Now we modify the term $(m^2 - p^2_\para)\{\tan(e\Be s) + \tan(e\Be
s')\}$ in the first line of the right hand side of the above
equation. As the trace of $T^{(o)}_{\mu \nu}$ is related to
$\mbox{R}^{(o)}_{\mu\nu}$ by Eq.~(\ref{deftmunu}), and as
$\mbox{R}^{(o)}_{\mu\nu}$ is inside the momentum integrals in
Eq.~(\ref{compl}) in chapter \ref{chap4} so we can use the operations
under the integral signs. The operations were explicitly written down
in appendix \ref{muis}. Using Eq.~(\ref{msqr}) we can write,
\begin{eqnarray}
(m^2 - p^2_\para)\{\tan(e\Be s) &+& \tan(e\Be s')\} e^{[\Phi(p,s)
+ \Phi(p',s')]}\nonumber\\
&\stackrel\circ= & \{tan(e\Be s) + \tan(e\Be s')\}
\left[i\frac{d}{ds} - \sec^2(e\Be s) p^2_\pr\right]
 e^{[\Phi(p,s)+ \Phi(p',s')]}\,,\nonumber
\end{eqnarray}
where $e^{[\Phi(p,s)+ \Phi(p',s')]}$ is the overall phase factor
appearing inside the momentum integrals in  Eq.~(\ref{compl}).
Using Eq.~(\ref{psq}) the above equation becomes,
\begin{eqnarray}
(m^2 - p^2_\para)\{\tan(e\Be s) &+& \tan(e\Be s')\} 
e^{[\Phi(p,s)+ \Phi(p',s')]}\nonumber\\
& \stackrel\circ= & \left[\{\tan(e\Be s) + \tan(e\Be s')\}(i\frac{d}{ds})
\right.\nonumber\\
&-&\left.\sec^2(e\Be s)\left\{-ieB + \frac{\tan^2(e\Be s')}
{\tan(e\Be s) + \tan(e\Be s')} k^2_\pr\right\}\right]
e^{[\Phi(p,s)+ \Phi(p',s')]}\,.\nonumber\\
\label{9th step}
\end{eqnarray}
The above equation is valid under the momentum integrals and the
parametric integrals in Eq.~(\ref{compl}), so we can do some of the
parametric integrals right now to modify Eq.~(\ref{9th step}). First
of all,
\begin{eqnarray}
\int_{-\infty}^\infty \tan(e\Be s')\frac{d}{ds} 
e^{[\Phi(p,s)+ \Phi(p',s')]}\,\,ds &=& \tan(e\Be s')
e^{\Phi(p',s')}\left[e^{\Phi(p,s)}
\right]_{-\infty}^\infty\,,\nonumber\\
&=& 0\,,
\label{10th step}
\end{eqnarray}
then,
\begin{eqnarray}
i\int_{-\infty}^\infty \tan(e\Be s)\frac{d}{ds} 
e^{[\Phi(p,s)+ \Phi(p',s')]}\,\,ds &=& -ie \Be \int_{-\infty}^\infty
\sec^2(e\Be s)  
e^{[\Phi(p,s)+ \Phi(p',s')]}\,\,ds\,,
\label{11th step}
\end{eqnarray}
and so combining the above steps we get from Eq.~(\ref{9th step}),
\begin{eqnarray}
(m^2 - p^2_\para)\{\tan(e\Be s) + \tan(e\Be s')\} 
e^{[\Phi(p,s)+ \Phi(p',s')]} &\stackrel\circ=&
-\frac{\sec^2(e\Be s)\tan^2(e\Be s')}{\tan(e\Be s) + \tan(e\Be s')}
k^2_\pr\nonumber\\ 
&\times& e^{[\Phi(p,s)+ \Phi(p',s')]}\,.
\label{12th step}
\end{eqnarray}
Now from Eq.~(\ref{8th step}) and Eq.~(\ref{12th step}), we can write
\begin{eqnarray}
\Tr[T^{(o)}_{\mu \nu}] &\stackrel{\circ}{=}&
4i\left[-\varepsilon_{\mu \nu 1 2} 
\left\{ \frac{\sec^2(e{\mathcal B}s)\tan^2(e{\mathcal
B}s')}{\tan(e{\mathcal B}s) 
+ \tan(e{\mathcal B}s')}
k_{\pr}^2 
+ (k\cdot p)_\para \{\tan(e{\mathcal B}s) +
\tan(e{\mathcal B}s')\}\right\}\right.\nonumber\\ 
&+&\left. 2\varepsilon_{\mu 1 2 \alpha_\para}(p'_{\nu_\para}
p^{\alpha_\para}\tan(e{\mathcal B}s) +
p_{\nu_\para}p'^{\alpha_\para}\tan(e{\mathcal B}s'))\right.\nonumber\\
&+&\left. g_{\mu\alpha_\para} k_{\nu\pr}\left\{p^{\widetilde
\alpha_\para}\{\tan(e{\mathcal B}s)
 - \tan(e{\mathcal B}s')\}
- k^{\widetilde \alpha_\para}\,
{\sec^2(e{\mathcal B}s)\tan^2(e{\mathcal B}s')\over{\tan(e{\mathcal B}s) +
\tan(e{\mathcal B}s')}}\right\}\right.\nonumber\\
&+&\left.\{g_{\mu\nu}(p\cdot \widetilde k)_\para + g_{\nu \alpha_\para}
p^{\widetilde \alpha_\para} k_{\mu\pr}\} 
\{\tan(e{\mathcal B}s) - \tan(e{\mathcal B}s')\}\right.\nonumber\\
&+&\left. g_{\nu \alpha_\para} 
k^{\widetilde\alpha_\para}p_{\mu\pr}\sec^2(e{\mathcal
B}s)\tan(e{\mathcal B}s')\right]. 
\label{13th step}
\end{eqnarray}
From the above equation and Eq.~(\ref{deftmunu}) we immediately see
that $\mbox{R}^{(o)}_{\mu\nu}$ matches with Eq.~(\ref{crmunu}) in
chapter \ref{chap4}.
\section{Derivation of the expression of $\Pi^5_{3 0}$}                
First of all it is shown why the other components of $\Pi^5_{\mu 0}$
except $\Pi^5_{3 0}$ turns out to be zero in the limit when the
external momentum goes to zero. First the contributions from 
$\Pi^{5(o)}_{\mu 0}$ are considered. From Eq.~(\ref{compl})
and Eq.~(\ref{crmunu}) we can write
\begin{eqnarray}
\Pi^{5(o)}_{0 0}&=&4e^2 \int {{d^4 p}\over {(2\pi)^4}}
\int_{-\infty}^\infty ds\, e^{\Phi(p,s)}\nonumber\\
&\times&\int_0^\infty
ds'e^{\Phi(p',s')}\eta_+(p_0)\left[2\varepsilon_{0 1 2 3}
\{p'_0 p^3 \tan(e\Be s) + p_0 p'^3 \tan(e\Be s')\}\right.
\nonumber\\
&+&\left. g_{0 0}(p\cdot \widetilde k)_\para 
\{\tan(e\Be s) - \tan(e\Be s')\}\right]\,.  
\end{eqnarray}
$e^{[\Phi(p,s) + \Phi(p',s')]}$ is an even function of $p_0$ when
$k_0=0$.  $\eta_+(p_0)$ inside the integrand is also an even function
of $p_0$. From these observations it can be said that when $k_0 = 0$
the integrand on the right hand side of the above equation becomes an
odd function of $p_0$ and as a result the integral vanishes.

Then comes $\Pi^{5(o)}_{1 0}$ and it is,
\begin{eqnarray}
\Pi^{5(o)}_{1 0}&=&-4e^2 \int {{d^4 p}\over {(2\pi)^4}}
\int_{-\infty}^\infty ds\, e^{\Phi(p,s)}\nonumber\\
&\times&\int_0^\infty
ds'e^{\Phi(p',s')}\eta_+(p_0)\left[p^3 k^1\{\tan(e\Be s) - 
\tan(e\Be s')\} + k^3 p^1 \sec^2(e\Be s)\tan(e\Be s')\right]\,.\nonumber\\
\end{eqnarray}
$\Pi^{5(o)}_{1 0}$ can exist only when the parametric integrals can
yield terms which cancels terms like $k^1$ or $k^3$. In the present
case as this does not happen so due to the presence of the terms like
$k^3$ and $k^1$, the integral vanishes in the limit $\vec{k}\to 0$.
Similarly $\Pi^{5(o)}_{2 0}$ will also be zero.

Now comes the contributions from $\Pi^{5(e)}_{\mu 0}$. From the
expression of $\mbox{R}^{(e)}_{\mu\nu}$ in Eq.~(\ref{reven1}) we can
immediately say that $\Pi^{5(e)}_{\mu_\para 0}$ must be
zero. 
If on the other hand $\mu=1$, then
\begin{eqnarray} 
\Pi^{5(e)}_{1 0}&=&4e^2 \int {{d^4 p}\over {(2\pi)^4}}
\int_{-\infty}^\infty ds\, e^{\Phi(p,s)}\nonumber\\
&\times&\int_0^\infty
ds'e^{\Phi(p',s')}\eta_+(p_0)\varepsilon_{1 0 \alpha \beta}\{
p^{\alpha_\para} p'^{\beta_\pr} \sec^2(e\Be s') + 
p^{\alpha_\pr} p'^{\beta_\para} \sec^2(e\Be s)\}\,, 
\end{eqnarray}
and the above equation shows that each term in the right hand side
inside the momentum integrals contain one perpendicular component of
loop momentum, and so it must integrate out to perpendicular
components of the external momentum $\vec{k}$. $\Pi^{5(e)}_{1 0}$ can
exist when the parametric integrals can produce terms which will
cancel the perpendicular components of $\vec{k}$. As in the present
case it does not happen as a result the integral vanishes when
$\vec{k} \to 0$. Similar result is obtained for $\Pi^{5(e)}_{2
0}$. Ultimately the only component of $\Pi^5_{\mu 0}$ which is non
vanishing in the specified external momentum limit is $\Pi^{5(o)}_{3
0}$, which is simply written as $\Pi^5_{30}$. 

Now the relevant steps are given by which we get Eq.~(\ref{charge1})
from Eq.~(\ref{pi5k0}). To begin with we explicitly write down the
sum of the functions $\Phi(p,s)$ and $\Phi(p',s')$ in terms of the
momentums $p$ and $p'$. 
\begin{eqnarray}
\Phi(p,s) + \Phi(p',s')&=& is p^2_\para + is'p'^2_\para - i(s+s')m^2
\nonumber\\
&-&\frac{i}{e\Be}\{\tan(e\Be s)p^2_\pr + \tan(e\Be s')p'^2_\pr\}
-\varepsilon|s| -\varepsilon|s'|\,.
\label{phasesum}
\end{eqnarray}

Now we concentrate on the integration of the perpendicular components
of the loop momentum in Eq.~(\ref{pi5k0}), and try to see what happens
when $|k_\pr|\to 0$. It must be noted that we have thrown away some of
the terms in deriving the expression of $\Pi^5_{\mu 0}$ by assuming
that $k_0=0$ and $\vec{k}\to 0$. The terms still existing in
Eq.~(\ref{pi5k0}) do not vanish in the above specified momentum limit.
First we integrate out the perpendicular components of momenta in the
integral appearing in Eq.~(\ref{pi5k0}). As the integrand in the right
hand side of Eq.~(\ref{pi5k0}) does not contain any term involving the
perpendicular momentum components, except the phase like terms, so
here the integration is straight forward.  From Eq.~(\ref{phasesum})
if we only take the terms containing the squares of the perpendicular
components of $p$ and $p'$, and try to integrate them out, using the
results given in appendix \ref{relint}, we will get
%
\begin{eqnarray}
& &\int\frac{d^2 p_\pr}{(2\pi)^2} \exp{\{-\frac{i}{e\Be}[\tan(e\Be s)p^2_\pr
 + \tan(e\Be s')p'^2_\pr]\}}\nonumber\\ 
&=&\exp{\Big\{-\frac{ik^2_\pr}{e\Be}\left[
\frac{\tan(e\Be s)\tan(e\Be s')}{\tan(e\Be s) + \tan(e\Be s')}\right]\Big\}}
\nonumber\\
&\times& \int\frac{d^2 p_\pr}{(2\pi)^2} \exp{\Big\{-\frac{i}{e\Be}
[\tan(e\Be s) + \tan(e\Be s')]\left[p_\pr + k_\pr 
\frac{\tan(e\Be s')}{\tan(e\Be s) + \tan(e\Be
s')}\right]^2\Big\}}\,,\nonumber\\   
\end{eqnarray}
and as $k^2_\pr \to 0$ the above equation simplifies to,
\begin{eqnarray}
\int\frac{d^2 p_\pr}{(2\pi)^2} \exp{\{-\frac{i}{e\Be}[\tan(e\Be s)p^2_\pr
 + \tan(e\Be s')p'^2_\pr]\}} = -\frac{1}{4\pi}\frac{ie\Be}{\tan(e\Be s)
+ \tan(e\Be s')}\,.
\end{eqnarray}
Using the above equation we get Eq.~(\ref{charge1}) from
Eq.~(\ref{pi5k0}).
\section{Effective charge calculation}
This section gives the details of the calculations by which the
expression of neutrino effective charge was calculated in chapter
\ref{chap4}.

The $s$ integral in Eq.~(\ref{charge1}) gives
\begin{eqnarray}
\int^{\infty}_{-\infty} ds\, e^{is(p^2_\para - m^2) - \varepsilon|s|} =
2\pi \delta(p^2_\para - m^2)
\label{delta}
\end{eqnarray}
and the $s'$ integral gives
\begin{eqnarray}
\int^{\infty}_0 ds'\, e^{is'(p'^{\,2}_\para - m^2) - \varepsilon|s'|} =
{i\over{(p'^{\,2}_\para - m^2) + i\varepsilon}}.
\label{divergent}
\end{eqnarray}
Using the above results in Eq.(\ref{charge1}) and using the delta
function constraint, we arrive at,
\begin{eqnarray}
\Pi^5_{3 0} &=&
\lim_{k_0=0,\vec{k}\rightarrow 0} 
- 2(e^3 {\mathcal B}) \int{d^2 p_\para \over{(2\pi)^2}}
{\delta(p^2_\para - m^2)}
\eta_+(p_0) 
\Bigg[{2 p^2_0\over{(k^{\,2}_\para +2(p.k)_\para)}}
 -  \frac{1}{2}\Bigg].
\label{charge2}
\end{eqnarray}
In deriving Eq.~(\ref{charge2}), pieces proportional to $k^2_\para$ in
the numerator were neglected. Now if one makes the substitution, $
p'_\para \to (p_\para + k_\para/2) $ and sets $k_0 =0$ one arrives at,
\begin{eqnarray}
\Pi^5_{3 0} 
&=&
\lim_{k_0=0,\vec{k}\rightarrow 0} 
2(e^3 {\mathcal B}) \int{d p_3 \over{(2\pi)^2}}
\left(n_+(E'_p)+ n_-(E'_p)\right) 
\Bigg[ \frac{E'_p}{p_3k_3}
 + \frac{1}{2E'_p}\Bigg].
\label{charge3}                
\end{eqnarray}
Here  $n_{\pm}(E'_p)$ are the functions $f_F(E'_p,\mu,\beta)$,
and $f_F(E'_p,-\mu,\beta)$, as given in Eq.(\ref{distrib}), which are
nothing but the Fermi-Dirac distribution functions of the particles
and the antiparticles in the medium with a modified energy $E'_p$.
The new term $E'_p$  is defined as follows,
\begin{eqnarray}
E'^2_p=[(p_3-k_3/2)]^2 + m^2, \nonumber
\end{eqnarray}
and it can be expanded for small external
momenta in the following way
\begin{eqnarray}
 E'^2_p\simeq p^2_3+m^2 - p_3 k_3 = E^2_p - p_3 k_3\nonumber
\end{eqnarray}
where $E^2_p = p^2_3+m^2$.
Noting, that
\begin{eqnarray}
E'_p= E_p -  {p_3 k_3 \over 2 E_p} + O(k^2_3),
\label{epprime}
\end{eqnarray}
one can use this expansion in Eq.~(\ref{charge3}), to arrive at:
\begin{eqnarray}
\Pi^5_{3 0} 
&=&
\lim_{k_0=0,\vec{k}\rightarrow 0} 
2(e^3 {\mathcal B}) \int{d p_3 \over{(2\pi)^2}}
\left(n_+(E'_p)+ n_-(E'_p)\right) 
\Bigg[ \frac{E_p}{p_3 k_3}
 \Bigg].
\label{charge5}
\end{eqnarray}
In the classical gas limit 
the expression for for 
$\eta_+(E'_p) = n_+(E'_p)+ n_-(E'_p)$ when
expanded in powers of the external momentum $k_3$ is given by
\begin{eqnarray}
\eta_+(E'_p) = ( 1 + \frac{1}{2} \frac{\beta p_3 k_3}{E_p}) \eta_+(E_p)\,,
\label{neweta}
\end{eqnarray}
up to first order terms in the external momentum $k_3$.

From Eq.~(\ref{neweta}) and Eq.~(\ref{charge5}) we get,
\begin{eqnarray}
\Pi^5_{3 0} 
&=&
\lim_{k_0=0,\vec{k}\rightarrow 0} 
2(e^3 {\mathcal B}) \int{d p_3 \over{(2\pi)^2}}
\left(\frac{E_p}{p_3 k_3} + \frac{\beta}{2}\right)\eta_+(E_p) .
\label{charge6}
\end{eqnarray}
The first term in the integral on the right hand side of the above
equation contains a factor of $p_3$ in the denominator and so is an odd
function of $p_3$. As a result the integral vanishes by
symmetry. But the second integral survives and $\Pi^5_{3 0}$ is given by,
\begin{eqnarray}
\Pi^5_{3 0} = \frac{e^3 {\mathcal B}}{
2\pi} \beta \int \frac{dp}{2\pi} \eta_+(E_p)\,.
\label{charge7}
\end{eqnarray}
In the classical limit:
\begin{eqnarray}
\Pi^5_{3 0} =  \frac{e^3 {\mathcal B}}{
2\pi^2} \beta \cosh(\beta\mu) \int dp \,\,e^{-\beta E_p}.
\end{eqnarray}
Using the standard result~\cite{Gradch},
\begin{eqnarray}
\int_0^{\infty}e^{-a\sqrt{b^2 + x^2}}\,\,\cos(cx)\,dx =
\frac{ab}{\sqrt{a^2 + c^2}} K_1(b\sqrt{a^2 + c^2})\,,
\end{eqnarray}
where $a$, $b$, $c$ are real constants here, and putting $c=0$, $\Pi^{5}_{3 0}$ becomes,
\begin{eqnarray}
\Pi^5_{3 0} =  \frac{e^3 {\mathcal B}}{
\pi^2} m\beta \cosh(\beta\mu) K_1(m\beta)\,.
\end{eqnarray}
$K_1(m\beta)$ is the modified Bessel function.

\chapter{Gauge invariance}
\label{gauin}
The axialvector-vector amplitude has an electromagnetic vertex and as
a result the electromagnetic current must be conserved in that vertex.
In our case the $\nu$-vertex is the electromagnetic vertex and the
current conservation condition is given in Eq.~(\ref{ngi}). 
The following sections explicitly verifies the current conservation
condition, which is also called the gauge invariance condition here.
\section{Gauge invariance for $\Pi^5_{\mu \nu}$ to even
orders in the external field}
The  axialvector-vector amplitude even in the external field is given by
\begin{eqnarray}
i\Pi^{5(e)}_{\mu \nu}&=& -e^2 \int {{d^4 p}\over {(2\pi)^4}}
\int_{-\infty}^\infty ds\, e^{\Phi(p,s)}  \int_0^\infty
ds'\,e^{\Phi(p',s')}\mbox{R}^{(e)}_{\mu\nu}(p,p',s,s')\,.
\label{a1}
\end{eqnarray}
Noting that, 
\begin{eqnarray}
q^{\alpha}p_{\alpha} = q^{\alpha_\para}p_{\alpha_\para} + 
q^{\alpha_\pr}p_{\alpha_\pr}\,,\nonumber
\end{eqnarray}
we can write Eq.(\ref{reven1}) as,  
\begin{eqnarray}
\mbox{R}^{(e)}_{\mu \nu}&\stackrel{\circ}{=}&
4i\eta_{-}(p_0)\left[(\varepsilon_{\mu \nu \alpha \beta}
p^{\alpha} p'^{\beta} - \varepsilon_{\mu \nu \alpha \beta_\pr}
p^{\alpha} p'^{\beta_\pr}
-\varepsilon_{\mu \nu \alpha_\pr \beta}
p^{\alpha_\pr} p'^{\beta})(1 + \tan(e{\cal B}s)\tan(e{\cal B}s'))\right.\nonumber\\
&+&\left. \varepsilon_{\mu \nu \alpha
\beta_\pr} p^{\alpha} p'^{\beta_\pr} \sec^2 (e{\cal
B}s')
+\varepsilon_{\mu \nu \alpha_\pr \beta} p^{\alpha_\pr} 
p'^{\beta} \sec^2 (e{\cal B}s)\right].
\label{a3}
\end{eqnarray}
Here throughout we have omitted terms such as $\varepsilon_{\mu \nu
\alpha_\pr \beta_\pr} p^{\alpha_\pr}  
p'^{\beta_\pr}$, since  by the application of Eq.~(\ref{pperpint}) we
have
\begin{eqnarray}
\varepsilon_{\mu \nu \alpha_\pr \beta_\pr} p^{\alpha_\pr} 
p'^{\beta_\pr} &=&\varepsilon_{\mu \nu \alpha_\pr \beta_\pr}
p^{\alpha_\pr} p^{\beta_\pr} + \varepsilon_{\mu \nu \alpha_\pr
\beta_\pr} p^{\alpha_\pr}  
k^{\beta_\pr} \nonumber\\
&\stackrel{\circ}{=}& - \frac{\tan(e{\cal B}s')}{\tan(e{\cal B}s')+\tan(e{\cal B}s')}
\varepsilon_{\mu \nu \alpha_\pr 
\beta_\pr} k^{\alpha_\pr}  
k^{\beta_\pr}\,,\nonumber
\end{eqnarray}
which is zero.

After rearranging the terms appearing in Eq.~(\ref{a3}), and by the
application of Eq.~(\ref{pperpint}) and Eq.~(\ref{primeperpint}) we
arrive at the expression
\begin{eqnarray}
\mbox{R}^{(e)}_{\mu
\nu}&\stackrel{\circ}{=}&4i\eta_{-}(p_0)\Bigg[\varepsilon_{\mu \nu 
\alpha \beta} 
p^{\alpha} k^{\beta}(1 + \tan(e{\cal B}s)\tan(e{\cal B}s'))\nonumber\\
&+& \varepsilon_{\mu \nu \alpha
\beta_\pr} k^{\alpha} k^{\beta_\pr} 
\tan(e{\cal B}s) \tan(e{\cal B}s') \frac{\tan(e{\cal B}s)-\tan(e{\cal
B}s')}{\tan(e{\cal B}s) + \tan(e{\cal B}s')}\Bigg]. 
\label{evenpart}
\end{eqnarray}
Because of the presence of terms like $\varepsilon_{\mu \nu
\alpha \beta} 
 k^{\beta}$ and $ \varepsilon_{\mu \nu \alpha
\beta_\pr} k^{\alpha} $ if we contract $\mbox{R}^{(e)}_{\mu \nu}$  by
$k^\nu$, it vanishes.
\section{Gauge invariance for $\Pi^5_{\mu \nu}$ to odd
orders in the external field}
The  axialvector-vector amplitude odd in the external field is given by
\begin{eqnarray}
i\Pi^{5(o)}_{\mu \nu}&=& -e^2 \int {{d^4 p}\over {(2\pi)^4}}
\int_{-\infty}^\infty ds\, e^{\Phi(p,s)} \int_0^\infty
ds'\,e^{\Phi(p',s')}\mbox{R}^{(o)}_{\mu\nu}(p,p',s,s')\,,
\end{eqnarray}
where $\mbox{R}^{(o)}_{\mu\nu}(p,p',s,s')$ is given by Eq.~(\ref{crmunu}).
The general gauge invariance condition in this case
\begin{eqnarray}
k^{\nu} \Pi^{5(o)}_{\mu \nu} &=& 0\,, 
\end{eqnarray}
can always be written down in terms of the following two equations,
\begin{eqnarray}
k^{\nu} \Pi^{5(o)}_{\mu_\para \nu} &=& 0\,, \label{gipara}\\
k^{\nu} \Pi^{5(o)}_{\mu_\pr \nu} &=& 0\,,
\label{gipr}
\end{eqnarray}
where $ \Pi^{5(o)}_{\mu_\para \nu}$  is that part of $
\Pi^{5(o)}_{\mu \nu}$ where the index $\mu$ can take the values $0$ and 
$3$ only. Similarly 
 $ \Pi^{5(o)}_{\mu_\pr \nu}$ stands for the part of $
\Pi^{5(o)}_{\mu \nu}$ where  $\mu$  can take the values $1$ and $2$
only.  $ \Pi^{5(o)}_{\mu_\para \nu}$ contains $\mbox{R}^{(o)}_{\mu_\para
\nu}(p,p',s,s')$ which from Eq.~(\ref{crmunu}) is as follows,
\begin{eqnarray}
\mbox{R}^{(o)}_{\mu_\para \nu}&\stackrel{\circ}{=}&
4i\eta_+(p_0)
\left[
-\varepsilon_{\mu_\para \nu 1 2} 
\left\{ \frac{\sec^2(e{\cal B}s)\tan^2(e{\cal B}s')}{\tan(e{\cal B}s)
+ \tan(e{\cal B}s')} 
k_{\pr}^2 + (k\cdot p)_\para (\tan(e{\cal B}s) +
\tan(e{\cal B}s'))\right\}\right.\nonumber\\ 
&+&\left. 2\varepsilon_{\mu_\para 1 2 \alpha_\para}\,(p'_{\nu_\para}
p^{\alpha_\para}\tan(e{\cal B}s) +
p_{\nu_\para}p'^{\alpha_\para}\tan(e{\cal B}s'))\right.\nonumber\\
&+&\left. g_{\mu_\para \alpha_\para} k_{\nu_\pr}\left\{p^{\widetilde
\alpha_\para}(\tan(e{\cal B}s)
 - \tan(e{\cal B}s')) - k^{\widetilde \alpha_\para}\,
{\sec^2(e{\cal B}s)\tan^2(e{\cal B}s')\over{\tan(e{\cal B}s) +
\tan(e{\cal B}s')}}\right\}\right.\nonumber\\
&+&\left. g_{\mu_\para\nu}(p\cdot \widetilde k)_\para(\tan(e{\cal B}s) -
\tan(e{\cal B} s'))
\right]\,,
\label{para}
\end{eqnarray}
and $ \Pi^{5(o)}_{\mu_\pr \nu}$ contains
$\mbox{R}^{(o)}_{\mu_\pr \nu}(p,p',s,s')$  which is 
\begin{eqnarray}
{\mbox R}^{(o)}_{\mu_\pr
\nu}&\stackrel{\circ}{=}&4i\eta_+(p_0)\left[\{g_{\mu_\pr \nu}(p\cdot
\widetilde k)_\para + g_{\nu \alpha_\para} p^{\widetilde \alpha_\para}
k_{\mu_\pr}\}
(\tan(e{\cal B}s) - \tan(e{\cal B}s'))\right.\nonumber\\ 
&+& \left.g_{\nu \alpha_\para}
k^{\widetilde\alpha_\para}p_{\mu_\pr}\sec^2(e{\cal B}s)\tan(e{\cal
B}s')\right]\,.  
\label{perp}
\end{eqnarray}
Eq.~(\ref{gipara}) and Eq.~(\ref{gipr}) implies one should have the
following relations satisfied,
\begin{eqnarray}
k^{\nu} \int {{d^4 p}\over {(2\pi)^4}} 
\int_{-\infty}^\infty ds\, e^{\Phi(p,s)} \int_0^\infty
ds'\,e^{\Phi(p',s')} \, 
{\mbox R}^{(o)}_{\mu_\pr \nu}=0\,,
\nonumber \\
\label{gipr1}
\end{eqnarray}
and
\begin{eqnarray}
k^{\nu}\int {{d^4 p}\over {(2\pi)^4}} 
\int_{-\infty}^\infty ds\, e^{\Phi(p,s)} \int_0^\infty
ds'\,e^{\Phi(p',s')}\, {\mbox R}^{(o)}_{\mu_\para \nu}=0.
\label{gipara1}
\end{eqnarray}
Out of the two above equations,  Eq.~(\ref{gipr1}) can be verified 
easily since
\begin{eqnarray}
k^{\nu}{\mbox R}_{\mu_\pr \nu}=0.
\end{eqnarray}

Now we look at Eq.~(\ref{gipara1}). 
We explicitly consider the case $\mu_{\parallel}=3$ (the
$\mu_{\parallel}=0$ case lead to similar result). For
$\mu_{\parallel}=3$
\begin{eqnarray}
k^{\nu}{\mbox R}^{(o)}_{3 \nu}&\stackrel{\circ}{=}&-p_0\left[(p'^{\,2}_\para -
p^2_\para)(\tan(e{\cal B}s) + \tan(e{\cal B}s')) 
- k^2_\pr(\tan(e{\cal B}s) - \tan(e{\cal B}s'))\right](4i
\eta_+(p_0)).\nonumber\\ 
\label{a_1}
\end{eqnarray}

Apart from the small convergence factors, 
\begin{eqnarray}
{i \over e{\cal B}}\left(\Phi(p,s) + \Phi(p',s')\right)
&=&\left( p_\parallel^{\prime2} + p_\parallel^2 - 2m^2 \right) \xi 
- \left(p_\parallel^{\prime2} - p_\parallel^2 \right) \zeta
- p_\pr^{\prime2} \tan (\xi-\zeta)\nonumber\\ 
&-& p_\pr^2 \tan (\xi+\zeta) \,,
\label{newa}
\end{eqnarray}
where we have defined the parameters
\begin{eqnarray}
\xi &=& \frac12 e{\cal B}(s+s') \,, \nonumber\\*
\zeta &=& \frac12 e{\cal B}(s-s') \,.
\label{xizeta}
\end{eqnarray}
From the last two equations we can write
\begin{eqnarray}
{ie{\cal B}} \; {d\over d\zeta} e^{\Phi(p,s) + \Phi(p',s')} 
&=& e^{\Phi(p,s) + \Phi(p',s')}
\left(p_\parallel^{\prime2} - p_\parallel^2 - p_\pr^{\prime2} \sec^2
(\xi-\zeta) + p_\pr^2 \sec^2 (\xi+\zeta) \right) \,,\nonumber\\ 
\label{C1par}
\end{eqnarray}
which implies
\begin{eqnarray}
p'^{\,2}_\para - p^2_\para \stackrel\circ=
ie{\cal B}{d\over{d\zeta}} + 
\left[ p'^{\,2}_\pr
\sec^2(e{\cal B}s') - p^2_\pr \sec^2(e{\cal B}s)\right].       
\label{a_parsqrdiff}
\end{eqnarray}
The equation above is valid in the sense that both sides of it
actually acts upon $e^{\widetilde \Phi(p,s,p',s')}$ inside the
momentum integrals, where
\begin{eqnarray}
\widetilde \Phi(p,p',s,s') = \Phi(p,s) + \Phi(p',s').
\label{a_Phi}
\end{eqnarray}
From Eq.~(\ref{a_1}) and Eq.~(\ref{a_parsqrdiff}) we have
\begin{eqnarray}
k^{\nu} \, {\mbox R}_{3 \,\nu}
&\stackrel{\circ}{=}&-4i\eta_+(p_0)p_0\left[(p'^{\,2}_\pr
\sec^2(e{\cal B}s') - p^2_\pr 
\sec^2(e{\cal B}s)) (\tan(e{\cal B}s)+\tan(e{\cal
B}s'))\right.\nonumber\\ 
&-&\left. k^2_\pr(\tan(e{\cal B}s)-\tan(e{\cal B}s'))
+ie{\cal B}(\tan(e{\cal B}s) + \tan(e{\cal B}s'))\,{d
\over{d \zeta}}\right]\,.\nonumber\\ 
\label{a_kdotR1}
\end{eqnarray}
Now using the the expressions for $p^2_\pr$ and $p'^2_\pr$ from
Eq.~(\ref{psq}) and Eq.~(\ref{p'sq}) we can write
\begin{eqnarray}
k^{\nu} \, {\mbox R}_{3 \,\nu}
&\stackrel{\circ}{=}&4e{\cal B}\eta_+(p_0) 
p_0\left[(\sec^2(e{\cal B}s) - \sec^2(e{\cal B}s'))\right.\nonumber\\
&+&\left. (\tan(e{\cal B}s) + \tan(e{\cal B}s')){d
\over{d \zeta}}\right]\,, \nonumber\\
\label{a_kdotR2}
\end{eqnarray}
Explicitly writing $e^{\widetilde\Phi}$ in its proper place, the above
equation can also be written as
\begin{eqnarray}
k^{\nu} \, {\mbox R}_{3 \,\nu}\, e^{\widetilde
\Phi}
= 4e{\cal B} \eta_+(p_0)p_0 
{d \over{d \zeta}} \left[ e^{\widetilde \Phi} (\tan(e{\cal B}s) +
\tan(e{\cal B}s'))\right].
\end{eqnarray}
Transforming to $\xi, \zeta$ variables and using the above
equation we can write the parametric integrations (integrations over
$s$ and $s'$) on the left hand side of Eq.~(\ref{gipara1}) as
\begin{eqnarray}
\int_{-\infty}^{\infty} ds \int_0^{\infty} ds' k^{\nu} \, {\mbox R}_{3\,
\nu}\, e^{\widetilde \Phi}
=\frac{8\eta_+(p_0)p_0}{e{\cal B}}\int_{-\infty}^{\infty} d\xi
\int_{-\infty}^{\infty} d\zeta \Theta 
(\xi - \zeta) {d\over{d\zeta}}{\cal F}(\xi,\zeta)
\label{odd}
\end{eqnarray}
where
\begin{eqnarray}
{\cal F}(\xi,\zeta)=e^{\widetilde \Phi} (\tan(e{\cal B}s) +
\tan(e{\cal B}s'))\,. 
\end{eqnarray}
The integration over the $\xi$ and $\zeta$ variables in Eq.~(\ref{odd}) 
can be represented as,
\begin{eqnarray}
& &\int_{-\infty}^{\infty} d\xi \int_{-\infty}^{\infty} d\zeta \Theta
(\xi - \zeta) {d\over{d\zeta}}{\cal
F}(\xi,\zeta)\nonumber\\
&=&\int_{-\infty}^{\infty} d\zeta \int_{-\infty}^{\infty}
d\zeta \left[{d\over{d\zeta}}\{ \Theta(\xi - \zeta) {\cal F}(\xi,\zeta)\}
- \delta(\xi - \zeta) {\cal F}(\xi,\zeta)\right]\nonumber\\
&=&-\int_{-\infty}^{\infty} d\xi {\cal F}(\xi,\xi)
\label{end1}
\end{eqnarray}
here the second step follows from the first one as the first integrand
containing the $\Theta$ function vanishes at both limits of the
integration. The remaining integral is now only a function of $\xi$
and is even in $p_0$. But in Eq.~(\ref{odd}) we have $\eta_+(p_0)p_0$
sitting, which makes the the integrand odd under $p_0$ integration in
the left hand side of Eq.~(\ref{gipara1}), as $\eta_+(p_0)$ is an even
function in $p_0$. So the $p_0$ integral as it occurs in the left hand
side of Eq.~(\ref{gipara1}) vanishes as expected, yielding the
required result shown in Eq.~(\ref{gipara}).


\plist
\label{plist}

\begin{enumerate}
\item  K.~Bhattacharya and P.~B.~Pal, {\bf Neutrino Scattering in 
       Strong Magnetic Fields}, conference proceedings, 
       COSMO 99: 3rd International Conference on Particle Physics
       and The Early Universe. Trieste, Italy 27th September-3rd
       October 1999 [arXiv: hep-ph/0001077].

\item  K.~Bhattacharya, A.~K.~Ganguly and S.~Konar, {\bf Effective 
       Neutrino Photon Interaction in a Magnetized Medium},
       Physical Review D, {\bf 65}, 013007, 2002 
       [arXiv: hep-ph/0107259].

\item  K.~Bhattacharya and A.~K.~Ganguly, {\bf The Axialvector-vector 
       Amplitude and Neutrino Effective Charge in a Magnetized Medium},
       Physical Review D, {\bf 68}, 053011, 2003
       [arXiv: hep-ph/0308063].

\item  K.~Bhattacharya and P.~B.~Pal, {\bf Inverse Beta Decay of 
       Arbitrarily Polarized Neutrons in a
       Magnetic Field}, Pramana, {\bf 62}, 1041, 2004 
       [arXiv: hep-ph/0209053]. 

\item  K.~Bhattacharya and P.~B.~Pal, {\bf Neutrinos and Magnetic 
       Fields: A Short Review}, to be published in the special issue of 
       the Indian National Science Academy (INSA) proceedings on 
       Neutrino Physics 
       [arXiv: hep-ph/0212118].
 


\end{enumerate}

\end{document}